\documentclass[fleqn,usenatbib]{mnras}

\usepackage{newtxtext,newtxmath}

\usepackage[T1]{fontenc}

\DeclareRobustCommand{\VAN}[3]{#2}
\let\VANthebibliography\thebibliography
\def\thebibliography{\DeclareRobustCommand{\VAN}[3]{##3}\VANthebibliography}

\usepackage{graphicx}	
\usepackage{amsmath}
\usepackage{ulem}

\usepackage{orcidlink}
\usepackage{multirow}
\usepackage[capitalize]{cleveref}
\usepackage{xspace}
\usepackage{subcaption} 
\usepackage{booktabs}
\usepackage{stfloats}

\newcommand{\healpix}{\texttt{HEALPix}\xspace}

\newcommand{\hbtherons}{\texttt{HBT-HERONS}\xspace}

\newcommand{\borg}{\texttt{BORG}\xspace}
\newcommand{\cola}{\texttt{COLA}\xspace}
\newcommand{\slow}{\texttt{SLOW}\xspace}
\newcommand{\swift}{\texttt{SWIFT}\xspace}
\newcommand{\tmpp}{\texttt{2M++}\xspace}
\newcommand{\flamingo}{\texttt{FLAMINGO}\xspace}
\newcommand{\lcdm}{$\Lambda$CDM\xspace}

\newcommand{\msolh}{$h^{-1}$~M$_\odot$\xspace}
\newcommand{\mpch}{$h^{-1}$~Mpc\xspace}
\newcommand{\hmpc}{$h$~Mpc$^{-1}$\xspace}
\newcommand{\dquotes}[1]{``#1''}
\newcommand{\squotes}[1]{`#1'}
\newcommand{\masscrit}{$M_{200\mathrm{c}}$\xspace}

\newcommand{\tentoninety}{$10^{\mathrm{th}}-90^{\mathrm{th}}$\xspace}
\newcommand{\manticore}{\texttt{Manticore}\xspace}
\newcommand{\manticorelocal}{\texttt{Manticore-Local}\xspace}

\newcommand{\manticoredeep}{\texttt{Manticore-Deep}\xspace}

\newcommand{\pvalue}{$p$-value\xspace}
\newcommand{\numrealisations}{15\xspace}

\newcommand{\randomlcdm}{\texttt{Random-$\Lambda$CDM}\xspace}
\newcommand{\ncontrolsims}{10\xspace}
\newcommand{\lensingsnr}{$7.4\,\sigma$\xspace}
\newcommand{\lensingsnrfilled}{$11.2\,\sigma$\xspace}

\Crefname{figure}{Figure}{Figures}
\Crefname{equation}{Equation}{Equations}
\Crefname{table}{Table}{Tables}
\Crefname{section}{Section}{Sections}

\title[\manticoredeep]{The \manticore\ Project II: Bayesian digital twins of cosmic structure across the SDSS and BOSS volumes}

\author[S. McAlpine et al.]{
Stuart McAlpine\,\textsuperscript{\orcidlink{0000-0002-8286-7809}},$^{\,1}$\thanks{E-mail: \url{stuart.mcalpine@fysik.su.se}}
Jens Jasche,$^{\,1}$
Guilhem Lavaux,$^{\,2}$
Ludvig Doeser,$^{\,1,3}$
and Arthur Loureiro$^{\,1,4}$
\\
$^{1}$The Oskar Klein Centre, Department of Physics, Stockholm University, Albanova University Center, 106 91 Stockholm, Sweden\\
$^{2}$CNRS \& Sorbonne Université, Institut d'Astrophysique de Paris (IAP), UMR 7095, 98 bis bd Arago, F-75014 Paris, France\\
$^{3}$Center for Computational Astrophysics, Flatiron Institute, 162 5th Avenue, New York, NY 10010, USA
\\
$^{4}$Astrophysics Group, Blackett Laboratory, Imperial College London, London SW7 2AZ, UK
}

\date{Accepted XXX. Received YYY; in original form ZZZ}

\pubyear{2026}

\begin{document}
\label{firstpage}
\pagerange{\pageref{firstpage}--\pageref{lastpage}}
\maketitle

\begin{abstract}
We present \manticoredeep, a high-resolution Bayesian field-level reconstruction of cosmic large-scale structure over a comoving volume of $(4~h^{-1}\mathrm{Gpc})^{3}$ to $z\approx0.7$ at ${\sim}4$~\mpch resolution. Extending the companion \manticorelocal analysis (Paper~I), \manticoredeep jointly constrains five galaxy redshift surveys---2M++, 6dFGS, 2dFGRS, SDSS, and BOSS---within a single hierarchical Bayesian framework using the \borg algorithm. The inference reconstructs primordial initial conditions evolved under gravity, yielding a posterior ensemble of three-dimensional density and velocity fields that causally reproduce the observed large-scale structure. A novel tiled inference strategy extends the reconstructed volume by more than an order of magnitude beyond Paper~I. Posterior realisations are consistent with \lcdm, reproducing Gaussian isotropic initial conditions and the expected $z=0$ matter power spectrum, bispectrum, and halo mass function over the resolved scales. We validate the reconstruction using two independent template-free posterior-predictive tests against observations excluded from the inference. Cross-correlation with the \textit{Planck} PR3 CMB lensing map yields a cumulative detection significance of \lensingsnr, while velocity-weighted stacking of $64{,}750$ galaxy clusters on the \textit{Planck} 217~GHz map detects the kinetic Sunyaev--Zel'dovich effect at $3.5\sigma$, with a model-independent approach--recession split confirming the inferred velocities. Together, these tests validate both the projected-density and three-dimensional velocity fields recovered by \manticoredeep. The BOSS Great Wall is recovered as a ${\sim}3\sigma$ overdensity consistent with \lcdm across the posterior ensemble. \manticoredeep establishes a benchmark for survey-depth constrained cosmological digital twins and reproducible field-level validation of large-scale structure reconstructions.
\end{abstract}




\section{Introduction}

The large-scale structure (LSS) of the Universe—the cosmic web of galaxies, clusters, filaments, and voids—is one of the richest observational probes of cosmology and structure formation \citep[e.g.][]{1980lssu.book.....P, 2011ARA&A..49..409A}. Yet most analyses of the galaxy distribution rely on low-order summary statistics, principally the two-point correlation function or power spectrum, which capture only Gaussian information and discard the higher-order correlations and phase structure encoded in the evolved matter field \citep[e.g.][]{2024arXiv240502252B, 2025arXiv250716590C}. The cost of this compression is substantial: both theoretical work \citep{2019A&A...621A..69R, 2021JCAP...11..049M,2021MNRAS.506L..85L,2021MNRAS.502.3035P,2023MNRAS.520.5746A,2024PhRvL.133v1006N} and empirical analyses exploiting non-Gaussian galaxy statistics \citep{2024NatAs...8.1457H} have demonstrated significantly tighter cosmological constraints than those obtainable from the power spectrum alone.

Field-level inference offers a fundamentally different approach: rather than compressing observations into summary statistics, it seeks to recover the full three-dimensional density and velocity fields that gave rise to the observed galaxy distribution. The primary object of inference is the initial density field, modelled as a Gaussian random field in a \lcdm cosmology. A forward model encoding gravitational evolution, galaxy bias, and observational effects maps these primordial fluctuations to present-day observables, and the posterior distribution over both initial and evolved fields is sampled directly. The result is a set of physically self-consistent reconstructions of cosmic structure with rigorously propagated uncertainties \citep[e.g.][]{Jasche2013,Lavaux2016,Jasche2019}.

Over the past decade, Bayesian field-level analyses have moved from proof of concept to mature applications on galaxy survey data. Early applications of the \textit{Bayesian Origin Reconstruction from Galaxies} (\borg) algorithm \citep{Jasche2015,Jasche2019,2019arXiv190906396L} demonstrated that initial conditions could be constrained from realistic survey data, and subsequent work by the Aquila Consortium\footnote{\hyperlink{https://aquila-consortium.org/}{https://aquila-consortium.org/}} has progressively incorporated more accurate gravity solvers \citep{Jasche2019,Stopyra2024_COLA}, hierarchical galaxy bias models, and treatments of survey systematics \citep[e.g.][]{Lavaux2016,Jasche2019,McAlpine2025}. Applied to the \tmpp all-sky catalogue \citep{Lavaux2011} and the Sloan Digital Sky Survey III (SDSS-III) Baryon Oscillation Spectroscopic Survey (BOSS) spectroscopic sample \citep{2013AJ....145...10D}, these methods have recovered prominent clusters, filaments, and velocity flows across the nearby Universe \citep{Lavaux2016,2019arXiv190906396L,Jasche2019}. Complementary frameworks that use perturbative effective field theory as a forward model \citep[e.g.][]{2020JCAP...04..042C,2021JCAP...04..033S,2023JCAP...10..069S} have demonstrated the viability of field-level forward modelling in complementary dynamical regimes. Crucially, both these approaches do not yield a single deterministic map but a \textit{posterior ensemble} of physically consistent realizations, enabling uncertainties on individual structures to be quantified and providing a rigorous foundation for cosmological interpretation.

The nearby Universe ($\lesssim 200~h^{-1}\mathrm{Mpc}$), where dense, nearly all-sky redshift surveys and rich peculiar velocity compilations provide ideal data, has naturally become the primary testing ground for field-level inference. \borg-derived initial conditions have been resimulated at higher resolution to produce posterior ensemble suites, including \textsc{CSiBORG} \citep[e.g.][]{2021PhRvD.103b3523B,2022MNRAS.516.3592H} and \textsc{SIBELIUS} \citep[e.g.][]{Sawala2021b,2022MNRAS.512.5823M}, which propagate reconstruction uncertainty into the resolved halo population. The \manticorelocal inference \citep{McAlpine2025} established a new benchmark by delivering a high-resolution, fully Bayesian posterior ensemble of the nearby Universe. Subsequent studies using the \manticorelocal posterior have demonstrated its physical fidelity across a range of independent probes, including cluster masses and positions \citep{McAlpine2025,2025arXiv251016574M}, the peculiar velocity field \citep{Stiskalek2025,2026MNRAS.546f2260S}, the thermal Sunyaev--Zel'dovich signal \citep{2026arXiv260115935S}, and the cosmic void population \citep{2026A&A...705A.160M}. This extensive validation programme provides the foundation upon which the present work builds.

Complementary approaches based on constrained realizations rather than full posterior sampling have also targeted the local Universe. The \textit{Constrained Local UniversE Simulations} (\textsc{CLUES}) project \citep{Gottloeber2010} pioneered the use of Hoffman--Ribak constrained realizations \citep{Hoffman1991} tied to Cosmicflows peculiar velocity data, with subsequent extensions to targeted zoom simulations of the Virgo and Coma clusters \citep[\textsc{CLONES},][]{Sorce2021}. The \slow simulations \citep{Dolag2023,SLOW2024} develop this further, using Cosmicflows constraints to produce hydrodynamical realizations of the $500~h^{-1}\mathrm{Mpc}$ local region. While these efforts have yielded valuable insights into local structure, they typically produce single or small numbers of constrained realizations and thus offer limited uncertainty quantification. More recently, the \textsc{HamletPM} framework \citep{ValadeHAMLETPM} has moved toward full posterior sampling, using Cosmicflows-4 peculiar velocities as constraints within a Bayesian particle-mesh forward model of the local Universe.

Beyond the local ($\lesssim 200~h^{-1}\mathrm{Mpc}$) region, reconstruction techniques have been extended to the much larger volumes probed by SDSS. The \textit{Exploring the Local Universe with the Reconstructed Initial Density Field} (\textsc{ELUCID}) project \citep{Wang2014,Wang2016} reconstructs the initial density field across the SDSS DR7 footprint and evolves it with particle--mesh or $N$-body dynamics to produce constrained simulations on $\sim 500~h^{-1}\mathrm{Mpc}$ scales. While these reconstructions reproduce key features of the observed cosmic web \citep[e.g.][]{2022ApJ...936...11L}, they remain single-best-fit realizations without posterior uncertainty quantification. Bayesian field-level methods have also been applied at these scales: extensions of \borg to the BOSS and SDSS main samples \citep{Jasche2015,Jasche2019,2019arXiv190906396L} and the \textsc{COSMIC BIRTH} framework \citep{2021MNRAS.502.3456K} demonstrated that forward modelling can recover three-dimensional density fields and initial conditions across gigaparsec-scale volumes, yielding valuable cross-correlations with cosmic microwave background (CMB) lensing. In a companion analysis, \citet{2025arXiv260202363A} extend \borg to the Quaia quasar catalogue, delivering Bayesian field-level reconstructions across a $(10~h^{-1}\mathrm{Gpc})^{3}$ volume, albeit at coarser spatial resolution. However, in all these cases computational constraints have required coarse spatial resolution and approximate dynamics, limiting sensitivity to supercluster-scale structure. A Bayesian reconstruction at high spatial resolution across these volumes has therefore remained an open challenge.

In this work we present \manticoredeep, the first high-resolution, fully Bayesian field-level inference across the combined volumes of 2M++, the Six-degree Field Galaxy Survey (6dFGS), the Two-degree Field Galaxy Redshift Survey (2dFGRS), SDSS, and BOSS. The inference spans a comoving cube of $4096~h^{-1}\mathrm{Mpc}$ on a side and reaches redshifts of $z \approx 0.7$, extending \manticorelocal by more than an order of magnitude in volume. Building upon the modelling foundations established by \citet{McAlpine2025}, \manticoredeep incorporates heterogeneous multi-survey data into a single coherent Bayesian framework while retaining high spatial fidelity, with a voxel size of $\sim 4~h^{-1}\mathrm{Mpc}$ and a forward model capable of resolving cluster-scale structure throughout the SDSS footprint. The inference yields a full posterior ensemble over both initial conditions and evolved density fields, enabling probabilistic analyses of clusters, filaments, voids, and the velocity field across multi-gigaparsec scales with quantitatively controlled uncertainties.

This work forms part of the wider scientific programme of the Simons Collaboration on \textit{Learning the Universe} (LTU)\footnote{\hyperlink{https://learning-the-universe.org/}{https://learning-the-universe.org/}}, whose aim is to infer the initial conditions, cosmological parameters, and uncertain astrophysical processes governing structure formation through Bayesian forward modelling of the observable Universe. Within this programme, high-resolution field-level reconstructions play a foundational role by providing physically self-consistent posterior ensembles of initial conditions and evolved density fields, with rigorously quantified uncertainties, against which fast emulators, neural simulation models, and likelihood-free inference techniques can be developed and tested. Extending such reconstructions from the local volume to multi-gigaparsec scales is therefore an essential step toward LTU's broader goal of performing cosmological inference directly at the level of the full matter and galaxy fields.

The paper is laid out as follows. \Cref{sect:method} describes the inference framework, including the \manticore forward model and tiling strategy, the $N$-body resimulation of posterior initial conditions, and the methodology for CMB lensing cross-correlation and kinetic Sunyaev--Zel'dovich stacking. \Cref{sect:results} presents the results: statistical consistency tests against \lcdm, a visual tour of the reconstructed large-scale structure, cross-correlation with \textit{Planck} CMB lensing, detection of the kinetic Sunyaev--Zel'dovich effect, and a case study of the BOSS Great Wall. We discuss our findings in \cref{sect:discussion} and conclude in \cref{sect:summary}.

\section{method}
\label{sect:method}

\subsection{The \manticore model}
The \manticore model, introduced in \citet{McAlpine2025}, extends the \borg Bayesian field-level inference framework \citep{Jasche2013,Jasche2015,Jasche2019}, which jointly infers initial conditions and galaxy bias parameters from spectroscopic galaxy redshift surveys through forward modelling. A core aim of the \manticore extension is to impose physically motivated priors on the initial conditions, regularising the inference toward statistically homogeneous, isotropic, and Gaussian fields.
This is achieved through a suite of physics-informed priors that guide the posterior toward statistically consistent Gaussian white-noise realisations. These include Gaussianity- and power-spectrum–based consistency conditions on the initial field, as well as moment-based regularisation on the weighted mean, variance, and skewness of the final density field. Together, these priors confine the inference to the manifold of valid \lcdm initial conditions without imposing overly rigid constraints.

At each step of the inference, the forward model maps a candidate Gaussian white--noise field to a predicted galaxy distribution through the following chain:
\begin{enumerate}
    \item the white--noise field is modulated by the primordial power spectrum, computed using \textsc{CLASS} \citep{Lesgourgues2011}, and evolved to the target epoch using a non-linear \textsc{COLA}-based gravity solver \citep[][]{Tassev2013,Izard2016,Stopyra2024_COLA};
    \item particle positions are displaced into redshift space to account for redshift-space distortions;
    \item the resulting dark matter density field is transformed into an expected galaxy count field via a local non-linear bias model\footnote{The bias relation is applied to the redshift-space matter density rather than in the galaxy rest frame, following the standard \borg implementation in which the local bias model acts on the mesh density field constructed from particles already displaced into redshift space. At linear order this modifies the anisotropic (Kaiser) response; a more physical rest-frame treatment is given by \citet{2023JCAP...10..069S}.} and a three-dimensional survey response operator;
    \item the predicted counts are compared with the observed data through a likelihood function.
\end{enumerate}
Gradients of the resulting log-posterior with respect to the $O(10^{7})$ white--noise amplitudes are computed analytically, enabling Hamiltonian Monte Carlo (HMC) to efficiently explore the high-dimensional posterior. The overall sampling follows a Gibbs scheme that alternates between updating the initial conditions via HMC and updating the nuisance parameters via slice sampling \citep{Neal2000}, conditioned on the current density field realisation. The nuisance parameters comprise the galaxy bias and overdispersion coefficients for each galaxy subcatalogue---a subset of the input galaxies defined by specific luminosity, flux, and redshift cuts, allowing different tracer populations to be modelled independently.

The galaxy bias mapping employs a Sigmoid-Truncated Double Power-Law (STDP) model, consisting of a sigmoid suppression at low densities, a primary power-law at intermediate densities, and a secondary power-law at high densities. Each galaxy subcatalogue carries an independent set of five STDP parameters plus a mean galaxy count, and the bias function is normalized at every step to preserve the ergodic mean of the galaxy field. The likelihood adopts a Generalized Poisson distribution that introduces a density-dependent overdispersion parameter, modelled as a power-law in local dark matter density, to accommodate the super-Poisson variance observed in galaxy counts at high densities. Together, the STDP bias and Generalized Poisson likelihood provide a flexible yet physically motivated description of the galaxy--density connection. The survey response operator models angular completeness, spectroscopic selection, magnitude limits, and radial selection effects following the Aquila methodology \citep{Jasche2019}.

Combining these elements, the inference targets the posterior of the initial white-noise field $\mathbf{x}$ and the galaxy-bias (nuisance) parameters $\boldsymbol{\theta}$ given the observed galaxy counts $\mathbf{N}$,
\begin{equation}
P(\mathbf{x}, \boldsymbol{\theta} \mid \mathbf{N}) \;\propto\; P(\mathbf{N} \mid \mathbf{x}, \boldsymbol{\theta})\; P(\mathbf{x})\; P(\boldsymbol{\theta}),
\label{eq:posterior}
\end{equation}
where $P(\mathbf{N} \mid \mathbf{x}, \boldsymbol{\theta})$ is the Generalized Poisson likelihood evaluated through the forward model and STDP bias relation, $P(\mathbf{x})$ combines the unit-variance Gaussian white-noise prior with the \manticore power-spectrum and Gaussianity consistency priors and the moment-based regularisation of the final field, and $P(\boldsymbol{\theta})$ collects the priors on the bias and overdispersion parameters. Equivalently, \borg samples from $\exp(-H)$, where $H$ is the sum of the corresponding likelihood and prior energy terms. A complete mathematical specification of the priors, bias prescription, likelihood, and forward model is presented in Appendix~A of \citet{McAlpine2025}.

\subsection{Tiling strategy}
\label{sect:tiling}

In \manticorelocal, the inference operated within a single $681$~\mpch\ domain at $256^{3}$ resolution. \manticoredeep extends this to a substantially larger and deeper cosmological volume, incorporating data from multiple wide-field redshift surveys---2M++, 6dFGS, 2dFGRS, SDSS Main, and BOSS---spanning a combined footprint out to $z \approx 0.7$. Accommodating this dataset requires a parent domain of $4096$~\mpch, which, to maintain a quasi-non-linear resolution of $4$~\mpch in the inference grid, would necessitate a single $1024^{3}$-voxel inference. Table~\ref{tab:manticore_comparison} summarises the key differences between the two configurations.

\begin{table}
\centering
\caption{
Summary of the implementation differences between \manticorelocal and \manticoredeep, highlighting the changes required to extend field-level inference from the local Universe to a deep, cosmologically representative volume.}
\label{tab:manticore_comparison}
\begin{tabular}{lcc}
\hline
 & \manticorelocal & \manticoredeep \\
\hline
Parent volume & $681\,h^{-1}\mathrm{Mpc}$ & $4096\,h^{-1}\mathrm{Mpc}$ \\
Resolution & $2.66\,h^{-1}\mathrm{Mpc}$ & $4\,h^{-1}\mathrm{Mpc}$ \\
Grid size & $256^3$ & $1024^3$ (effective) \\
Inference strategy & Single domain & 64 tiles (27 with data) \\
Redshift range & $z \lesssim 0.1$ & $z < 0.7$ \\
Scale factor treatment & Constant ($a = 1$) & Variable ($a \approx 0.55$--$0.9$) \\
Input catalogues & 2M++ & 2M++, 6dFGS \\
 &  & 2dFGRS, SDSS, BOSS \\
\hline
\end{tabular}
\end{table}

\begin{figure}
    \centering
    \includegraphics[width=\columnwidth]{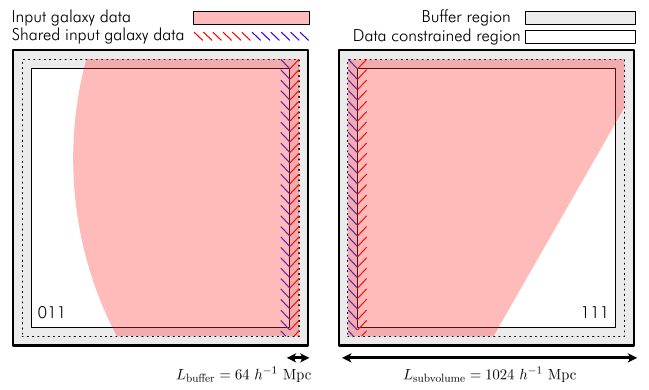}
    \caption{Schematic illustration of the tiling and buffer strategy, shown as a 2D cross-section through two adjacent subvolumes. Each subvolume has side length $L_{\mathrm{subvolume}} = 1024$~\mpch\ and includes a buffer of width 
    $L_{\mathrm{buffer}} = 64$~\mpch\ extending inward from all sides. The outer half of the buffer (32~\mpch) is excluded from the likelihood and is almost entirely driven by the $\Lambda$CDM prior, mitigating artefacts from periodic boundary conditions. The inner half (32~\mpch) marks the region where input galaxies are duplicated across neighbouring tiles; in this example, the red hatched region corresponds to galaxies shared between these two 
    adjacent subvolumes. After sampling, both buffer layers are removed, leaving a $224^{3}$ signal-carrying core per tile.}
    \label{fig:tiles_buffer}
\end{figure}

Inferring a full $4096$~\mpch cosmological volume at $4$~\mpch resolution would require a $1024^{3}$ grid, which is beyond the memory and runtime limits of current field-level inference methods. To render the problem tractable, we partition the parent domain into a $4 \times 4 \times 4$ set of 64 cubic \textit{tiles} (hereafter also \textit{subvolumes}), each with side length $L_{\mathrm{subvolume}} = 1024$~\mpch\ and inferred independently on a $256^{3}$ grid (corresponding to $4$~\mpch\ voxels). This strategy is conceptually related to spatial-COLA approaches for perfectly parallel cosmological simulations, which also use independently evolved buffered subvolumes to reduce memory and wall-clock costs \citep{2020A&A...639A..91L}; however, here the decomposition is applied to the Bayesian field-level inference itself rather than to a single forward simulation. The fiducial observer is positioned at $[1850,1800,1350]$~\mpch within the parent volume, chosen to centre the combined survey footprint within the tiled domain while ensuring that all data-containing tiles fall within the forward lightcone. The survey footprint occupies only a subset of this layout, and consequently only 27 tiles contain galaxy data; the remaining 37 tiles exist solely to ensure complete spatial coverage of the parent volume and are populated with random Gaussian phases drawn from the $\Lambda$CDM prior, resampled independently for each posterior realization.

Each tile includes a buffer of width $L_{\mathrm{buffer}} = 64$~\mpch\ (16 voxels) extending inward from all sides. This buffer plays two roles. Its outer half (8 voxels, or 32~\mpch) is excluded from the likelihood and is almost entirely driven by the $\Lambda$CDM prior,\footnote{Although the buffer region is excluded from the likelihood, these modes are not strictly prior draws: the non-local gravitational forward model weakly couples them to the data-constrained interior.} absorbing artefacts arising from the periodic boundary conditions applied within each subvolume. The inner half (8 voxels, or 32~\mpch) provides a small, controlled overlap in the input galaxy data between adjacent tiles: galaxies located near the edge of one tile also enter the likelihood of its neighbours (see \cref{fig:tiles_buffer}). This duplication prevents artificial discontinuities at tile boundaries. Although sharing galaxies across neighbouring tiles means the inferences are not strictly statistically independent, the duplicated galaxy-data layer occupies a modest shell (comprising $\sim\!15$ per cent of the tile volume) and is discarded after sampling, so in practice the correlation introduced between adjacent posterior chains is negligible.

We tested multiple buffer widths in lower-resolution ($8$~\mpch) pilot inferences and found that $L_{\mathrm{buffer}} = 64$~\mpch\ offered the best compromise between accuracy and computational efficiency: smaller buffers hindered the exploration of the inference, and in extreme cases produced visible edge artefacts, whereas larger buffers increased the number of required tiles, adding substantially to the computational cost.

After sampling, both buffer layers are removed from every tile, yielding a $224^{3}$ voxel core containing only the signal-carrying white--noise modes. These trimmed cores are inserted into their corresponding locations within a global $1024^{3}$ grid through direct concatenation. We validate that this concatenation does not introduce artificial power on the tiling scale through power spectrum analysis presented in \cref{sect:results}. Since $4 \times 224 = 896$ voxels span only $3584$~\mpch in each dimension rather than the full $4096$~\mpch, the remaining volume is filled with unconstrained white--noise realizations drawn from the \lcdm prior, resampled independently for each posterior sample. This procedure ensures that the entire $4096$~\mpch domain remains statistically consistent with \lcdm expectations, while the data-constrained regions preserve the inferred structure. All input galaxy data falls within this inner $(3584$~\mpch)$^3$ data-constrained region once the tiles are recombined, as can be seen visually in the survey coverage of \cref{fig:tiles_mean_density}. The resulting field constitutes a complete and statistically consistent initial--condition realisation spanning the full $4096$~\mpch parent volume, ready for subsequent $N$-body resimulation.

This tiling strategy limits the explicit representation of Fourier modes within any individual subvolume to wavelengths smaller than the subvolume scale ($L_{\mathrm{subvolume}} = 1024$~\mpch, corresponding to $k \gtrsim 0.006$~\hmpc). The constraint on the largest scales is further weakened by the inference setup: the per-tile nuisance parameters $\boldsymbol{\theta}_t$ (\cref{eq:tile_posterior}) include a freely sampled mean number density for each input galaxy sub-catalogue, and this amplitude is degenerate with the mean (monopole) density of the subvolume, which it largely absorbs. The very largest, super-tile modes are therefore expected to be prior-dominated rather than data-constrained, reverting toward draws from the $\Lambda$CDM prior. We emphasise that this represents a loss of \textit{constraint} rather than a source of \textit{bias}: because the prior is statistically unbiased, these modes introduce no systematic distortion of the recovered structure, even though they are largely decorrelated from the true realisation, a decorrelation that would be directly measurable in a simulation test with known ground truth.

This is not necessarily a sharp cutoff at the tile scale, however. The mean-density degeneracy is exact only for the per-tile monopole; modes with wavelengths moderately larger than a subvolume still imprint gradients and higher-order structure \textit{within} each tile, and because the galaxy data extend continuously across tile boundaries, some partial constraining power on large-scale coherence may survive as the tile scale is approached, even though such modes are not explicitly parameterised. We do not attempt to quantify this residual constraint here; a rigorous measurement would require the ground-truth simulation test noted above, which we leave to future work. In any case, the cosmological signal exploited in this work resides at sub-tile scales that remain well constrained by the data, while fluctuations genuinely larger than the subvolume are supplied by independent $\Lambda$CDM prior realisations when completing the parent volume. Consistent with this picture, validation through CMB cross-correlation tests and statistical comparisons of resimulated parent volumes reveals no clear evidence for substantial systematic bias (see \cref{sect:results}).

Future extensions could explore sCOLA-like boundary treatments, in which large-scale tidal fields are supplied analytically through Lagrangian perturbation theory and non-periodic boundary conditions \citep{2020A&A...639A..91L}, potentially reducing sensitivity to the finite tile size while preserving the parallel structure of the inference. Collectively, these tests support the robustness of the tiling approach for field-level inference on cosmological volumes of this scale.

\subsection{\textsc{BORG} inference setup}

We now describe the inference procedure applied within each of the 27 data-containing tiles. Each tile is inferred independently using the \borg framework under the \manticore model, following the configuration established in \manticorelocal. All subvolumes adopt the same fixed cosmological parameters as \manticorelocal, corresponding to the DES~Y3 `3×2pt + All Ext.' \lcdm cosmology \citep{DEScosmo}: $h = 0.681$, $\Omega_{\mathrm{m}} = 0.306$, $\Omega_{\Lambda} = 0.694$, $\Omega_{\mathrm{b}} = 0.0486$, $A_{\mathrm{s}} = 2.099\times10^{-9}$, and $n_{\mathrm{s}} = 0.967$, ensuring consistency with previous work and with the fiducial parameters of the \flamingo simulation suite \citep{Schaye2023}. Across all tiles, the total number of input galaxies is 1,664,630, with per-tile counts ranging from 2,670 to 336,454. We employ the standard Aquila implementation of the \borg selection-function machinery \citep{Jasche2019}, modelling angular completeness, spectroscopic selection, magnitude limits, and radial selection functions; these operators are applied independently in each tile using the appropriate masks and radial selection functions described in \cref{sect:input_galaxy_catalogs}. Each tile is also assigned its own set of free galaxy-bias parameters, inferred jointly with the initial conditions of that subvolume; these parameters are not shared across tiles, allowing for spatial variations in bias across the footprint. Formally, each tile $t$ is sampled from the posterior of \cref{eq:posterior} restricted to that subvolume,
\begin{equation}
P(\mathbf{x}_t, \boldsymbol{\theta}_t \mid \mathbf{N}_t) \;\propto\; P(\mathbf{N}_t \mid \mathbf{x}_t, \boldsymbol{\theta}_t)\;P(\mathbf{x}_t)\;P(\boldsymbol{\theta}_t),
\label{eq:tile_posterior}
\end{equation}
where $\mathbf{x}_t$, $\boldsymbol{\theta}_t$ and $\mathbf{N}_t$ are the white-noise field, nuisance parameters, and galaxy counts of tile $t$. Sampling \cref{eq:posterior} directly on the parent volume (that is, with a single global white-noise field $\mathbf{x}$ evolved through one $1024^3$ forward model) is precisely the computation rendered intractable by the memory and runtime limits noted above, and is the object the tiling is designed to circumvent. In its place, each tile samples its own posterior, \cref{eq:tile_posterior}, with no parameters shared between tiles. We stress that this is not the factorisation of a single global posterior: the tiles are statistically independent inferences, so although their joint distribution trivially factorises as $\prod_t P(\mathbf{x}_t,\boldsymbol{\theta}_t\mid\mathbf{N}_t)$, this product pools no information across tiles and constrains no shared modes; it expresses independence, not a combination of evidence over a common field. The only residual coupling between neighbours arises because a thin shell of galaxies in the overlap buffer enters the likelihood of more than one tile (\cref{sect:tiling}), introducing a weak statistical dependence between adjacent chains but, again, no shared parameters. This overlap concerns the galaxy data alone: the white-noise voxels in the shared shell are inferred independently within each tile and, lying inside the buffer, are trimmed before reassembly (\cref{eq:assembly}), so they are never carried into the final volume or double-counted. The residual dependence is therefore confined to the inferred cores, whose neighbouring values were constrained using partly overlapping galaxy data.

\begin{figure}
    \centering
    \includegraphics[width=\columnwidth]{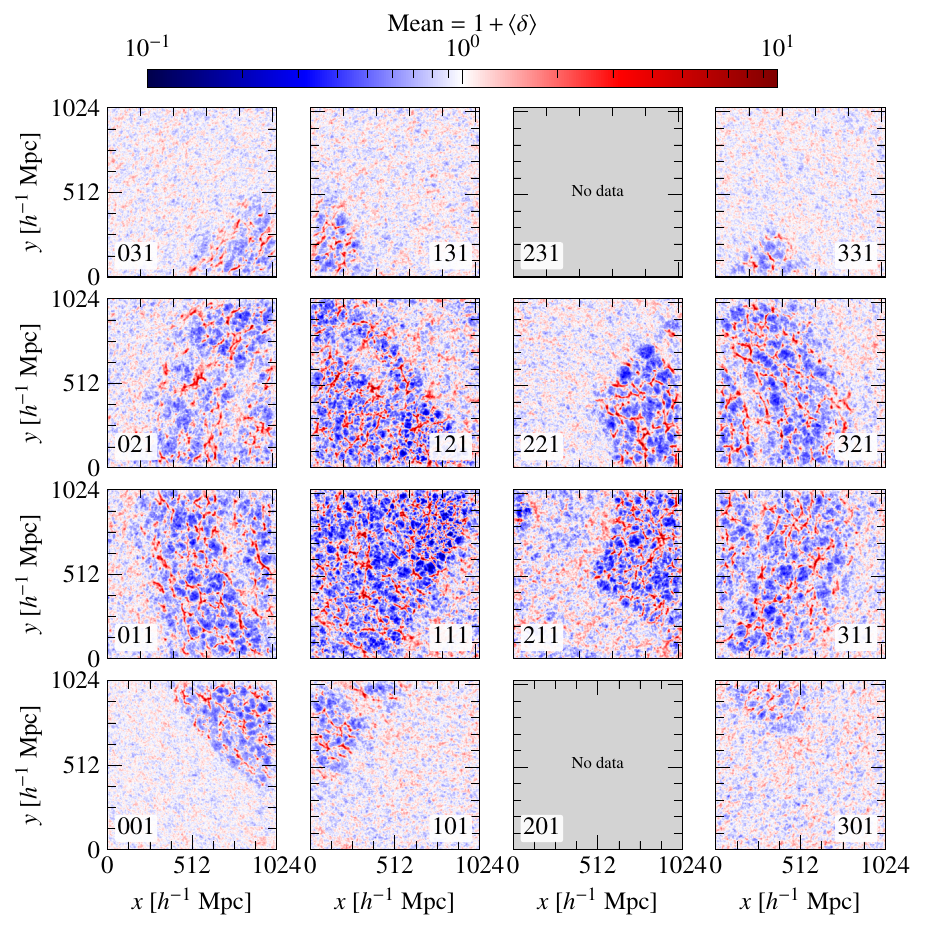}
    \caption{Posterior mean final density field (from the \cola forward model used inside \borg), averaged over all \numrealisations independent posterior samples, for a $20$~\mpch slice through the tile layer with index $z_{\mathrm{idx}} = 1$. Each panel corresponds to a single subvolume labelled by its $(x,y,z)$ tile index (e.g.\ 011, 111), where each index ranges from 0 to 3. Tiles containing survey data display coherent large-scale structure, while tiles without data (grey panels) were not inferred and are omitted. In data-containing tiles, regions beyond the survey footprint revert toward the $\Lambda$CDM prior mean, making the survey coverage visually apparent.}
    \label{fig:tiles_mean_density}
\end{figure}

A key distinction from \manticorelocal lies in the treatment of the observational light cone. The tiles span a broad redshift range ($0<z<0.7$), so galaxies at different comoving distances correspond to different evolutionary epochs. Although some \cola implementations---and earlier \borg forward models---can assign the density directly onto a continuous lightcone during sampling, the \borg implementation of \cola used here does not support this: it can only evolve each box to a single fixed redshift. We are therefore restricted to approximating the observational lightcone, evolving each tile to a fixed scale factor determined by the comoving distance between the observer and the geometric centre of that tile, computed using the DES~Y3 cosmology. This introduces a coarse lightcone treatment, yielding scale factors in the range $a \approx 0.55$--$0.9$. While this approach results in adjacent tiles being evolved to slightly different cosmic times, we assume minimal evolution across the 1024~\mpch\ extent of individual tiles, allowing the inferred white-noise fields from different epochs to be directly concatenated. We verify in \cref{sec:stat_consistency} that this approximation introduces no detectable artefacts in the statistical properties of the reconstructed fields. A full lightcone construction accounting for continuous cosmic evolution is recovered during the subsequent $N$-body resimulations (see \cref{sect:resimulations}).

Each tile is sampled independently following the burn-in and convergence procedures established in \manticorelocal, though the computational cost of \manticoredeep ($0.5$--$1.5\times10^{6}$ CPU hours per tile; $\sim$30 million CPU hours in total) prohibits running multiple chains per tile. Convergence is instead assessed through internal diagnostics, detailed in \cref{sect:appendix_convergence}: we monitor the likelihood trace until it reaches a stable plateau, and then measure autocorrelation lengths for the inferred field itself (individual voxels of the initial white-noise field and the modes of the final density power spectrum), continuing the MCMC until \numrealisations effectively independent posterior samples are obtained for each subvolume. Depending on the data content of a given tile, median autocorrelation lengths range from approximately 50 to 400 steps, with data-rich tiles exhibiting longer correlation times (\cref{fig:mcmc_diagnostics}). \Cref{fig:tiles_mean_density} shows the posterior mean final density for each tile: tiles with survey coverage display coherent large-scale structure, while unconstrained regions revert to the $\Lambda$CDM prior and average out across realisations, confirming that the per-tile inferences are producing genuinely independent samples. Following the reassembly procedure described in \cref{sect:tiling}, the buffer regions are removed and the trimmed cores are stitched into the global $1024^{3}$ grid. To construct full-volume posterior samples, we randomly draw one realisation from each tile's posterior set and concatenate their trimmed cores into a single full-volume initial-condition field,
\begin{equation}
\mathbf{x}^{(i)} = \bigoplus_{t}\, \mathcal{T}\!\left[\mathbf{x}^{(i)}_t\right],
\label{eq:assembly}
\end{equation}
where $\mathbf{x}^{(i)}_t$ is the $i$-th posterior draw of tile $t$, $\mathcal{T}[\cdot]$ denotes removal of the buffer layers to the $224^3$ signal core, and $\bigoplus_t$ is concatenation across the $4\times4\times4$ layout (with prior-filled padding completing the parent grid). Because $\mathcal{T}$ removes the buffer layers of every tile, each location in the parent volume is supplied by exactly one tile's core, so the white-noise modes duplicated across overlapping buffers never coexist in a full-volume realisation. Repeating this across the \numrealisations posterior samples per tile yields \numrealisations full-volume initial-condition realisations for the parent domain.

Equations~\eqref{eq:tile_posterior} and~\eqref{eq:assembly} together fix the statistical object that our full-volume ensemble represents. Each tile retains its own free parameters---no bias parameter is formally shared between tiles---and the unconstrained padding is supplied by $\Lambda$CDM prior modes. The subvolumes are, however, deliberately coupled through the data: the overlap shell (\cref{sect:tiling}) allows a common set of galaxies to enter the likelihoods of adjacent tiles, so that the initial phases inferred near tile boundaries---and, more loosely, perhaps even the per-tile bias parameters---are shaped by shared information rather than reconstructed in isolation. This is a deliberate choice. It does mean the assembled ensemble cannot be written as a product of independent per-tile posteriors, and in this sense the tiles are not statistically independent; but that coupling is precisely what draws the tiled inference toward the exact full-box posterior $P(\mathbf{x}\mid\mathbf{N})$, keeping structure coherent across tile boundaries in a way that a set of wholly independent subvolumes could not. The result is best understood as an \textit{induced} approximation to the full-volume posterior: on sub-tile scales, where the data constrain the field, it should closely approximate the true posterior, while on super-tile scales the field reverts to independent $\Lambda$CDM prior draws, since wavelengths larger than a subvolume are absent from each tile's finite periodic basis and the per-tile mean density absorbs the residual monopole (\cref{sect:tiling}). We expect this to represent a loss of constraint on the largest scales rather than a source of bias, though we do not prove this directly here.

\subsection{Resimulation of the posterior initial conditions}
\label{sect:resimulations}

High-fidelity $N$-body \squotes{resimulations} of the inferred initial conditions serve two complementary purposes: they evolve the posterior initial conditions at higher numerical resolution than was feasible during inference, providing access to halo catalogues, merger trees, lightcone outputs, and full-sky maps; and they provide an independent validation of the inference itself. Because \borg fits the data using a fast but approximate \cola gravity solver, any systematic bias in that solver would be partially absorbed into the inferred white--noise field, which is adjusted during sampling to reproduce the observed galaxy distribution. Re-evolving the same initial conditions with an independent, higher-fidelity $N$-body solver (\swift) and verifying that the resulting fields remain statistically consistent with \lcdm (\cref{sec:stat_consistency}) therefore tests that no such bias has been imprinted. In this context, \citet{Stopyra2024_COLA} showed that adopting a COLA-based gravity solver during inference, as used in both \manticorelocal and \manticoredeep, eliminates the systematic halo mass biases present in earlier \borg analyses that relied on simpler solvers \citep[see also][]{2015JCAP...06..015L,Jasche2019,2021PhRvD.103b3523B,Stopyra2021,2022MNRAS.516.3592H,2022MNRAS.512.5823M}.

The \textsc{Manticore-Deep} inference yields a posterior ensemble of \numrealisations independent white noise fields for the full parent volume, each defined on a $1024^3$ grid. 
These white-noise realizations are transformed into cosmological initial conditions at $z = 69$ using the \textsc{Monofonic} initial conditions generator\footnote{\url{https://github.com/cosmo-sims/monofonIC}} \citep{Hahn2020, Michaux2021}, following the procedure established in \textsc{Manticore-Local} \citep{McAlpine2025}. 
To enable higher-resolution structure formation beyond the inference grid, we oversample the initial conditions by a factor of four in each dimension: each realization is sampled by $4096^3$ dark matter particles, corresponding to a mean inter-particle spacing of $1$~\mpch and a particle mass of $m_{\mathrm{DM}} = 8.49 \times 10^{10}$~\msolh. 
As the inference is defined only up to the Nyquist frequency of the $1024^3$ grid, \textsc{Monofonic} augments the constrained large-scale modes by injecting Gaussian random small-scale fluctuations beyond the inference resolution. 
These small-scale modes are drawn from the $\Lambda$CDM power spectrum and are held fixed across all \numrealisations posterior samples, ensuring that differences between realizations reflect only the uncertainty in the large-scale, data-constrained structure. 
Initial displacements and velocities are computed using second-order Lagrangian perturbation theory (2LPT), providing accurate initial conditions for subsequent gravitational evolution.

The initial conditions are evolved forward to $z=0$ using the \swift $N$-body solver\footnote{\url{https://github.com/SWIFTSIM/SWIFT}} \citep{Schaller2024}, with gravitational forces softened using a Plummer-equivalent kernel with comoving softening length $\epsilon = 0.04$~\mpch (1/25 of the mean inter-particle spacing). Although the tiled inference evolved each subvolume to a single fixed scale factor, the resimulations start from the combined white-noise field at $z = 69$ and evolve continuously forward in time. This recovers a complete past lightcone spanning $0 < z \lesssim 0.7$, constructed on-the-fly by recording particle positions and velocities as they cross the lightcone surface. Snapshot outputs are produced at the same 79 scale-factor intervals adopted in the fiducial \flamingo simulations \citep{Schaye2023}. Each resimulation required approximately $6.5\times10^{5}$ CPU hours.

Halos are identified at each snapshot using the \textit{Hierarchical Bound-Tracing} algorithm \hbtherons\footnote{\url{https://github.com/SWIFTSIM/HBT-HERONS}} \citep{Han2012, Han2018, 2025arXiv250206932F}, which follows halo progenitors forward in time to build consistent merger trees and subhalo hierarchies. 
Structures containing at least 32 bound particles are retained as haloes (corresponding to a minimum mass of $M_{\mathrm{min}} \approx 2.72 \times 10^{12}$~\msolh), while subhaloes are required to contain at least 20 bound particles. 
Central halo positions are defined by the centre of mass of these bound particles. 
The resulting catalogs are post-processed with the \textit{Spherical Overdensity and Aperture Processor} (\textsc{SOAP})\footnote{\url{https://github.com/SWIFTSIM/SOAP}} \citep{2025JOSS...10.8252M}, which computes halo properties such as $M_{200\mathrm{c}}$, concentration, spin, and substructure content.
Throughout this work, $R_{\Delta\mathrm{c}}$ denotes the spherical-overdensity radius enclosing a mean density $\Delta$ times the critical density at the halo redshift, and $M_{\Delta\mathrm{c}}$ denotes the corresponding enclosed mass; catalogue quantities written without an explicit ``c'', such as $R_{500}$ and $M_{500}$, follow the same critical-density convention.
We reserve lowercase $r$ for generic three-dimensional or projected radial coordinates, such as $r/R_{500}$. 
The combination of \hbtherons and \textsc{SOAP} provides physically motivated, dynamically stable halo catalogs suitable for forward-modelling galaxy and cluster observables. 

In parallel, \swift generates full-sky \healpix maps \citep{2005ApJ...622..759G, Zonca2019} of the projected dark matter surface density in concentric 200~Mpc-thick shells centered on the observer, providing coverage out to $z \sim 0.7$ with angular resolution $\texttt{NSIDE} = 512$ (corresponding to $\sim6.9$ arcmin pixels). Since the full particle lightcone is stored, these maps can be recomputed at arbitrary angular resolution and shell width as needed for specific analyses. These projected surface-density maps underpin the CMB lensing cross-correlation presented in \cref{sec:cross_corr}, and more generally support density- and potential-based probes such as weak gravitational lensing and the integrated Sachs--Wolfe effect. They are not used for Sunyaev--Zel'dovich predictions: the thermal SZ effect requires the gas pressure, which is unavailable in these gravity-only resimulations, while the kinetic SZ analysis of \cref{sec:ksz} instead draws directly on the inferred three-dimensional velocity field through velocity-weighted cluster stacking, rather than on these projected maps.

As a control sample for validating the posterior resimulations, we additionally generate two complementary sets of unconstrained $\Lambda$CDM simulations with the same domain size and cosmology as the \manticoredeep posterior resimulations, each seeded with random, unconstrained phases and following the identical pipeline used for the posterior ensemble.
The first set consists of \ncontrolsims simulations at $N = 2048^3$ dark matter particles (eight times fewer than the posterior resimulations), corresponding to a particle mass of $m_{\mathrm{DM}} = 6.8 \times 10^{11}$~\msolh.
The second is a single simulation at the full posterior resolution of $N = 4096^3$ particles.
Running a full suite of controls at $4096^3$ resolution would be prohibitively expensive, so an ensemble of lower-resolution controls is used to characterise the statistical expectations of a random $\Lambda$CDM Universe for field-level comparisons.
The \ncontrolsims lower-resolution controls are well-suited for computing statistics such as the matter power spectrum and CMB lensing cross-correlations, where the scales of interest are well sampled at this resolution.
For the halo mass function, where increased dynamic range in halo mass is essential, we instead use the single full-resolution control simulation, which resolves the same mass range as the posterior halo catalogues.
Collectively, these simulations form our control sample and are denoted as \randomlcdm throughout this work.

\subsection{CMB Lensing Cross-Correlation Methodology}
\label{sect:cmb_methodology}

We validate the \manticoredeep posterior reconstructions by projecting the three-dimensional inferred matter density into predicted CMB lensing convergence maps and cross-correlating them with the observed \textit{Planck} 2018 lensing convergence map \citep{Planck2018,2022JCAP...09..039C}. This field-level test complements conventional galaxy–CMB lensing analyses \citep[e.g.][]{2017MNRAS.464.2120S,2022MNRAS.511.3548S}: it goes beyond two-point statistics, using the entire inferred matter field to predict the lensing convergence in each pixel rather than fitting a bias-dependent cross-power template. Galaxy bias is modelled within the \manticoredeep inference itself, so no additional bias assumption, template, or free amplitude enters at the cross-correlation stage.

We compute the predicted convergence map using ray-tracing in the Born approximation, accurate at the percent level for \textit{Planck} angular resolutions \citep{2006PhR...429....1L}. Each posterior realization includes forward-evolved matter fields from \swift, output as \healpix \citep{2005ApJ...622..759G} lightcone shells during runtime. These spherical shells, each $200\,\mathrm{Mpc}$ thick in comoving distance, yield 13 shells that sample the observer’s past light cone from $z = 0$ to $z = 0.7$, matching the depth of the BOSS galaxy samples that dominate the observational constraints in our inference. The shells are output at $\texttt{NSIDE} = 512$, yielding a native pixel size of $\sim$6.9~arcmin, which adequately samples the underlying simulation voxels of $4\,h^{-1}\,\mathrm{Mpc}$ that subtend angular scales of $\sim$47~arcmin at $z = 0.1$, $\sim$24~arcmin at $z = 0.2$, and $\sim$8~arcmin at $z = 0.7$. The convergence maps are computed natively at $\texttt{NSIDE} = 512$, then upgraded to $\texttt{NSIDE} = 2048$ using \healpix pixel replication prior to cross-correlation, matching the resolution of the \textit{Planck} lensing map; the effective angular resolution of $\kappa_{\mathrm{Manticore}}$ remains that of the native $\texttt{NSIDE} = 512$ grid.

For each shell $i$ we define the representative comoving distance as the arithmetic midpoint of the shell boundaries, $\chi_i = (r_{\mathrm{min},i} + r_{\mathrm{max},i})/2$, which is adequate for the $200\,\mathrm{Mpc}$ shell width given the slowly varying lensing kernel across an individual shell.
We define the shell thickness as $\Delta\chi_i = r_{\mathrm{max},i} - r_{\mathrm{min},i}$, and the corresponding redshift $z_i$ via the cosmological distance relation.
The shell overdensity field is then defined as
\begin{equation}
\delta_{\text{shell}}(\hat{n}, \chi_i) = \frac{m_{\text{DM}}(\hat{n}, \chi_i)}{\langle m_{\text{DM}}(\hat{n}, \chi_i) \rangle_{\hat{n}}} - 1,
\end{equation}
where $m_{\text{DM}}(\hat{n}, \chi_i)$ is the dark matter mass in \healpix pixel $\hat{n}$ of shell $i$, and $\langle \cdot \rangle_{\hat{n}}$ denotes the angular mean over all pixels in that shell.
The CMB convergence field, $\kappa(\hat{n})$, is then accumulated over shells via a discretized version of the standard weak-lensing line-of-sight integral \citep{2006PhR...429....1L}:
\begin{equation}
\kappa(\hat{n}) = \frac{3}{2}\,\Omega_\mathrm{m}\!\left(\frac{H_0}{c}\right)^2 \sum_i \frac{\Delta\chi_i}{a(\chi_i)} \frac{\chi_i(\chi_\mathrm{CMB} - \chi_i)}{\chi_\mathrm{CMB}}\,\delta_{\mathrm{shell}}(\hat{n}, \chi_i),
\label{eq:cmb_lensing}
\end{equation}
where $\chi_\mathrm{CMB} \approx 13{,}954\,\mathrm{Mpc}$ is the comoving distance to the CMB last-scattering surface, $a(\chi_i) = 1/(1+z_i)$ is the scale factor at shell $i$, and $\Omega_\mathrm{m}$ and $H_0$ are the same matter density parameter and Hubble constant adopted in both the inference and posterior resimulations.

To compare the predicted convergence maps with the observed \textit{Planck} data, we compute the decoupled angular cross-power spectrum using the MASTER pseudo-$C_\ell$ algorithm as implemented in \textsc{NaMaster} \citep{2019MNRAS.484.4127A}, which inverts the mask-induced mode-coupling matrix to recover an unbiased estimate of the underlying $C_\ell$ from the partial-sky bandpowers.
Our conservative fiducial cross-correlation uses a raw composite \borg--\textit{Planck} mask formed by multiplying the \textit{Planck} lensing mask with the joint \borg\ survey footprint built from the BOSS CMASS northern and southern masks (\cref{fig:borg_masks}).
The composite mask is apodised with a $0.2^\circ$ Gaussian kernel before the \textsc{NaMaster} mode-coupling matrix is computed.
As a variance-reduction test, we also evaluate a second pipeline in which small-scale holes in the BOSS survey footprint are regularised before this footprint is multiplied by the \textit{Planck} lensing mask and apodised; we describe this construction and quantify its impact on the recovered signal-to-noise in \cref{sect:appendix_mask}.
Within each mask choice, the same mask and mode-coupling matrix are applied to all posterior realizations and to the ensemble of null-control simulations---random universes with matched geometry and \healpix resolution but unconstrained by data---so that signal and null bandpowers are decoupled identically and remain directly comparable.

The cross-spectrum is computed over multipoles $\ell = 8$–$1250$. We report the full range but focus our validation on $\ell \leq 400$, where all modes are well resolved by the reconstruction grid and unaffected by numerical aliasing; this threshold is consistent with the voxel Nyquist limit near $\ell\!\sim\!500$ at $z\!\sim\!0.2$. Detection significance is assessed by comparing the cross-spectrum of the reconstructed matter field with \textit{Planck} to that of the null ensemble, providing a rigorous posterior-predictive test of the fidelity of the reconstructed matter distribution, analogous to simulation-based validation approaches such as \citet{2022JCAP...07..041C}.

\subsection{Kinetic Sunyaev--Zel'dovich stacking}
\label{sect:ksz_stacking}

The kinetic Sunyaev--Zel'dovich (kSZ) effect is the inverse-Compton scattering of CMB photons off free electrons in a galaxy group or cluster moving with a bulk peculiar velocity along the line of sight \citep{Sunyaev1980}.
Unlike the thermal SZ (tSZ) effect, the kSZ signal is spectrally indistinguishable from the primary CMB; its amplitude is proportional to the electron optical depth $\tau$ and the line-of-sight (LOS) peculiar velocity $v_\mathrm{LOS}$:
\begin{equation}
  \frac{\Delta T_\mathrm{kSZ}}{T_\mathrm{CMB}} = -\tau\,\frac{v_\mathrm{LOS}}{c},
  \label{eq:ksz}
\end{equation}
where $c$ is the speed of light.
Because $v_\mathrm{LOS}$ is equally likely to be positive or negative across a statistical ensemble of clusters, a naive mean stack of CMB temperature patches centred on cluster positions is expected to average to zero.
We therefore follow \citet{Tanimura2021} and adopt a velocity-weighted stacking estimator that coherently aligns contributions from clusters moving toward and away from the observer:
\begin{equation}
  \hat{T}(\boldsymbol{r}) =
    \frac{\displaystyle\sum_{i}\,T_{i}(\boldsymbol{r})\;v_i\,/\,\sigma^2_{T,i}}
         {\displaystyle\sum_{i}\,|v_i|\,/\,\sigma^2_{T,i}},
  \label{eq:ksz_estimator}
\end{equation}
where $T_i(\boldsymbol{r})$ is the CMB temperature at projected radius $\boldsymbol{r}$ (in units of $R_{500}$) in the patch centred on cluster $i$, $v_i$ is its signed LOS peculiar velocity, and $\sigma^2_{T,i}$ is the CMB temperature variance measured over the full $10\,R_{500} \times 10\,R_{500}$ patch, serving as an inverse-noise weight that captures both instrumental noise and residual primary CMB fluctuations.
The velocity sign in the numerator ensures coherent stacking: receding clusters contribute negative $T_\mathrm{kSZ}$ weighted by positive $v_i$, and approaching clusters contribute positive $T_\mathrm{kSZ}$ weighted by negative $v_i$, so both enter with the same sign. The denominator normalises by $|v_i|$, making the estimator proportional to $\tau$ in the limit of a known velocity field.
Any component of the CMB map that is statistically uncorrelated with $v_i$, including the primary CMB, tSZ signal, cosmic infrared background (CIB), and instrumental noise, is expected to average to zero in the stack \citep[see appendix~B of][]{Tanimura2021}.
Since $\sigma^2_{T,i}$ is measured from the same patch that contains the kSZ signal, the weight is in principle correlated with the quantity being estimated. However, the kSZ contribution from any individual cluster is $\ll 1$ per cent of the total patch variance (which is dominated by the primary CMB and instrumental noise), so this self-noise bias is entirely negligible.

We use the \textit{Planck} 2018 High Frequency Instrument map at 217~GHz \citep{Planck2018}, the null frequency of the tSZ effect, thereby avoiding contamination from the thermal SZ signal.
The map is provided in \healpix format at $\texttt{NSIDE} = 2048$ (pixel size $\simeq 1.7$~arcmin) and converted to $\mu\mathrm{K}$ units.
We apply a composite \borg--\textit{Planck} sky mask formed by multiplying the \textit{Planck} lensing mask with the raw joint \borg\ survey footprint from the BOSS CMASS northern and southern masks (\cref{fig:borg_mask_raw}), retaining approximately a quarter of the sky. This is the same conservative raw footprint used for the fiducial CMB lensing cross-correlation of \cref{sect:cmb_methodology}; the hole-filled footprint is used only as a variance-reduced comparison in that analysis.
Prior to stacking, we remove the dipole and apply a smooth spherical-harmonic high-pass filter to the full-sky map to suppress large-scale primordial CMB fluctuations, which otherwise dominate the kSZ signal by roughly two orders of magnitude.
Following \citet{Tanimura2021} (see their Section~4.1 and Figure~2 therein), the filter is applied by multiplying the map's $a_{\ell m}$ coefficients by an $\ell$-space window with zero response for $\ell < 360$, unit response for $\ell > 720$, and a smooth cosine ramp between these limits, corresponding approximately to angular scales of 30 and 15~arcmin.

We stack on galaxy groups and clusters from the Wen, Han \& Liu \citep[WHL;][]{Wen2012,Wen2015} catalogue, which identifies $158{,}103$ systems from SDSS photometry in the redshift range $0.05 < z < 0.8$, of which 89~per~cent have spectroscopic redshifts. Cluster masses are estimated from total optical luminosity calibrated against X-ray and tSZ measurements \citep{Wen2015}, and each system is characterised by a richness parameter $R_L$ proportional to the total $r$-band luminosity of member galaxies within $R_{500}$. Within the redshift range $0 < z < 0.7$ accessible to \manticoredeep, the WHL catalogue contains $119{,}196$ clusters, drawn from the same parent catalogue used by \citet{Tanimura2021}. We explore a range of cluster selections, varying the richness threshold $R_L$ between 10 and 30 (corresponding to approximate halo masses of ${\sim}2\times10^{13}$--$7\times10^{13}$~\msolh following the mass calibration of \citealt{Wen2015}) and the redshift window within $0 < z < 0.7$.

The key difference from \citet{Tanimura2021} lies in how the LOS peculiar velocity $v_i$ is assigned to each cluster. \citet{Tanimura2021} reconstruct $v_i$ from the SDSS galaxy density field via linear perturbation theory, an approach that requires assumptions about galaxy bias and introduces non-negligible per-cluster velocity uncertainties. Here we instead draw velocities directly from the \manticoredeep posterior, obtained from the full \swift $N$-body resimulations of each posterior realization. For each realization, we construct a mass-weighted velocity field by depositing the three Cartesian velocity components from the \swift particle lightcone onto a comoving grid using a cloud-in-cell (CIC) scheme. We explore three grid resolutions with cell widths of $4$, $8$, and $16$~\mpch (corresponding to grids of $1024^3$, $512^3$, and $256^3$ cells, respectively). Each WHL cluster is placed at its comoving Cartesian position, computed from its sky coordinates and spectroscopic redshift assuming the same cosmology as the resimulation, and assigned a radial peculiar velocity by interpolating the gridded velocity field and projecting onto the line of sight to the observer. This procedure is repeated independently for each of the \numrealisations posterior realizations, propagating the full inference uncertainty into the velocity assignments and treating each realization as an independent stack.

For each cluster $i$, we extract a square CMB patch centred on its Galactic coordinates using a gnomonic (tangent-plane) projection from the \healpix map, flagging masked pixels as missing data throughout.
Following \citet{Tanimura2021}, patches in which more than 20~per~cent of pixels within $10\,\theta_{500}$ of the cluster centre are masked are excluded from the stack, where $\theta_{500}$ is the angular radius corresponding to $R_{500}$.
Each patch is rescaled from native angular coordinates into dimensionless projected radius $r/R_{500}$ using the cluster's individual $\theta_{500}$ (both $R_{500}$ and $\theta_{500}$ are provided in the WHL catalogue), with bilinear interpolation applied to the temperature values and nearest-neighbour interpolation applied to the mask to avoid introducing spurious signal at mask edges.
All rescaled patches are placed onto a common $64 \times 64$ pixel grid spanning $10\,R_{500} \times 10\,R_{500}$, and the velocity-weighted stack is computed per pixel on this grid.
A circularly averaged one-dimensional profile is then extracted in 15 concentric annular bins, with uncertainties estimated from 1000 bootstrap resamplings of the cluster sample.

We assess detection significance against a null distribution constructed by randomly permuting the velocity assignments $v_i$ among the cluster sample, destroying any correlation between the inferred velocity field and the CMB temperature while preserving the angular clustering of objects on the sky. The null distribution is built from 1000 velocity-permuted bootstrap resamplings. Detection significance is quantified by integrating the stacked profile within $r \leq 4\,R_{500}$ and comparing the integrated signal to the distribution of null values.

\section{RESULTS}
\label{sect:results}

We present the results of the \manticoredeep posterior resimulations, following the validation strategy established for \manticorelocal \citep{McAlpine2025} and extending it to the significantly larger volume and higher redshift reach of the present work. Our analysis proceeds in three stages, progressing from internal consistency to qualitative inspection to quantitative comparison with independent observations.

We begin in \cref{sec:stat_consistency} by verifying that the inferred posterior volumes are statistically consistent with \lcdm expectations, examining the Gaussianity and isotropy of the initial conditions, the evolved matter power spectrum and bispectrum, and the halo mass function. These diagnostics are essential to confirm that the novel tiled inference methodology introduces no systematic artefacts into the reconstructed fields. Having established this statistical foundation, we turn in \cref{sec:lss} to a qualitative examination of the large-scale structure recovered by \manticoredeep, visually inspecting the density and velocity fields from the full posterior volume down to individual clusters and comparing the inferred halo populations directly against the observed galaxy distribution. In \cref{sec:cross_corr,sec:ksz} we then quantitatively benchmark the physical fidelity of the reconstruction against two independent observational probes not used in the inference: cross-correlation of the reconstructed matter field with the \textit{Planck} CMB lensing map, and detection of the kinetic Sunyaev--Zel'dovich effect via velocity-weighted stacking of galaxy clusters. Having confirmed the alignment of the \manticoredeep posterior with the observed Universe across these diagnostics, we conclude in \cref{sec:bgw} with an example case study of the BOSS Great Wall, demonstrating the utility of constrained digital twins for investigating one of the most massive known superstructures within a \lcdm-consistent framework.

\subsection{Statistical Consistency with \lcdm}
\label{sec:stat_consistency}

The \lcdm framework provides the most successful description of the large-scale structure and evolution of the Universe. Verifying that data-constrained posterior samples conform to this framework is, however, non-trivial. For an unconstrained \lcdm simulation, the white-noise field underpinning the initial conditions is produced simply by drawing each Fourier mode independently from a unit Gaussian, so that the ensemble expectation satisfies $\langle P_\xi(k)\rangle = 1$ by construction. In a field-level Bayesian inference such as \borg, by contrast, ${\sim}10^7$ white-noise amplitudes must be simultaneously constrained by observational data while remaining statistically consistent with the Gaussian prior \citep{Bertschinger1987, Hoffman1991}, thereby keeping the inferred field consistent with the white-noise prior throughout the sampling. This requirement is both computationally demanding and difficult to verify: a true Gaussian random field is fully specified by its two-point function, with all cumulants, equivalently all connected $n$-point functions, vanishing for $n>2$. Departures from Gaussianity, arising from likelihood miscalibration, numerical artefacts in the Hamiltonian Monte Carlo gradient, or boundary effects from the tiled inference ($k_{\mathrm{tile}} = 2\pi/L_{\mathrm{tile}} \simeq 0.006\,h\,\mathrm{Mpc}^{-1}$), can in principle generate non-zero connected higher-order correlations without obviously distorting the measured power spectrum alone. We therefore perform a series of post-inference consistency checks to assess whether the parent volumes of the posterior resimulations, spanning the full $4096\,h^{-1}$\,Mpc domain, are statistically compatible with \lcdm expectations, following the validation framework established for \textsc{Manticore-Local} \citep{McAlpine2025}.

These evaluations were not directly enforced during the inference itself (beyond the Gaussianity prior on initial conditions), but instead serve as post-hoc validation of the physical plausibility and statistical consistency of our inferred realisations across both the initial and evolved cosmic fields. The novel tiled reconstruction strategy makes such checks especially important, as we detail below.

We evaluate the following diagnostics:
\begin{itemize}
\item \textit{Gaussianity and isotropy of the initial conditions}:
We examine the power spectra of the inferred white-noise fields to test for departures from the expected $P_\xi(k)\equiv 1$. Because the inference tiles are Cartesian sub-domains aligned with the coordinate axes, any boundary mismatch or spurious mode introduced by the stitching would preferentially appear in axis-aligned Fourier modes; the direction-dependent power ratio $R_i(k)=P_i/P_{\mathrm{iso}}-1$ is therefore a targeted diagnostic specifically sensitive to this class of artefact.

\item \textit{Evolved matter power spectrum and bispectrum}:
The power spectrum measures the amplitude of density fluctuations as a function of scale, while the bispectrum captures the mode coupling generated by nonlinear gravitational evolution and is the lowest-order statistic sensitive to non-Gaussian phase correlations in the evolved field. Together they test whether the reconstructed density field reproduces both the correct clustering amplitude and the nonlinear structure expected in \lcdm. Spurious power at the tile scale or incorrect boun dary conditions would manifest as deviations in either statistic. We compare both to an ensemble of control \lcdm simulations at $z=0$.

\item \textit{Halo mass function (HMF)}:
The halo mass function provides a well-characterised prediction of hierarchical structure formation in \lcdm cosmology. We compare the HMF across the full parent volumes at $z=0$ to the single high-resolution control \lcdm simulation run at the full $4096^3$ particle resolution.
\end{itemize}

\begin{figure}
\centering
\includegraphics[width=\columnwidth]{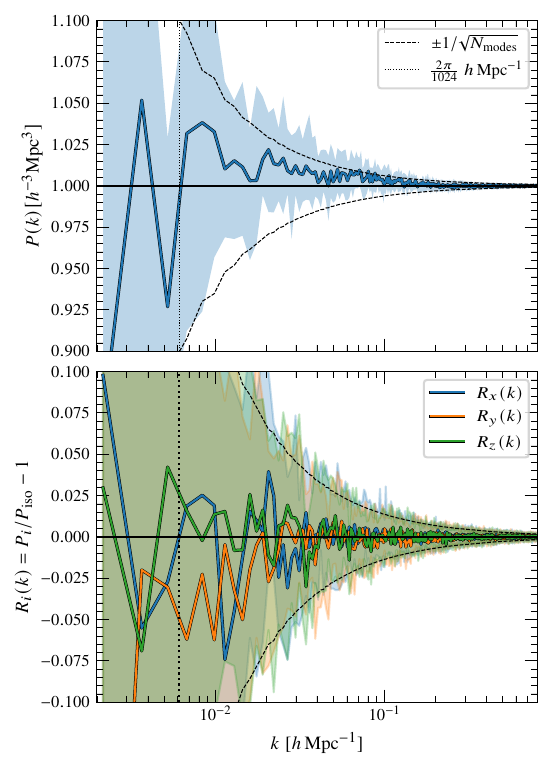}
\caption{
Power spectrum diagnostics for the initial white-noise fields inferred by \textsc{Manticore-Deep}.
\textit{Top panel:} median shell-averaged power spectrum across the posterior ensemble, normalised to the expected flat spectrum $P(k)=1$.
Shaded regions indicate the 10$^{\mathrm{th}}$--90$^{\mathrm{th}}$ percentile scatter, while the dashed line shows the expected $\pm1/\sqrt{N_{\mathrm{modes}}}$ sampling variance and the dotted line marks the tile scale ($k_{\mathrm{tile}} = 2\pi/L_{\mathrm{tile}} \simeq 0.006\,h\,\mathrm{Mpc}^{-1}$).
\textit{Bottom panel:} direction-dependent ratios $R_i(k)=P_i/P_{\mathrm{iso}}-1$ for modes oriented along the $x$ (blue), $y$ (orange), and $z$ (green) tile axes, with shaded bands denoting ensemble scatter.
All three components fluctuate randomly around zero within the predicted variance, demonstrating statistical isotropy and confirming that the tiled inference introduces no measurable directional artefacts in the reconstructed Gaussian initial conditions.
}
\label{fig:wnf_power_spectrum}
\end{figure}

\Cref{fig:wnf_power_spectrum} presents the power spectrum diagnostics for the inferred initial white-noise fields. As expected for Gaussian random initial conditions, the shell-averaged power is consistent with $P(k)=1$ across all scales resolved by our $4\,h^{-1}$\,Mpc grid, spanning wavenumbers from $k\simeq0.002$ to $0.8\,h\,\mathrm{Mpc}^{-1}$. The median power spectrum across posterior samples closely follows this expectation, with fluctuations remaining within the Poisson variance envelope predicted for independent Fourier modes.

To test for residual anisotropies potentially introduced by the tiled inference, the bottom panel of \Cref{fig:wnf_power_spectrum} shows the direction-dependent power ratios $R_i(k)=P_i/P_{\mathrm{iso}}-1$ for the three principal tile axes. Here $P_{\mathrm{iso}}(k)$ denotes the standard isotropic, spherically averaged power spectrum computed by averaging over all Fourier modes in a shell of radius $k$, while $P_i(k)$ is the power spectrum measured using only the subset of modes whose wavevectors are closely aligned with the $i$-th Cartesian axis, defined by $|k_i|/|\mathbf{k}| \geq 0.8$. All three directions exhibit fluctuations consistent with the expected sampling variance, with no coherent deviations from zero and no systematic power excess or deficit near the tile scale ($k\lesssim0.006\,h\,\mathrm{Mpc}^{-1}$). These results confirm that the stitched posterior fields preserve the isotropic Gaussian statistics assumed by the \lcdm model and that the tiled inference introduces no detectable directional bias or spurious large-scale correlations.

\begin{figure}
\centering
\includegraphics[width=\columnwidth]{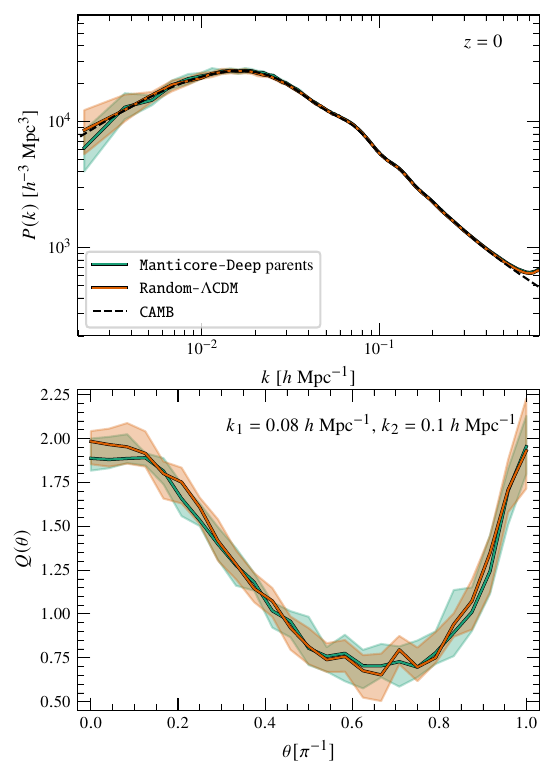}
\caption{
Matter power spectrum and reduced bispectrum at $z=0$ for the \textsc{Manticore-Deep} posterior resimulations.
\textit{Upper panel:} matter power spectrum, with the median across posterior realisations compared to \randomlcdm simulations and the \textsc{CAMB} theoretical prediction. Shaded regions indicate the 10$^{\mathrm{th}}$--90$^{\mathrm{th}}$ percentile spread. The vertical dotted line marks the Nyquist frequency of the $N=1024$ measurement grid, $k_\mathrm{nq}=\pi N/L \simeq 0.8\,h$\,Mpc$^{-1}$.
\textit{Lower panel:} reduced bispectrum $Q(\theta)$ as a function of opening angle $\theta$, for the configuration $k_1 = 0.08\,h$\,Mpc$^{-1}$, $k_2 = 0.1\,h$\,Mpc$^{-1}$.
Both diagnostics show excellent agreement with \lcdm expectations.
}
\label{fig:final_power_spectrum}
\end{figure}

Having validated the statistical properties of the initial conditions, we now assess whether the gravitationally evolved density field at $z=0$ reproduces the expected clustering properties of a \lcdm universe. Both statistics are computed from the cloud-in-cell (CIC) mass-assigned density field on an $N=1024$ grid, using the \textsc{Pylians} library \citep{Pylians}. \Cref{fig:final_power_spectrum} presents the matter power spectrum and reduced bispectrum measured from the \textsc{Manticore-Deep} posterior ensemble, compared against the \randomlcdm simulation suite. The power spectrum is consistent with \lcdm across all resolved scales, from the largest modes of the box ($k \sim 0.002\,h$\,Mpc$^{-1}$) up to the Nyquist frequency of the measurement grid ($k_\mathrm{nq} = \pi N/L \simeq 0.8\,h$\,Mpc$^{-1}$). The median posterior power spectrum closely tracks the control ensemble, with fluctuations well within the cosmic variance expected for a volume of this size. We observe no systematic suppression or enhancement of power at large scales that might indicate residual artefacts from the tiling procedure, confirming that the stitched initial conditions evolve self-consistently under gravitational dynamics. The reduced bispectrum $Q(\theta)$, which captures non-Gaussian features introduced by nonlinear gravitational evolution, similarly demonstrates strong consistency with \lcdm predictions across all opening angles $\theta$.

\begin{figure}
\centering
\includegraphics[width=\columnwidth]{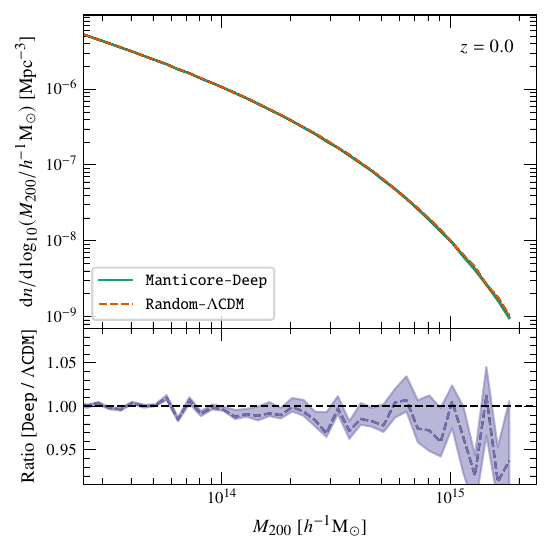}
\caption{
Halo mass function at $z=0$ for the \textsc{Manticore-Deep} posterior resimulations of the full $L=4096$~\mpch parent volume.
\textit{Upper panel:} median differential HMF from the posterior ensemble (solid) compared to a single control \lcdm simulation at the same $4096^3$ particle resolution (dashed).
\textit{Lower panel:} ratio of the posterior median to the control simulation, with the shaded band indicating the 10$^{\mathrm{th}}$--90$^{\mathrm{th}}$ percentile spread across posterior realisations. The ratio remains consistent with unity to within ${\sim}1\%$ over the well-sampled mass range, with increasing scatter at the high-mass end due to Poisson statistics.
}
\label{fig:hmf_z0}
\end{figure}

The halo mass function provides a complementary test of structure formation, directly probing the abundance of gravitationally collapsed objects as a function of mass. \Cref{fig:hmf_z0} compares the differential halo mass function at $z=0$ from the \textsc{Manticore-Deep} posterior ensemble to a single control \lcdm simulation run at the same $4096^3$ particle resolution, which resolves the same halo mass range as the posterior catalogues. Across the well-sampled mass range ($M_{200\mathrm{c}} \gtrsim 10^{13.5}\,{\rm M}_\odot$), the median posterior HMF closely tracks the control simulation, with the ratio remaining consistent with unity to within ${\sim}1\%$. At the highest masses ($M_{200\mathrm{c}} \gtrsim 10^{15}\,{\rm M}_\odot$), the scatter increases as expected from Poisson statistics in both the posterior and control samples. We note that the control here is a single realisation, so residual differences at the per cent level are consistent with sample variance between any two independent \lcdm volumes of this size.

Taken together, the diagnostics presented in this section---Gaussian initial conditions with flat, isotropic power spectra; an evolved matter power spectrum and bispectrum consistent with \lcdm at $z=0$; and a halo mass function matching a control simulation to within ${\sim}1\%$---demonstrate that the \textsc{Manticore-Deep} posterior volumes are statistically consistent with \lcdm expectations. While we cannot exclude all possible subtle effects introduced by the tiling procedure, the tests examined here reveal no measurable deviations, providing confidence that our Bayesian inference framework, extended to cosmologically significant volumes through a tiled reconstruction strategy, recovers physically plausible realisations of cosmic structure. This validation establishes a foundation for the comparisons with observational data presented in the following sections.

\subsection{The Large-Scale Structure of the \textsc{Manticore-Deep} Universe}
\label{sec:lss}

\begin{figure*}
\centering
\includegraphics[width=\textwidth]{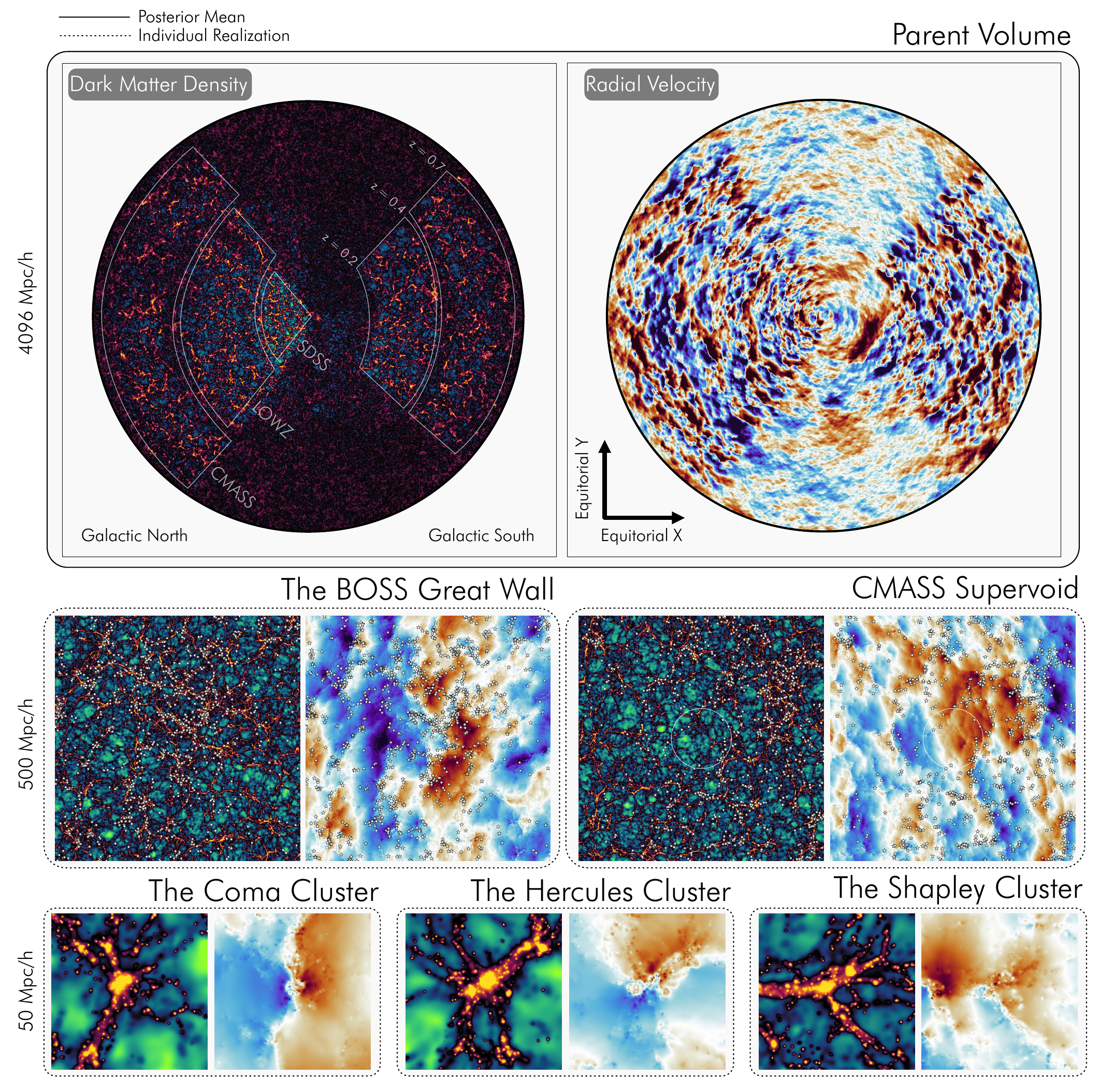}
\caption{
\textit{Top:} A $15~h^{-1}\mathrm{Mpc}$ slice through the full posterior volume, centred on the observer and shown as an orthographic projection in Cartesian–equatorial coordinates.
The left-hand map displays the posterior mean of the dark matter density field, and the right-hand map the corresponding posterior mean of the radial peculiar velocity field.
Approximate survey footprints are outlined in the density map; within these regions the inference recovers a strong, coherent signal tracing the observed cosmic web, while outside it smoothly reverts to the isotropic $\Lambda$CDM prior.
    \textit{Middle:} $500~h^{-1}\mathrm{Mpc}$ zooms centred on the BOSS Great Wall \citep{2016A&A...588L...4L} and a large supervoid \citep[void~5079,][]{2017ApJ...835..161M} with a reported effective radius $R_\mathrm{eff} = 93.8\,h^{-1}\,\mathrm{Mpc}$ (highlighted as a solid circle).
Individual CMASS galaxies used in the inference are overlaid, illustrating the close spatial correspondence between the data and the reconstructed dark matter field.
\textit{Bottom:} $50~h^{-1}\mathrm{Mpc}$ regions centred on the Coma, Hercules, and Shapley clusters, each embedded within its surrounding filamentary environment.
Throughout, panels enclosed by a \textit{solid} border show the posterior mean of the field, while those enclosed by a \textit{dashed} border show a single posterior realisation.
Together, these panels illustrate the range of scales and environments captured by the reconstruction, from the full posterior volume down to individual cluster environments.
}
\label{fig:lss_overview}
\end{figure*}

The top panel of \Cref{fig:lss_overview} presents a thin $15~h^{-1}\mathrm{Mpc}$ slice through the full \textsc{Manticore-Deep} posterior volume, centred on the observer and shown as an orthographic projection in the natural Cartesian–equatorial reference frame of the simulation.
The left-hand map displays the posterior mean of the dark matter density field, while the right-hand map shows the corresponding posterior mean of the radial peculiar velocity field (this left–right layout of density and radial velocity is used consistently throughout the figure).
Approximate survey footprints are outlined in the density map for reference.
Within these data-constrained regions, the posterior mean exhibits a coherent and spatially structured signal that traces the observed cosmic web, whereas outside the survey boundaries the field naturally reverts to the isotropic $\Lambda$CDM prior, producing a mean closer to zero.
This smooth transition highlights the balance between data-driven and prior-dominated regions achieved by the inference. The velocity field shows the same qualitative pattern, with coherent large-scale flows within the observational volume and a progressive loss of structure beyond it. Because peculiar velocities are dominated by large-scale gravitational modes, their mean field appears intrinsically smoother than that of the density: in Fourier space, the relation
\begin{equation}
\boldsymbol{v}(\boldsymbol{k}) \propto \frac{i\,\boldsymbol{k}\,\delta(\boldsymbol{k})}{k^{2}}
\label{eq:vel_fourier}
\end{equation}
suppresses small-scale power, yielding more extended, slowly varying features. The resulting picture highlights the complementary nature of the two fields—density revealing the fine filamentary skeleton of the cosmic web, and velocity tracing its broad dynamical environment—both reconstructed with high fidelity inside the data-constrained domain.

The middle panel of \Cref{fig:lss_overview} focuses on two interesting $500~h^{-1}\mathrm{Mpc}$ regions: one centred on a prominent superstructure \citep[the BOSS Great Wall;][]{2016A&A...588L...4L} at $\mathrm{RA} \approx 162^{\circ}$, $\mathrm{Dec} \approx +50^{\circ}$, $z \approx 0.47$, and one centred on a large supervoid \citep[void~5079 of the BOSS DR12 void catalogue,][]{2017ApJ...835..161M} at $\mathrm{RA} = 14.7^{\circ}$, $\mathrm{Dec} = 20.6^{\circ}$, $z = 0.494$, with an effective radius of $R_\mathrm{eff} = 93.8\,h^{-1}\,\mathrm{Mpc}$, among the largest voids in that catalogue. In these panels, drawn from a single posterior realisation of \manticoredeep to highlight small-scale features, the individual galaxies from the CMASS sample used in the inference are overlaid on the reconstructed fields. Despite the relative sparsity and inhomogeneity of the CMASS data, the inferred dark matter distribution forms a high-fidelity, continuous map of the underlying large-scale structure that is consistent with both the observations and the $\Lambda$CDM\ model. The reconstruction recovers a clear visual underdensity at the reported location of void~5079, with extent broadly consistent with the catalogued effective radius; a detailed quantitative comparison is beyond the scope of this work. The large-scale velocity field further reinforces this picture: the major walls in these regions delineate the boundaries between coherent inflows and outflows, with matter streaming toward dense wall complexes and diverging into adjacent voids. These flows trace the gradient of the reconstructed gravitational potential, illustrating that \textsc{Manticore-Deep} captures not only the morphology but also the dynamical state of the cosmic web.

Such reconstructions demonstrate the power of field-level inference to recover self-consistent structure even in regimes where the data are sparse or incomplete. They provide a natural framework for studying extreme features that have been invoked as potential challenges to cosmic homogeneity, including the Sloan Great Wall \citep{2005ApJ...624..463G}, the BOSS Great Wall \citep{2016A&A...588L...4L,2022A&A...666A..52E}, and the Great Arc \citep{2022MNRAS.516.1557L}. By investigating these structures within a causally consistent inference that explicitly accounts for observational selection effects within its forward model, \textsc{Manticore-Deep} enables direct tests of whether such apparent anomalies can emerge within the statistical expectations of $\Lambda$CDM, or whether they hint at new physics beyond it. We carry out an example investigation for the BOSS Great Wall in \cref{sec:bgw}.

The bottom panel of \Cref{fig:lss_overview} illustrates three massive clusters in the Local Universe ($r < 200$~\mpch, Coma, Hercules, and Shapley), each embedded within their surrounding large-scale environments. Drawn from a single posterior realisation, these zooms reveal the filamentary network feeding each cluster, with the velocity field showing coherent infall toward the cluster cores. These examples highlight the unique combination of volume and resolution achieved by \textsc{Manticore-Deep}: within a single inference, we resolve structures spanning from multi–hundred–$h^{-1}\mathrm{Mpc}$ walls and filaments down to individual cluster members with halo masses $M_{200\mathrm{c}} \gtrsim 10^{14.5}$~\msolh. This continuous dynamic range allows direct investigation of the hierarchical growth of structure, linking the emergence of massive clusters to their position within the cosmic web.

\begin{figure*}
\centering
\includegraphics[width=\textwidth]{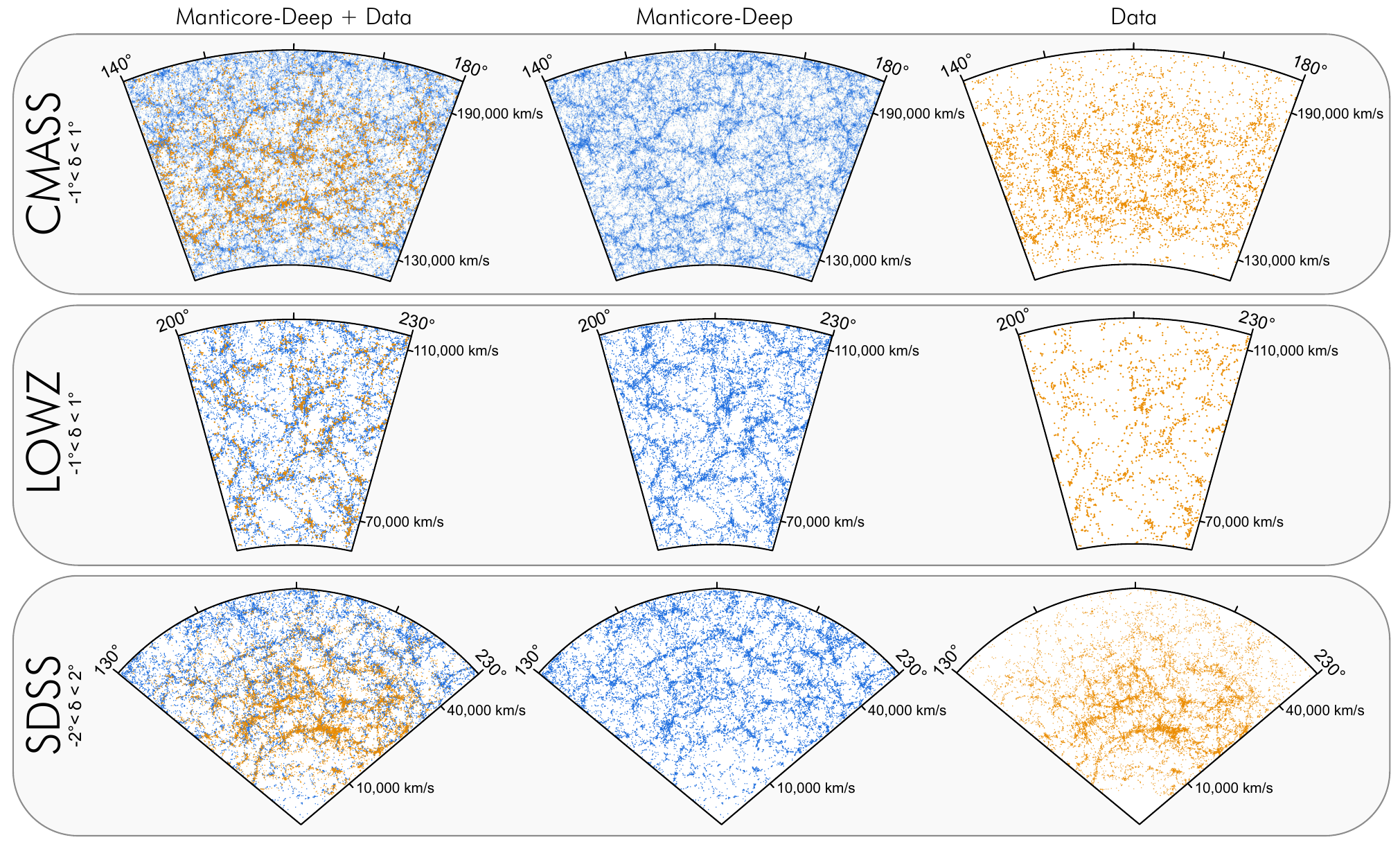}
\caption{
Qualitative comparison between all (sub)haloes from a single \textsc{Manticore-Deep} posterior realisation and the observed galaxy distribution in redshift space.
Each row shows a thin declination slice through one of the spectroscopic surveys used as input: SDSS Main ($-2^{\circ}<\delta<2^{\circ}$, $0 < z \lesssim 0.2$, bottom; haloes from the $z=0.1$ snapshot), LOWZ ($-1^{\circ}<\delta<1^{\circ}$, $0.15 < z < 0.43$, middle; $z=0.3$ snapshot), and CMASS ($-1^{\circ}<\delta<1^{\circ}$, $0.43 < z < 0.70$, top; $z=0.5$ snapshot). The radial coordinate is recession velocity.
From left to right, the panels show (i) \textsc{Manticore-Deep} haloes overlaid with the observed galaxies, (ii) the haloes alone, and (iii) the galaxies alone.
The strong spatial correspondence across the three samples shown demonstrates that the inferred (sub)halo distribution closely traces the observed galaxy positions across the full redshift range $0 < z \lesssim 0.7$.
No survey selection functions have been applied to the inferred halo catalogues, whereas the observational samples are subject to luminosity and completeness limits.
These selection effects enhance the apparent contrast of features such as the Sloan Great Wall \citep{2005ApJ...624..463G} in the galaxy data; the same structures are clearly visible in the halo field, which carries no imposed luminosity or completeness cuts.
The positional agreement between haloes and galaxies confirms that the inference extends from large-scale density fields down to the halo population, spanning the full redshift range of the input surveys.
}
\label{fig:halo_wedge}
\end{figure*}

\Cref{fig:halo_wedge} extends this validation to the halo scale, comparing the (sub)haloes from a single \textsc{Manticore-Deep} posterior realisation with the observed galaxy distributions in three representative input samples spanning low, intermediate, and high redshift. The close spatial correspondence across the full redshift range $0 < z \lesssim 0.7$ confirms that the inference extends coherently from the smoothly varying density field to the discrete halo population within a single cosmological model.

\subsection{Cross-correlation with \textit{Planck} lensing}
\label{sec:cross_corr}

\begin{figure}
    \centering
    \includegraphics[width=\columnwidth]{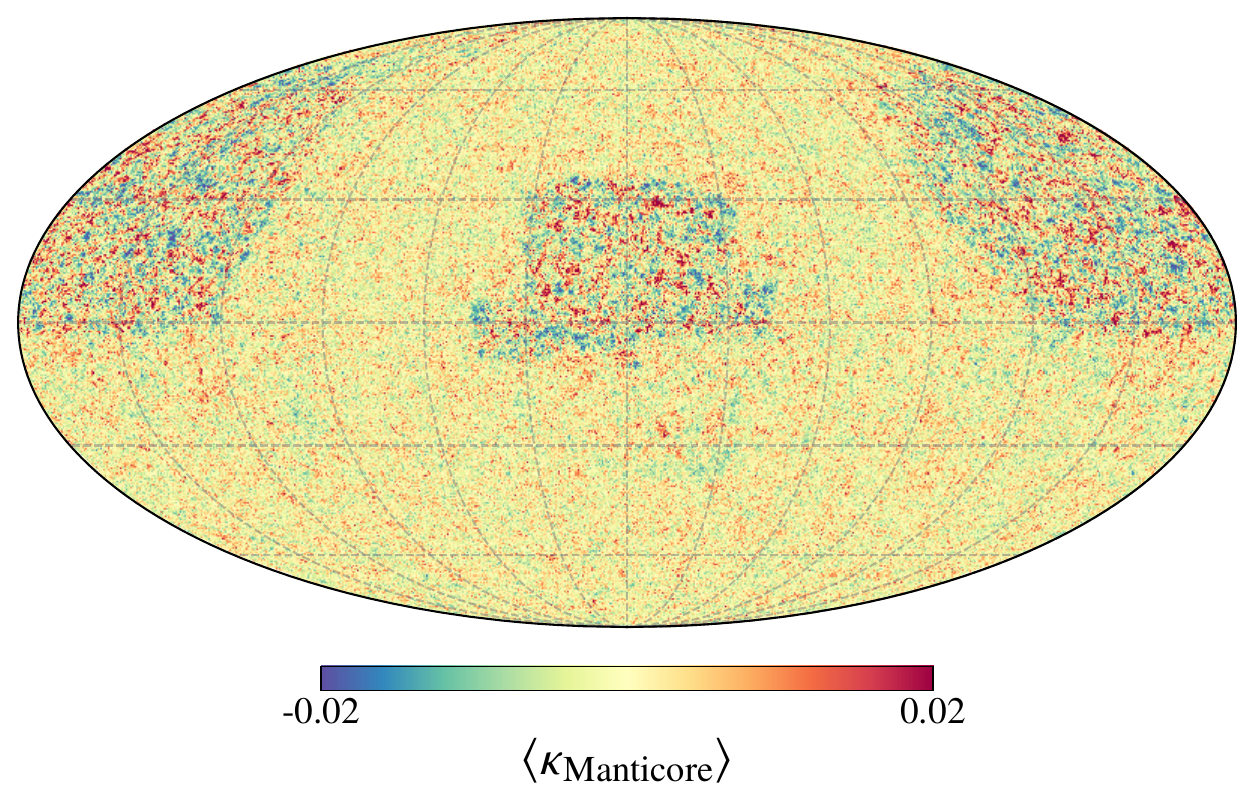}
    \caption{
    Posterior mean convergence field, $\langle \kappa_{\mathrm{Manticore}} \rangle$, from \manticoredeep\ in equatorial coordinates. 
    The map shows the integrated projected mass distribution out to $z=0.7$. 
    Large-scale overdensities and voids appear as coherent features in the posterior mean, while regions outside the survey mask naturally revert to the prior, highlighting the transition between data- and prior-dominated regimes.}
    \label{fig:kappa_map}
\end{figure}

The preceding section demonstrated that the \manticoredeep\ reconstruction visually reproduces the observed cosmic web from the full posterior volume down to individual haloes. We now test the physical fidelity of the inferred matter field against an entirely independent observable: gravitational lensing of the CMB. If the reconstructed three-dimensional mass distribution is correct, its line-of-sight projection should produce a convergence field correlated with the lensing signal measured by \textit{Planck}. \Cref{fig:kappa_map} shows the posterior mean convergence field, $\langle \kappa_{\mathrm{Manticore}} \rangle$, obtained by projecting the \manticoredeep\ matter reconstructions along the line of sight over $0<z<0.7$ (see \cref{sect:cmb_methodology} for methodological details). Prominent superclusters and extended voids are clearly imprinted within the data-constrained region, while outside the survey footprint the posterior mean reverts to the prior as expected, providing a two-dimensional summary of the three-dimensional mass distribution that can be compared directly against the \textit{Planck} lensing convergence.

The cross-correlation with the \textit{Planck} PR3 CMB lensing convergence map from \citet{Planck2018} is shown in \cref{fig:kappa_crosscorr}.
To quantitatively assess whether the reconstructed matter field lenses the CMB consistently with observations, we compute the angular cross-power spectrum between the predicted convergence, $\kappa_{\mathrm{Manticore}}$, and the \textit{Planck} convergence map for each posterior realisation:
\begin{figure}
    \centering
    \includegraphics[width=\columnwidth]{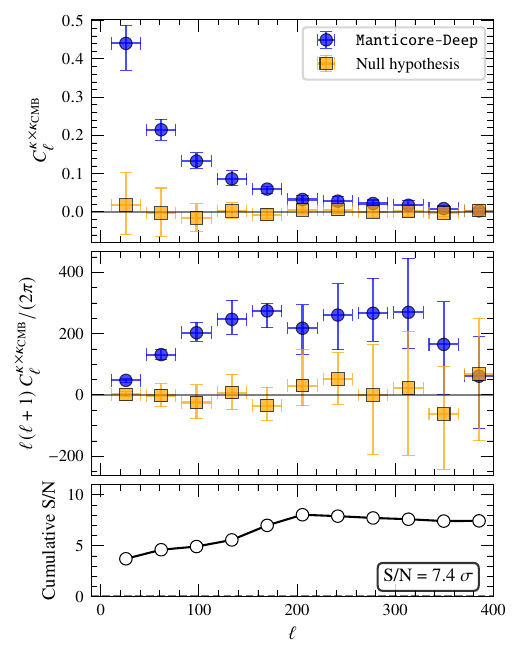}
    \caption{
    Cross-correlation between the \manticoredeep\ convergence field and the \textit{Planck} PR3 CMB lensing map \citep{Planck2018}.
    \textit{Top:} Binned angular cross-power spectrum $C_\ell^{\kappa_{\mathrm{Manticore}}\kappa_{\mathrm{CMB}}}$ (bin width $\Delta\ell = 30$; blue points with $16$--$84$th percentile error bars over posterior realizations) compared to the null-control simulations (orange; random, \lcdm data-unconstrained realisations processed identically).
    \textit{Middle:} Same cross-power spectrum scaled by $\ell(\ell+1)/(2\pi)$ to emphasize the separation between signal and null at intermediate multipoles.
    \textit{Bottom:} Cumulative detection significance relative to the null ensemble, computed via \cref{eq:sn_kappa}, saturating at the conservative raw-mask value of \lensingsnr.
    The strong positive signal confirms that the reconstructed three-dimensional matter field coherently lenses the CMB in a manner consistent with the true Universe.}
    \label{fig:kappa_crosscorr}
\end{figure}
\begin{equation}
C_\ell^{\kappa_{\mathrm{Manticore}}\kappa_{\mathrm{CMB}}}
    = \big\langle a_{\ell m}^{\kappa_{\mathrm{Manticore}}}
      \left(a_{\ell m}^{\kappa_{\mathrm{CMB}}}\right)^{*} \big\rangle ,
\end{equation}
where $a_{\ell m}$ are the spherical-harmonic coefficients in the expansion
$\kappa(\hat{\mathbf{n}})=\sum_{\ell m}a_{\ell m}Y_{\ell m}(\hat{\mathbf{n}})$.
We focus on $\ell\leq 400$ to ensure all included modes are well resolved by the reconstruction grid. The voxel Nyquist multipole scales with comoving distance as $\ell_{\mathrm{nq}} = \pi\chi(z)/d_{\mathrm{vox}}$, ranging from $\ell\!\sim\!230$ at $z\!=\!0.1$ to $\ell\!\sim\!1370$ at $z\!=\!0.7$; however, the CMB lensing kernel strongly upweights higher-redshift shells where the resolution limit is least restrictive, making $\ell\!\leq\!400$ a conservative threshold. The fiducial cross-correlation is computed within the raw composite \borg--\textit{Planck} sky mask, which combines the \textit{Planck} lensing mask with the joint BOSS CMASS northern and southern survey footprint (\cref{fig:borg_masks}), ensuring that only the data-constrained region contributes to the measurement. We also quote a variance-reduced comparison in which small-scale holes in the BOSS footprint are regularised before multiplication by the \textit{Planck} lensing mask and apodisation (see \cref{sect:appendix_mask} for mask construction details). The detection significance is then assessed from the difference vector
\begin{equation}
\Delta = \langle C_\ell^{\kappa_{\mathrm{Manticore}}\kappa_{\mathrm{CMB}},\,\mathrm{post}} \rangle - \langle C_\ell^{\kappa_{\mathrm{Manticore}}\kappa_{\mathrm{CMB}},\,\mathrm{null}} \rangle
\label{eq:delta_kappa}
\end{equation}
and the total covariance
\begin{equation}
\mathbf{\Sigma} = \mathbf{\Sigma}_{\mathrm{post}} + \mathbf{\Sigma}_{\mathrm{null}},
\label{eq:cov_kappa}
\end{equation}
where bold symbols denote covariance matrices of the binned bandpower estimates (bin width $\Delta\ell = 30$), giving
\begin{equation}
(S/N)^2 = \Delta^{\mathrm{T}} \, \hat{\mathbf{\Sigma}}^{-1} \, \Delta.
\label{eq:sn_kappa}
\end{equation}
The inverse covariance is corrected for finite-sample bias using the \citet{2007A&A...464..399H} factor, $\hat{\mathbf{\Sigma}}^{-1} = \alpha\,\mathbf{\Sigma}^{-1}$ with $\alpha = (N_\mathrm{s} - n_\mathrm{b} - 2)/(N_\mathrm{s} - 1)$, where $N_\mathrm{s}$ is the effective number of independent samples and $n_\mathrm{b}$ the number of bandpower bins.

The top panels of \cref{fig:kappa_crosscorr} show a clear positive cross-power spectrum over $30 \lesssim \ell \lesssim 300$, well above the null ensemble of unconstrained simulations, which quantify the level of spurious cross-correlation arising from chance alignments due to cosmic variance and the shared survey geometry alone. The bottom panel shows the cumulative signal-to-noise computed using \cref{eq:sn_kappa}, saturating at \lensingsnr\ for the conservative raw composite mask, constituting a highly significant detection of correlated structure between the \manticoredeep\ reconstructions and the observed \textit{Planck} lensing map. Repeating the measurement with a composite mask built from the hole-filled BOSS footprint raises the detection to \lensingsnrfilled, consistent with the reduction in pseudo-$C_\ell$ estimator noise produced by regularising the many small-scale survey-footprint holes before apodisation (see \cref{sect:appendix_mask}). The recovered $C_\ell^{\kappa_{\mathrm{Manticore}}\kappa_{\mathrm{CMB}}}$ is also consistent in shape with \lcdm\ expectations: the cross-power peaks at large angular scales and declines towards smaller scales, following the matter power spectrum weighted by the CMB lensing kernel across the survey volume \citep{2006PhR...429....1L}, confirming that the inferred structures lens the CMB with the correct spatial coherence across the full range of probed scales.

The detection level is fully consistent with expectations given the depth and resolution of the reconstruction. Because the CMB lensing kernel peaks at $z\simeq 2$, \manticoredeep\ captures only the low-redshift tail of the integrated lensing signal. We quantify this kernel weighting directly by recomputing the predicted convergence from subsets of the 13 lightcone shells and repeating the cross-correlation for each (quoted for the conservative raw mask configuration, for which the full projection yields \lensingsnr): the high-redshift half of the volume alone ($z\gtrsim0.3$) retains $6.8\,\sigma$, or 92 per cent of the full significance, and the outermost three shells alone ($z\gtrsim0.5$) retain $4.8\,\sigma$, whereas the low-redshift half ($z\lesssim0.35$) yields only $2.4\,\sigma$. These shell subsets also confirm that the $\ell\leq400$ threshold does not limit the detection: even for the high-redshift-only predictions, whose voxel Nyquist limits lie at $\ell_{\mathrm{nq}}\gtrsim600$, the cumulative signal-to-noise saturates by $\ell\simeq250$, indicating that the small-scale information content is bounded by the declining cross-power and the \textit{Planck} reconstruction noise rather than by the resolution of the reconstruction grid; excluding the low-redshift shells to extend the multipole range therefore reduces, rather than increases, the total significance. Earlier galaxy--CMB lensing cross-correlation analyses have often focused on template-based measurements using galaxy positions and redshift distributions, yielding higher reported signal-to-noise values \citep[$\sim 10$--$40\,\sigma$,][]{2017MNRAS.464.2120S,2022MNRAS.511.3548S,2022MNRAS.515.1993S}. Our test is different: it compares the inferred dark-matter convergence field directly to the observed CMB lensing map without fitting a template amplitude, making it a stricter field-level validation of the reconstructed mass distribution. Galaxy bias and the survey redshift distributions are accounted for within the inference that produces the matter field; the cross-correlation stage itself therefore introduces no further assumptions about galaxy bias, amplitude rescaling, or a fiducial cosmological model for the cross-power template, in contrast to forward-modelling approaches that fit $\kappa\times\delta_g$ using theoretical templates \citep[e.g.][]{Kitanidis2021,2022MNRAS.511.3548S} or amplitude-fitting methods applied to $\kappa\times\kappa_\mathrm{CMB}$ \citep[e.g.][]{2025arXiv260202363A}.

Within this field-level inference setting, \citet{2019arXiv190906396L} validated a BORG reconstruction of the BOSS data against \textit{Planck} lensing as a qualitative consistency check without quoting a detection significance, and \citet{2025arXiv260202363A} recently reported a detection from a BORG reconstruction of the Quaia quasar catalogue spanning $0.17 < z < 3.0$. That work employs two complementary estimators: a template-based amplitude fit to a fiducial \lcdm\ cross-power spectrum, yielding $\sim\!4\,\sigma$, and a model-independent phase-alignment test analogous to ours (\cref{eq:sn_kappa}), yielding $\sim\!3\,\sigma$. Our conservative \lensingsnr\ detection therefore exceeds the comparable model-independent Quaia result. The higher significance is driven by both the density and the spectroscopic precision of our galaxy samples: although the Quaia reconstruction extends to higher redshifts that more fully sample the CMB lensing kernel, the sparser photometric quasar catalogue and its broader redshift uncertainties leave the matter field less tightly constrained, producing a noisier convergence map. The strong correlation observed here therefore constitutes a direct, parameter-free, posterior-predictive confirmation that \manticoredeep\ accurately recovers the three-dimensional matter field responsible for CMB lensing.

\subsection{Detection of the kinetic Sunyaev--Zel'dovich effect}
\label{sec:ksz}

A further independent validation of the \manticoredeep\ reconstruction is provided by the kinetic Sunyaev--Zel'dovich (kSZ) effect.
A galaxy cluster moving with a line-of-sight peculiar velocity $v_\mathrm{LOS}$ imprints a temperature shift on the CMB proportional to $v_\mathrm{LOS}$ and the cluster's optical depth \citep{Sunyaev1972,Sunyaev1980}.
Because $v_\mathrm{LOS}$ is equally likely to be positive or negative across a statistical ensemble of clusters, a naive mean stack of CMB patches at cluster positions averages to zero.
A detection requires weighting the stack by an estimate of each cluster's peculiar velocity, so that approaching and receding contributions reinforce rather than cancel.
If the \manticoredeep\ velocity field is physically accurate, this weighted stack should produce a detectable signal at the observed cluster positions; if the inferred velocities were uncorrelated with the true motions, no signal would emerge.

Velocity-weighted kSZ stacking has a substantial history as a probe of cosmic baryons and large-scale flows. \citet{Lim2020} applied the technique to ${\sim}40{,}000$ galaxy groups using independently reconstructed peculiar velocities, detecting the kSZ signal across a wide range of halo masses to constrain their baryon content. \citet{Tanimura2021} introduced the velocity-weighted estimator we adopt here, reconstructing velocities from the SDSS galaxy density field via linear theory, while \citet{Tanimura2022} improved these estimates using a convolutional neural network trained on hydrodynamic simulations. In a complementary field-level approach, \citet{2020JCAP...12..011N} built a kSZ likelihood that marginalises over an ensemble of \borg-reconstructed velocity fields, propagating velocity-reconstruction uncertainties directly into the inferred signal. Methodologically we follow \citet{Tanimura2021}, adopting their velocity-weighted stacking estimator, but we assign velocities from a Bayesian field-level inference rather than from linear theory or a CNN---sharing with \citet{2020JCAP...12..011N} the use of \borg-style reconstructed velocities and the propagation of reconstruction uncertainty into the measurement. We differ from the latter in drawing velocities from full $N$-body resimulations of the \manticoredeep\ posterior, and in propagating the inference uncertainty by repeating the stack across posterior realisations rather than through a marginalised ensemble likelihood. We give detailed quantitative comparisons with these analyses later in this section.

We select galaxy groups and clusters from the WHL catalogue \citep{Wen2012,Wen2015} over the full redshift range of the \manticoredeep\ constraints ($0 < z < 0.7$), applying a richness threshold of $R_L > 20$ (approximate halo mass ${\sim}3\times10^{13}$~\msolh; \citealt{Wen2015}).
Of the $119{,}196$ WHL clusters in this redshift range, $77{,}901$ pass the richness cut, and $64{,}750$ ($83$~per~cent) survive our composite \borg--\textit{Planck} sky mask (\cref{fig:borg_mask_raw}) to enter the final stacked sample.
For each cluster we assign a line-of-sight peculiar velocity from the \manticoredeep\ posterior resimulations (using a velocity grid resolution of $256^3$ cells, i.e.\ $16$~\mpch\ per voxel), extract a square \textit{Planck} 217~GHz temperature patch centred on the cluster position (chosen as the null frequency of the tSZ effect; see \cref{sect:ksz_stacking}), and rescale it to units of $r/R_{500}$.
The patches are then combined using the velocity-weighted estimator of \citet{Tanimura2021}, reproduced in \cref{eq:ksz_estimator}, to produce a radial stacked profile.
For the fiducial analysis we use velocities sampled on a $256^3$ grid, matching the large-scale coherent flows that are robustly constrained by the inference and dominate the kSZ stacking signal.
Finer velocity grids probe smaller-scale modes that are less well constrained by the posterior and lead to weaker, noisier stacks, as shown in \cref{sect:appendix_ksz}.
The richness threshold reflects the physical requirement that clusters must host a sufficiently hot intracluster medium to produce a measurable kSZ signal; low-mass systems contribute negligible optical depth and act primarily as noise.
We find that the detection significance is sensitive to the cluster population and velocity-assignment scale, as explored through the redshift window, richness threshold, velocity grid resolution, and velocity S/N cuts in \cref{sect:appendix_ksz}.

\begin{figure*}
    \centering
    \begin{subfigure}{\textwidth}
        \centering
        \includegraphics[width=\textwidth]{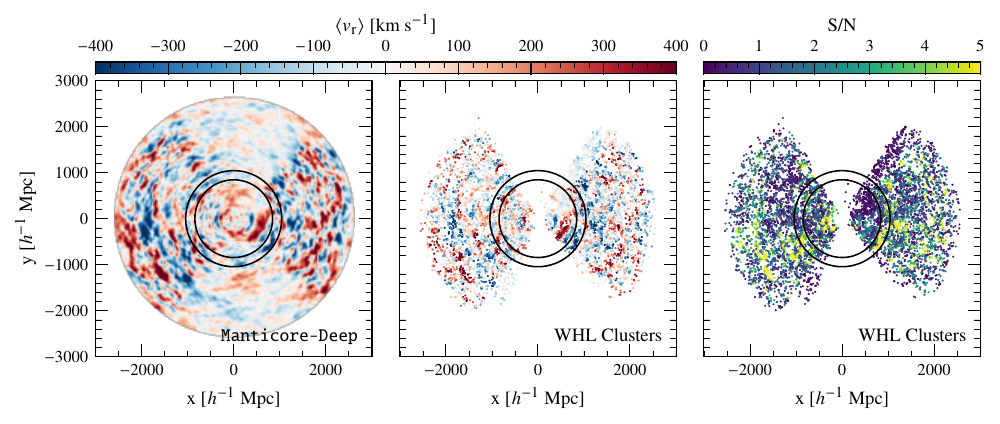}
        \caption{
        A $60$~\mpch thick slice ($\pm 30$~\mpch\ about the observer) through the \manticoredeep\ posterior mean radial velocity field and WHL cluster positions in the Cartesian--equatorial plane, containing $9{,}483$ clusters.}
        \label{fig:ksz_cart}
    \end{subfigure}
    \vspace{0.5em}
    \begin{subfigure}{\textwidth}
        \centering
        \includegraphics[width=\textwidth]{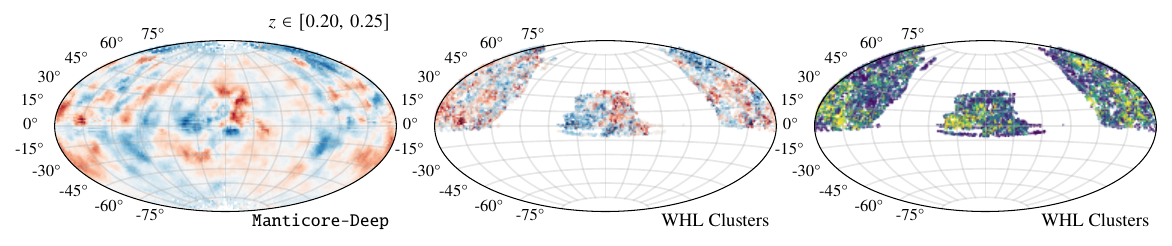}
        \caption{
        Full-sky Mollweide projection of the radial velocity field and WHL cluster positions for the redshift shell $0.20 < z < 0.25$ (indicated by the rings in the upper panel), containing $9{,}198$ clusters.
        }
        \label{fig:ksz_skymap}
    \end{subfigure}
    \caption{
        In both panels, columns show, from left to right: \textit{(i)} the posterior mean radial peculiar velocity field (positive/red = recession, negative/blue = approach); \textit{(ii)} WHL clusters coloured by their mean assigned velocity $\langle v_\mathrm{r} \rangle$, averaged across all \numrealisations\ posterior realisations; and \textit{(iii)} the same clusters coloured by the velocity S/N, $|\langle v_\mathrm{r} \rangle| / \sigma_{v}$, where $\sigma_v$ is the standard deviation across realisations. Colour bars are shared between the two panels. Two concentric rings in the upper panel mark the comoving boundaries of the redshift shell shown in the lower panel.
    }
    \label{fig:ksz_velocity_field}
\end{figure*}

\Cref{fig:ksz_velocity_field} shows the posterior mean radial velocity field from the \manticoredeep\ resimulations together with the positions of WHL clusters, coloured by their assigned radial velocity (WHL clusters falling within each slice are plotted without any richness or mask cuts applied).
The velocity field is coherent across the survey volume, with contiguous regions of approach and recession tracing the gravitational influence of overdensities in the reconstruction.
The centre panels show WHL clusters coloured by their mean assigned velocity, averaged over all \numrealisations\ posterior realisations.
Crucially, each cluster receives an independent velocity from each MCMC sample; we do not simply assign the posterior mean velocity at the cluster position.
This per-realisation assignment propagates the full inference uncertainty into the stack and enables meaningful covariance estimation across realisations.
The right panels display the per-cluster velocity S/N, defined as the absolute value of the mean assigned velocity divided by its standard deviation across realisations; a high S/N indicates that the posterior consistently assigns a similar velocity to that cluster across independent MCMC samples, reflecting greater confidence in the velocity estimate at that location.
The S/N is highest within the well-constrained central survey volume and decreases toward the survey boundaries, where fewer data constrain the density field.
Many low-S/N clusters are naturally excluded by the sky mask, but the S/N also serves as a useful diagnostic for additional quality cuts, which we explore in \cref{sect:appendix_ksz}.

\begin{figure}
    \centering
    \includegraphics[width=\columnwidth]{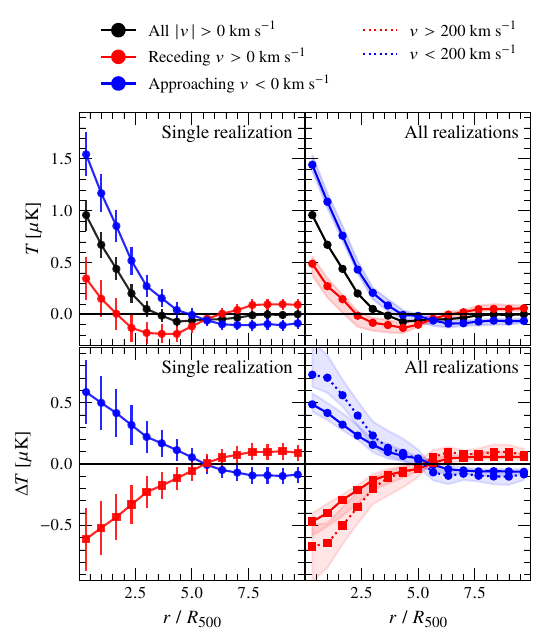}
    \caption{
    Unweighted mean stacked CMB temperature profiles and their differential for WHL clusters split by the sign of their assigned line-of-sight peculiar velocity.
    \textit{Left column:} Results from a single posterior realisation; error bars show $1\sigma$ uncertainties estimated from 1000 bootstrap resamplings of the cluster sample.
    \textit{Right column:} Median profiles across all \numrealisations\ posterior realisations; shaded bands show the 10th--90th percentile range of the per-realisation profiles.
    \textit{Top row:} Mean non-differenced stacked temperature $T$ for all clusters (black), approaching ($v_\mathrm{LOS} < 0$, blue), and receding ($v_\mathrm{LOS} > 0$, red).
    \textit{Bottom row:} The differential signal $\Delta T$ isolating the kSZ contribution; dotted curves in the right panel restrict to clusters with $|v| > 200~\mathrm{km\,s^{-1}}$, enhancing the amplitude by selecting clusters with the most confident velocity assignments.
    }
    \label{fig:ksz_approach_recede}
\end{figure}

Before applying the velocity-weighted estimator, we perform a model-independent test by splitting the cluster sample according to the sign of the assigned velocity and computing a simple unweighted mean stack for each subset.
Under \cref{eq:ksz}, receding clusters ($v_\mathrm{LOS} > 0$) imprint a cold signal ($\Delta T < 0$), while approaching clusters ($v_\mathrm{LOS} < 0$) imprint a hot signal ($\Delta T > 0$).
\Cref{fig:ksz_approach_recede} shows that approaching and receding clusters produce clearly separated profiles with equal and opposite amplitudes at the cluster centre, both for a single realisation (left column; error bars from bootstrap resampling) and in the median across all \numrealisations\ posterior realisations (right column; shaded bands showing the 10th--90th percentile spread).
Throughout, we show both a single posterior realisation and the full-ensemble summary because they capture complementary uncertainties: the single-realisation panels isolate the measurement error from the finite cluster sample, estimated by bootstrap resampling at fixed inference, and are directly comparable to kSZ stacking analyses built on a single reconstructed velocity field \citep[e.g.][]{Tanimura2021,Tanimura2022}; the ensemble panels add the posterior (inference) uncertainty as the 10th--90th percentile spread across realisations.
The top panels contain a common positive signal shared by all clusters, arising from sources uncorrelated with the velocity (e.g.\ residual cosmic infrared background emission); the bottom panels show the differential $\Delta T$, which isolates the antisymmetric kSZ contribution.
Restricting to clusters with $|v| > 200~\mathrm{km\,s^{-1}}$ (bottom-right, dotted) increases the amplitude, consistent with higher-confidence velocity assignments producing a cleaner kSZ signal.
This separation constitutes a direct, model-independent confirmation that the \manticoredeep\ velocity field is statistically aligned with the true line-of-sight motions of galaxy clusters, entirely independent of the velocity-weighted estimator presented next.

\begin{figure}
    \centering
    \includegraphics[width=\columnwidth]{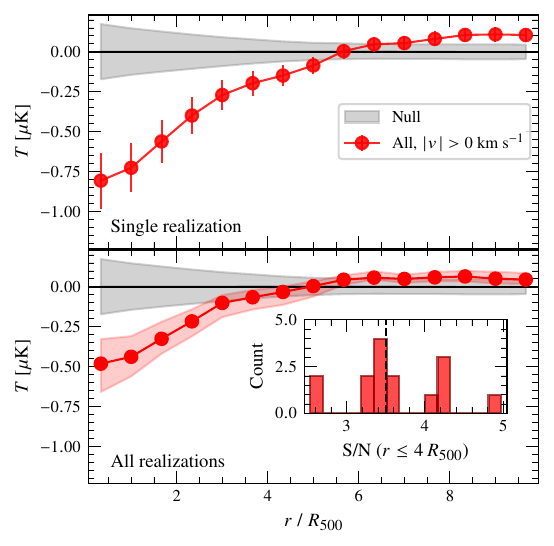}
    \caption{
    Velocity-weighted kSZ stacked temperature profile under the fiducial cluster selection (\cref{eq:ksz_estimator}).
    The red curve shows the stacked profile; the grey band shows the $1\sigma$ scatter of the null distribution from 1000 velocity-permuted realisations.
    \textit{Top:} A single posterior realisation; error bars show $1\sigma$ uncertainties from 1000 bootstrap resamplings of the cluster sample.
    \textit{Bottom:} The median profile across all \numrealisations\ posterior realisations; the shaded band shows the 10th--90th percentile range of the per-realisation profiles.
    The inset shows the distribution of detection S/N across realisations, evaluated by integrating the profile within $r \leq 4\,R_{500}$ and comparing to the velocity-permuted null; the dashed line indicates the median.
    }
    \label{fig:ksz_detection}
\end{figure}

\Cref{fig:ksz_detection} presents the main result: the velocity-weighted stacked profile under our fiducial selection.
The estimator produces a cold central decrement, with the signal returning to zero beyond ${\sim}4\,R_{500}$.
The negative sign is expected: receding clusters contribute negative $T_\mathrm{kSZ}$ weighted by positive $v$, and approaching clusters contribute positive $T_\mathrm{kSZ}$ weighted by negative $v$, so both enter the numerator of \cref{eq:ksz_estimator} with the same sign.
The signal stands clearly below the velocity-permuted null distribution, which confirms that the detection arises from the correlation between the \manticoredeep\ velocity field and the CMB temperature: randomising the velocity assignments destroys this correlation and produces no central feature.

The lower panel of \cref{fig:ksz_detection} shows the median profile across all \numrealisations\ posterior realisations, with the shaded band indicating the 10th--90th percentile range.
The detection is robust across the full posterior, with a median S/N of $3.5$ and a 10th--90th percentile range of $[2.9,\,4.3]$; every realisation exceeds $2.5\sigma$.
We quote a per-realisation detection significance, summarised by this median and spread, rather than a single value from an ensemble-combined profile: the current chain provides only \numrealisations\ realisations, too few to construct the ensemble-marginalised estimator that would ideally exploit the full posterior velocity covariance \citep[the approach of][]{2020JCAP...12..011N}. Treating each realisation as an independent measurement is the conservative alternative this sample size permits.
This significance is sensitive to the cluster selection criteria; in particular, restricting to clusters with the highest per-cluster velocity S/N yields median detection significances up to $3.9\sigma$ (see \cref{sect:appendix_ksz} and \cref{tab:ksz_sensitivity}).
This realisation-to-realisation consistency is a particularly stringent test: each MCMC sample is a statistically independent draw from the posterior, so the detection is a stable property of the \manticoredeep\ reconstruction rather than an artefact of any single realisation.

We compare our detection significance with previous applications of this estimator.
\citet{Tanimura2021} introduced the velocity-weighted kSZ stacking technique and reported a $3.5\sigma$ detection using the same WHL catalogue and \textit{Planck} 2018 217~GHz map, with line-of-sight velocities reconstructed from the SDSS galaxy density field via the linearised continuity equation.
\citet{Tanimura2022} subsequently improved the velocity estimates using a convolutional neural network (CNN) trained on hydrodynamic simulations, increasing the detection significance to ${\sim}4.7\sigma$ (${\sim}4.9\sigma$ with the more recent PR4 maps).

Our median detection significance of $3.5\sigma$ is comparable to the original \citet{Tanimura2021} result and lower than the CNN-based result of \citet{Tanimura2022}, though individual realisations span from $2.5\sigma$ to $5\sigma$, with the best cases competitive with or exceeding the CNN-based detection.
A direct quantitative comparison of detection levels is not straightforward, however, as the S/N is highly sensitive to the details of the cluster selection.
\citet{Tanimura2021} and \citet{Tanimura2022} applied stricter cuts---restricting to spectroscopic redshifts in the range $0.25 < z < 0.55$ with $M_{500} > 10^{13.5}$~\msolh---yielding a smaller but more massive sample of $30{,}431$ clusters.
Our fiducial selection uses a lower richness threshold ($R_L > 20$, corresponding to $M \sim 3 \times 10^{13}$~\msolh) over a broader redshift range ($0 < z < 0.7$), producing a sample of $64{,}750$ clusters that extends to lower masses and includes systems at the edges of the \manticoredeep\ volume where velocity constraints are weaker.
Furthermore, the three analyses differ fundamentally in how velocities are assigned: linearised perturbation theory \citep{Tanimura2021}, a simulation-trained CNN \citep{Tanimura2022}, and a full Bayesian field-level inference with $N$-body dynamics (this work).
Each method introduces different systematic uncertainties and noise properties in the velocity estimates, which propagate directly into the stacked signal.
We cannot exactly reproduce the cluster selections of \citet{Tanimura2021,Tanimura2022} within our framework, as our richness and redshift cuts differ by construction, but we explore the sensitivity of the signal to these choices in \cref{sect:appendix_ksz}.

Despite the comparable S/N, the key distinction of our approach lies in the nature of the velocity estimates: the \manticoredeep\ velocities are drawn directly from an $N$-body--based posterior that self-consistently infers the three-dimensional density and velocity fields, rather than assigning velocities through an assumed linear bias--velocity relation or an external training simulation.
Moreover, by repeating the stack independently for each of the \numrealisations\ posterior realisations, we naturally propagate the full inference uncertainty into the measurement---an advantage not available to point-estimate velocity reconstructions.
\citet{2020JCAP...12..011N} developed this idea into a rigorous statistical framework, constructing a kSZ likelihood that marginalises over an ensemble of \borg-reconstructed velocity fields, thereby folding velocity reconstruction uncertainties directly into the inferred signal amplitude.
Applying this to maxBCG clusters on the \textit{Planck} Spectral Matching Independent Component Analysis (SMICA) map, they reported ${\sim}2\sigma$ evidence for the kSZ effect---limited primarily by the small cluster sample (908 objects) and the coarse \textit{Planck} resolution---while demonstrating that neglecting velocity uncertainty can bias the measurement by ${\sim}15$~per~cent.
Ideally, we would adopt a similar ensemble-level estimator that exploits the full posterior distribution rather than treating each realisation independently.
In practice, however, such an approach requires many hundreds of independent posterior samples to robustly estimate the intra-chain velocity covariance, well beyond the $\numrealisations$ realisations available in the current \manticoredeep\ chain; we therefore defer this to future work.

Taken together, the results in this section demonstrate that the peculiar velocity field recovered by \manticoredeep\ is sufficiently accurate to coherently align the kSZ contributions of tens of thousands of galaxy groups and clusters across a cosmological volume.
The approach--recession split provides a model-independent confirmation that the inferred velocities track the true line-of-sight motions, while the velocity-weighted stack yields a statistically significant detection consistent across all posterior realisations.
This constitutes an independent validation complementary to the CMB lensing cross-correlation of \cref{sec:cross_corr}: whereas the lensing analysis tests the projected density field, the kSZ stacking directly probes the three-dimensional velocity field, providing a physically distinct test of the fidelity of the reconstruction.

\subsection{The BOSS Great Wall}
\label{sec:bgw}

We now turn to an example case study of one of the most massive structures within the reconstructed volume.
The BOSS Great Wall (BGW) is one of the largest known superstructure systems in the Universe, discovered by \citet{2016A&A...588L...4L} as an ensemble of four superclusters (labelled A--D) at $z \approx 0.47$, centred near $\mathrm{RA} \approx 162^{\circ}$, $\mathrm{Dec} \approx +50^{\circ}$.
With a reported diameter of $\sim$271~\mpch and an estimated total mass of $\sim 2 \times 10^{17}$~\msolh, the BGW ranks among the most massive and spatially extended structures identified in galaxy surveys.
Subsequent analyses by \citet{2017A&A...603A...5E} characterised the morphology and luminosity of its constituent superclusters, finding them to be exceptionally elongated compared to local counterparts, while \citet{2022A&A...666A..52E} identified eight high-density cores (HDCs) within the system and traced their dynamical evolution.
The existence of such a massive, elongated structure at $z \approx 0.5$ has been discussed as a potential tension with \lcdm, on the grounds that comparably rich and extended superclusters are rare in cosmological simulations \citep{2011MNRAS.417.2938S,2017A&A...603A...5E,2022A&A...666A..52E}.
Because each \manticoredeep\ posterior realisation is drawn from the \lcdm\ prior conditioned on the observed galaxy data, any structure recovered in the posterior is embedded in a \lcdm-consistent density field.
The posterior does not by itself provide a selection-function-corrected abundance of BGW analogues in unconstrained \lcdm; that would require a prior ensemble and a structure definition including morphology, survey volume, and look-elsewhere effects.
It does, however, allow us to ask whether the data-constrained \lcdm\ posterior recovers a coherent structure at the BGW location, and how overdense that region is compared to random lines of sight through the same reconstructed volume.

\begin{figure}
\centering
\includegraphics[width=\columnwidth]{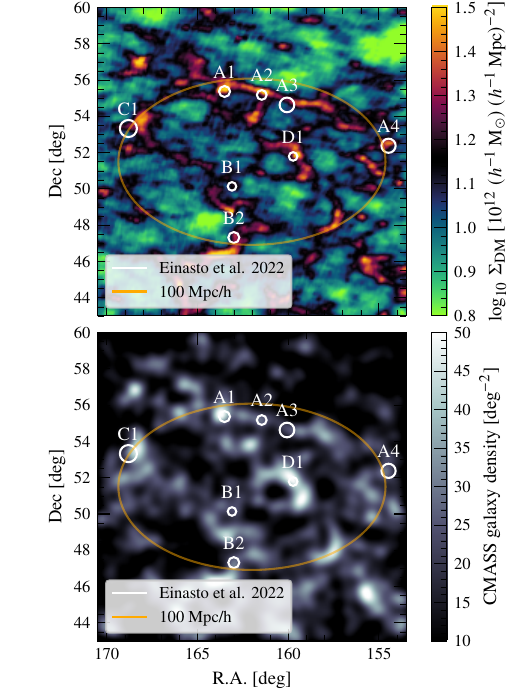}
\caption{
\textit{Top:} Projected dark matter surface mass density $\Sigma_{\rm DM}$ from the \manticoredeep posterior mean, integrated over the redshift shell $z = 0.45$--$0.50$ and shown in equatorial sky coordinates centred on the BOSS Great Wall region.
\textit{Bottom:} CMASS galaxy number density over the same angular window, smoothed with a Gaussian kernel of $\sigma = 0.3^{\circ}$.
In both panels, circles mark the positions and turnaround radii $R_{\rm T}$ of the eight HDCs identified by \citet{2022A&A...666A..52E}.
The orange ellipse indicates a projected aperture of 100~\mpch, the radius at which the BGW overdensity is most significant in our cylinder mass analysis (\cref{fig:bgw_sigma}).
}
\label{fig:bgw_image}
\end{figure}

\Cref{fig:bgw_image} presents the projected dark matter surface mass density from the \manticoredeep posterior mean alongside the smoothed CMASS galaxy density across the BGW region.
The reconstructed dark matter field reveals a coherent, elongated overdensity spanning the full extent of the reported BGW system, containing prominent filaments and density peaks.
The HDC positions catalogued by \citet{2022A&A...666A..52E} generally align well with the recovered density peaks; however, since the HDC centres are defined by the most massive galaxy in each core rather than by the local stellar mass density maximum, exact coincidence with the dark matter density peaks is not expected.
In particular, B1 does not coincide with an obvious density peak in either the reconstructed dark matter field or the CMASS galaxy map, suggesting that it may trace a less prominent concentration within the broader wall structure.
Despite this, the close spatial correspondence between the galaxy and dark matter distributions confirms that the BGW is a robust feature of the data-constrained posterior.

\begin{figure}
\centering
\includegraphics[width=\columnwidth]{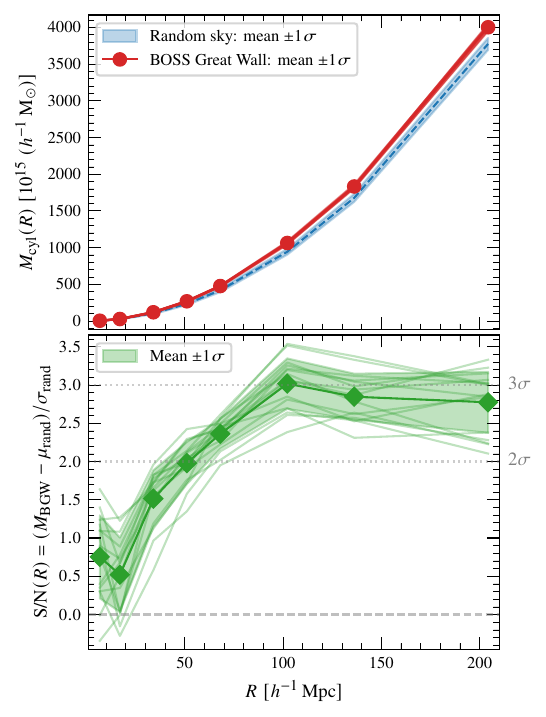}
\caption{
Cylinder mass analysis comparing the BGW region to random sky positions drawn from the same posterior realisations.
\textit{Upper panel:} Total enclosed mass $M_{\rm cyl}(R)$ within a cylinder of transverse radius $R$ and depth spanning $z = 0.43$--$0.53$, centred on the BGW (red) and averaged over 1000 random sky directions at the same redshift (blue dashed, with $\pm 1\sigma$ shading).
\textit{Lower panel:} $\mathrm{S/N}(R) = (M_{\rm BGW} - \mu_{\rm rand}) / \sigma_{\rm rand}$, the signal-to-noise ratio quantifying the number of standard deviations by which the BGW cylinder mass exceeds the random-position mean at each aperture radius.
Thin lines show individual posterior realisations; thick lines and shading denote the posterior mean and $\pm 1\sigma$ spread across \numrealisations realisations.
At $R \approx 100$~\mpch, the BGW is a $\sim\!3\sigma$ overdensity relative to the random sky.
}
\label{fig:bgw_sigma}
\end{figure}

To quantify the statistical significance of the BGW overdensity, we measure the total dark matter mass enclosed in a cylinder of varying transverse radius $R$ centred on the BGW sky position, integrating over the full reported redshift extent of the wall ($z = 0.43$--$0.53$, corresponding to a comoving depth of $\approx 230$~\mpch).
We compare this to the same measurement performed along 1000 random lines of sight drawn uniformly on the sphere, evaluated within each posterior realisation independently.
The results are summarised in \cref{tab:bgw_cyl} and shown graphically in \cref{fig:bgw_sigma}.

\begin{table}
\centering
\caption{Cylinder mass analysis of the BGW region. For each transverse aperture radius $R$, we report the total enclosed dark matter mass $M_{\rm cyl}$ for the BGW and for the mean of 1000 random sky positions at the same redshift depth ($z = 0.43$--$0.53$), together with the signal-to-noise ratio $\mathrm{S/N} = (M_{\rm BGW} - \mu_{\rm rand}) / \sigma_{\rm rand}$. Uncertainties denote the $1\sigma$ spread across \numrealisations posterior realisations.}
\label{tab:bgw_cyl}
\begin{tabular}{rccc}
\hline
$R$ & $M_{\rm BGW}$ & $M_{\rm rand}$ & S/N \\
{[\mpch]} & \multicolumn{2}{c}{[$10^{15}$~\msolh]} & \\
\hline
$6.8$   & $3.3 \pm 0.3$    & $2.9 \pm 0.6$    & $0.76 \pm 0.57$ \\
$17.0$  & $19.2 \pm 1.3$   & $17.9 \pm 2.6$   & $0.52 \pm 0.50$ \\
$34.0$  & $81.7 \pm 2.5$   & $71.4 \pm 6.8$   & $1.51 \pm 0.36$ \\
$51.1$  & $183.5 \pm 3.2$  & $160.6 \pm 11.5$  & $1.98 \pm 0.26$ \\
$68.1$  & $324.9 \pm 3.3$  & $285.5 \pm 16.7$  & $2.36 \pm 0.19$ \\
$102.1$ & $724.3 \pm 8.9$  & $642.8 \pm 27.0$  & $3.02 \pm 0.30$ \\
$136.2$ & $1248.1 \pm 11.2$ & $1142.7 \pm 37.0$ & $2.85 \pm 0.29$ \\
$204.3$ & $2724.8 \pm 22.2$ & $2570.3 \pm 55.7$ & $2.78 \pm 0.36$ \\
\hline
\end{tabular}
\end{table}

The BGW cylinder mass consistently exceeds the random-sky mean at all aperture radii.
The S/N rises from $\mathrm{S/N} \approx 1.5$ at $R \approx 35$~\mpch to a peak of $\mathrm{S/N} = 3.0 \pm 0.3$ at $R \approx 100$~\mpch, before declining gently at larger radii as the cylinder averages over increasingly uncorrelated volume.
The excess dark matter mass in the BGW cylinder relative to the random-sky mean, $\Delta M = M_{\rm BGW} - \mu_{\rm rand}$, provides a measure of the excess mass associated with the wall.
At $R = 136.2$~\mpch, closely matching the radius corresponding to the reported $\sim$271~\mpch diameter of the system, we find $\Delta M \approx 1.1 \times 10^{17}$~\msolh.
For comparison, \citet{2016A&A...588L...4L} estimated the total BGW mass as $\sim 1.6$--$2.4 \times 10^{17}$~\msolh (from stellar mass scaling and critical-density methods respectively), measured within the irregular three-dimensional supercluster boundary defined by a luminosity density threshold.
Our cylindrical excess mass is therefore of the same order, though somewhat lower, than the literature total-mass estimates.
A direct comparison is not straightforward because our measurement is an excess relative to random lines of sight in a regular cylinder, whereas \citet{2016A&A...588L...4L} measured the mass within an irregular threshold-defined supercluster boundary.
Nevertheless, the agreement in scale confirms that the \manticoredeep\ posterior recovers a substantial mass excess associated with the BGW region.

The BGW is therefore among the most overdense regions of the cosmic web at this scale and redshift, but all \numrealisations\ posterior realisations reproduce it as a coherent overdensity, confirming that such a structure is fully compatible with \lcdm.
This case study illustrates the power of constrained digital twins for evaluating claims of anomalously large structures: rather than asking whether an observed superstructure is \squotes{too big} for \lcdm---a question complicated by look-elsewhere effects and \textit{a posteriori} selection---we ask whether data-constrained \lcdm\ realisations recover it as a coherent overdensity, and for the BOSS Great Wall, they do.

\section{Discussion}
\label{sect:discussion}

A central challenge facing constrained cosmological simulations is the limited number of robust, universal benchmarks against which to assess the fidelity of a reconstruction.
In the local Universe ($z \lesssim 0.05$), the validation landscape is uniquely rich, with observations sufficiently abundant and well-resolved to permit object-by-object comparisons.
\manticorelocal\ \citep{McAlpine2025}, constrained by galaxy redshift surveys, and the velocity-based \textsc{HAMLET-PM} simulations \citep{ValadeHAMLETPM}, constrained by the Cosmicflows-4 peculiar velocity catalogue, have both validated their reconstructions against overlapping sets of individually identified galaxy clusters.
These comparisons span cluster masses, positions, and peculiar velocities, all benchmarked against observational estimates.
Both have also been benchmarked using the Bayesian evidence for the reconstructed velocity field via the ``Velocity Field Olympics'' framework \citep{Stiskalek2025} and cluster detection rates quantified probabilistically via the Local Universe Model \citep{Pfeifer2023}.
The \textsc{ELUCID} project \citep{Wang2014,Wang2016} has similarly demonstrated high halo-matching rates for the most massive clusters in the SDSS DR7 volume ($z < 0.12$).
That multiple independent groups, using fundamentally different input data and inference methodologies, can be evaluated against the same metrics enables direct cross-comparison of reconstruction fidelity, something that becomes increasingly scarce at greater depths.

The smaller volumes of local reconstructions also make hydrodynamical resimulations feasible, opening the door to physically enriched validation against thermal gas observables.
The \textsc{ELUCID} project has performed constrained hydrodynamical zoom-in simulations of the SDSS Great Wall, the Coma cluster, and a representative void region \citep{2022ApJ...936...11L}, predicting gas temperatures, warm-hot intergalactic medium distributions, and shock structures that can be tested against X-ray and Sunyaev-Zel'dovich observations.
A dedicated Coma resimulation \citep{2024ApJ...966..236L} was used to calibrate feedback models against observed intracluster medium profiles.
The \texttt{SLOW} project \citep{Dolag2023,SLOW2024,SLOW_V} has taken a complementary approach, comparing the thermodynamic profiles (temperature, pressure, entropy, and electron density) of 12 local galaxy clusters against X-ray and thermal SZ observations, and linking cluster formation histories to intracluster medium core classifications.
These represent Tier~3 validations in the framework of \citet{McAlpine2025}: physically enriched observational comparisons that are only possible when volumes are small enough and resolution high enough for hydrodynamics.

Each of these advantages diminishes at the depths probed by \manticoredeep.
At $z < 0.7$, direct observations of individual clusters are sparser and less precise, peculiar velocity measurements from distance indicators are unavailable, and the inference resolution relative to the characteristic scale of structures is lower.
For nearby clusters, \cref{sect:appendix_clusters} shows that \manticoredeep\ recovers their positions and masses with increased scatter relative to \manticorelocal, a direct consequence of the coarser constraints.
The volumes involved also make full hydrodynamical resimulations impractical, removing the possibility of enriched comparisons against thermal gas observables.
Validation must therefore shift from object-level to field-level diagnostics: statistical comparisons that test the aggregate properties of the reconstructed density and velocity fields against independent observations, rather than matching individual systems.
The fact that fewer reconstruction groups have attempted constrained simulations at these depths makes establishing standardised field-level benchmarks all the more important.

The two field-level validations presented in this work, CMB lensing cross-correlation (\cref{sec:cross_corr}) and kSZ stacking (\cref{sec:ksz}), represent the first application of such diagnostics to a high-resolution deep constrained simulation \citep[though see][]{2025arXiv260202363A}.
The lensing cross-correlation tests the projected matter density field integrated along the line of sight, while the kSZ stacking directly probes the three-dimensional peculiar velocity field.
Together they constrain both the scalar (density) and vector (velocity) degrees of freedom of the reconstruction, providing physically distinct and largely independent checks.
We decompose the lensing comparison into an angular cross-power spectrum $C_\ell^{\kappa_\mathrm{Manticore}\kappa_\mathrm{CMB}}$, rather than performing a single pixel-level correlation \citep[e.g.][]{2019arXiv190906396L}.
This reveals at which angular scales the reconstruction remains faithful to the observed lensing signal and where agreement begins to degrade, providing a more informative diagnostic than a single aggregate correlation coefficient.
A similar cross-power spectrum analysis was recently applied by \citet{2025arXiv260202363A} to a \borg\ reconstruction of the Quaia quasar catalogue ($0.17 < z < 3.0$), yielding a ${\sim}\,4\,\sigma$ detection via template fitting and ${\sim}\,3\,\sigma$ via a model-independent phase-alignment test, further demonstrating the value of CMB lensing as a field-level benchmark for constrained reconstructions across a wide range of redshifts.

The detection significances obtained in this work, a conservative \lensingsnr for the lensing cross-correlation and a median of $3.5\sigma$ ($[2.9,\,4.3]$ across the posterior) for the kSZ stacking, should not be interpreted as standalone pass/fail criteria.
Their primary value is as \textit{comparative} metrics.
Any constrained simulation can produce the same predicted fields, namely convergence maps from the inferred density field and line-of-sight velocities from the reconstructed velocity field, and be evaluated under identical conditions.
As long as the cluster catalogue, sky mask, CMB map, and analysis parameters are held fixed, different reconstructions can be directly and fairly compared using the same cross-power spectra and stacked profiles.
This makes these diagnostics natural candidates for standardised benchmarks, extending to the deep Universe the tiered validation framework proposed in \citet{McAlpine2025}.
At $z < 0.7$, the Tier~2 object-level tests available in the local Universe (cluster detection rates, velocity field comparisons) are less feasible, but the CMB lensing cross-correlation and kSZ stacking fill this gap, broadening the Tier~2 concept from object-level to field-level comparisons against independent external data.

These field-level metrics do, however, have inherent limitations.
Both are integrated or stacked quantities: the lensing cross-power spectrum averages over the entire survey volume in projection, while the kSZ signal averages over tens of thousands of clusters.
A reconstruction could therefore reproduce the correct large-scale correlations while failing on smaller scales or in specific environments, particularly where the effective resolution of the inference falls below the characteristic scale of the structures being tested.
The cumulative signal-to-noise as a function of multipole $\ell$ (\cref{fig:kappa_crosscorr}) begins to address this by revealing the angular scales at which the reconstruction remains faithful.
A more granular, per-bin detection analysis would provide a sharper diagnostic, and we defer this to future work.
As shown in \cref{sect:appendix_ksz}, the kSZ detection significance is particularly sensitive to the choice of richness threshold, redshift window, and velocity grid resolution, with both the amplitude and shape of the stacked profile changing appreciably across selections; meaningful cross-study comparisons therefore require identical cluster samples and selection cuts.
Similarly, systematics in the input galaxy data, such as foreground contamination, selection effects, and survey inhomogeneities, could in principle propagate through the inference into the predicted convergence and velocity fields, biasing the validation metrics themselves.
The robust treatment of such systematics during inference, as implemented in this work, is therefore essential not only for the reconstruction but also for the trustworthiness of the field-level diagnostics derived from it.

We encourage the community to adopt a common suite of field-level benchmarks for constrained simulations at cosmological depths.
In practice, this would require publishing standardised reference datasets, including cluster catalogues with uniform selection cuts, sky masks, and filtered CMB maps, alongside the reconstructions themselves, so that different groups can evaluate their models under identical conditions.
Cross-correlation with the \textit{Planck} Compton-$y$ map \citep[thermal SZ;][]{Planck2016_XXII} offers a further Tier~2 diagnostic that traces the hot gas in groups and clusters without requiring velocity information, and could be applied to any reconstruction with a predicted density field.
\textsc{ELUCID} has already demonstrated the value of such a comparison in the local SDSS volume \citep{Wang2016}, and extending it to deeper reconstructions is straightforward.
The internal consistency metrics proposed by \citet{2024MNRAS.534.3120S}, which quantify the stability of individual halo properties across posterior realisations, provide a complementary axis of validation that assesses the \textit{precision} of a reconstruction independently of its \textit{accuracy} relative to observations.

Looking forward, several current and near-term developments will substantially sharpen the validation landscape.
Higher-resolution CMB experiments such as the Atacama Cosmology Telescope \citep[ACT;][]{Qu2024} and SPT-3G, with arcminute-resolution beams, will resolve individual cluster kSZ profiles rather than relying on stacked averages, bridging the gap between field-level and object-level validation at cosmological depths.
Tomographic galaxy lensing from \textit{Euclid} \citep{2024arXiv240513491E} and the Vera C.\ Rubin Observatory Legacy Survey of Space and Time \citep[LSST;][]{2012arXiv1211.0310L} will enable redshift-binned cross-correlations that test the reconstruction shell-by-shell, moving beyond the single projected comparison presented here.
Deeper spectroscopic surveys such as DESI \citep{DESI2016} will simultaneously extend the input data for future reconstructions and enable direct velocity comparisons at higher redshifts.
On the methodological side, the \numrealisations\ posterior samples available in the current chain are too few to robustly estimate the intra-chain velocity covariance; longer MCMC chains will enable ensemble-level kSZ inference that marginalises over this covariance, reducing bias in the inferred signal amplitude \citep{2020JCAP...12..011N}.
As the quality and volume of observational data grow, establishing how reconstructions should be compared against these data, in a fair, standardised, and quantitative way, becomes as important as the reconstructions themselves.
The field-level metrics presented here provide a scalable template for validating the next generation of constrained cosmological digital twins.

\section{Conclusions}
\label{sect:summary}

We have presented \manticoredeep, the first high-resolution, fully Bayesian field-level inference of cosmic large-scale structure across a cosmological volume of $(4~h^{-1}\mathrm{Gpc})^{3}$, reaching redshifts of $z \approx 0.7$ at ${\sim}4~h^{-1}\mathrm{Mpc}$ resolution. By jointly constraining five galaxy redshift surveys within a single hierarchical framework, \manticoredeep\ delivers posterior realisations of the Universe within the survey footprint: three-dimensional density and velocity fields, together with the primordial initial conditions from which they arose, all self-consistently linked through \lcdm\ gravitational dynamics. A novel tiled inference strategy makes this computation feasible, extending the reconstructed volume by more than an order of magnitude beyond \manticorelocal\ \citep{McAlpine2025} while preserving the Gaussian statistics and isotropy of the inferred initial conditions. The physical fidelity of the reconstruction has been validated through two independent, template-free posterior-predictive tests against observations not used in the inference---cross-correlation with \textit{Planck} CMB lensing and detection of the kinetic Sunyaev--Zel'dovich effect---which together show that the inferred density and velocity fields are simultaneously consistent with independent projected-density and velocity-sensitive observables.

The key results of this study are as follows:

\begin{itemize}

    \item \textbf{The \manticoredeep\ posterior volumes are statistically consistent with \lcdm.} The inferred initial white-noise fields exhibit flat, isotropic power spectra consistent with $P(k) = 1$, with no detectable directional artefacts from the tiled inference. The evolved matter power spectrum and reduced bispectrum at $z = 0$ agree with control \lcdm\ simulations across all resolved scales, and the halo mass function at $z = 0$ reproduces the expected abundance of collapsed structures to within ${\sim}1$~per~cent across the well-sampled mass range. These diagnostics confirm that the tiled reconstruction strategy preserves the Gaussian statistics and hierarchical structure formation expected in \lcdm\ cosmology.

    \item \textbf{The \manticoredeep\ matter field is detected in cross-correlation with \textit{Planck} CMB lensing at a conservative \lensingsnr.} The posterior-mean convergence field predicted from the \manticoredeep\ matter reconstructions exhibits a strong, positive cross-power spectrum with the \textit{Planck} PR3 CMB lensing map over $30\lesssim\ell\lesssim 300$, well above a null ensemble of data-unconstrained simulations. The cumulative detection significance reaches \lensingsnr\ using the raw composite survey mask, constituting a highly significant detection of correlated structure. Crucially, beyond the galaxy bias and selection already modelled within the inference, the cross-correlation itself introduces no further assumptions---no bias template, amplitude rescaling, or fiducial cosmological model for the cross-power template---making it a direct, parameter-free, posterior-predictive test of the physical realism of the \manticoredeep\ reconstruction.

    \item \textbf{The \manticoredeep\ velocity field yields a statistically significant detection of the kinetic Sunyaev--Zel'dovich effect.} Velocity-weighted stacking of the \textit{Planck} 217~GHz map at the positions of $64{,}750$ WHL galaxy groups and clusters produces a clear kSZ signal, with a median detection significance of $3.5\sigma$ (10th--90th percentile range $[2.9,\,4.3]\sigma$) that is consistent across all \numrealisations\ posterior realisations. A model-independent approach--recession split confirms that the inferred velocities are statistically aligned with the true line-of-sight motions. The detection significance is sensitive to the cluster population selection (\cref{sect:appendix_ksz}), underscoring the need for standardised cluster samples when comparing results across studies.
This provides a physically distinct validation of the reconstruction complementary to the CMB lensing cross-correlation: while lensing probes the projected density field, kSZ stacking directly tests the three-dimensional velocity field.

    \item \textbf{The BOSS Great Wall is recovered as a coherent overdensity consistent with \lcdm.} A cylinder mass analysis centred on the BOSS Great Wall at $z \approx 0.47$ reveals a $\sim\!3\sigma$ overdensity at an aperture radius of $\sim$100~\mpch relative to 1000 random lines of sight drawn from the same posterior realisations. At $R = 136.2$~\mpch, closely matching the radius corresponding to the reported BGW diameter, the excess enclosed mass is $\Delta M \approx 1.1 \times 10^{17}$~\msolh, of the same order as independent literature estimates despite the different mass definition. Because each posterior realisation is drawn from the \lcdm prior conditioned on the data, the recovery of the BGW as a coherent structure across all realisations confirms its compatibility with standard cosmology, demonstrating the utility of constrained digital twins for evaluating claims of anomalously large superstructures while making the structure definition explicit.

\end{itemize}

The CMB lensing cross-correlation and kSZ stacking diagnostics developed in this work constitute the first field-level benchmarks applied to a deep constrained simulation (\cref{sect:discussion}), filling the validation gap that opens at cosmological depths where object-level tests become less practical. Because these diagnostics test physically distinct degrees of freedom, projected density and three-dimensional velocity, respectively, and require no reconstruction-specific assumptions, they can be applied identically to any constrained reconstruction once the observational datasets and analysis choices are fixed. We advocate for their adoption as standardised community benchmarks, enabling fair and transparent comparison across reconstruction approaches as the next generation of galaxy surveys comes online.

Beyond the validation presented here, the \manticoredeep\ posterior ensemble opens a broad range of scientific applications. The reconstructed density field provides a natural basis for constructing posterior void catalogues with full uncertainty quantification, building convergence sightlines for Integrated Sachs--Wolfe (ISW) and CMB anomaly studies, and cross-correlating with thermal Sunyaev--Zel'dovich, galaxy lensing, and 21\,cm observations. The velocity field enables targeted searches for signals that remain undetected or only marginally detected, including the moving lens effect, in which the transverse motion of massive structures deflects CMB photons \citep{2021PhRvD.104h3529H}, as well as refined kSZ analyses that exploit the full posterior covariance. The three-dimensional cosmic web recovered by the reconstruction further supports filament identification and stacking, and principled investigation of anomalously large superstructures, as demonstrated here for the BOSS Great Wall. Forthcoming surveys, including DESI \citep{DESI2016}, \textit{Euclid} \citep{2024arXiv240513491E}, and LSST \citep{2012arXiv1211.0310L}, will simultaneously extend the input data for future reconstructions and sharpen the validation landscape through higher-resolution CMB experiments and tomographic lensing.

The initial conditions, posterior resimulations, and reduced data products from \manticoredeep\ will be made publicly available after the publication of this work via the \href{https://www.cosmictwin.org/}{Manticore Project website}.\footnote{\href{https://www.cosmictwin.org/}{https://www.cosmictwin.org/}} We encourage the community to leverage these resources for resimulations, zoom-in studies, and cross-correlation analyses, and to contribute to the development of shared benchmarks for the next generation of constrained cosmological digital twins.

\section*{Acknowledgements}

The authors thank the anonymous referee for a careful reading of the manuscript and constructive comments that helped improve this work. The authors also thank Metin Ata, without whom this work would not have been possible. We thank Florent Leclercq for useful comments and discussions, and the Virgo Consortium and the Sibelius team for their collaboration and discussions. We thank Jeger Broxterman and Matthieu Schaller for their assistance with the gravitational lensing ray-tracing code.

We acknowledge computational resources provided by the National Academic Infrastructure for Supercomputing in Sweden (NAISS), partially funded by the Swedish Research Council through grant agreement no. 2022-06725. This work used the DiRAC@Durham facility managed by the Institute for Computational Cosmology on behalf of the STFC DiRAC HPC Facility (www.dirac.ac.uk). The equipment was funded by BEIS capital funding via STFC capital grants ST/K00042X/1, ST/P002293/1, ST/R002371/1 and ST/S002502/1, Durham University and STFC operations grant ST/R000832/1. DiRAC is part of the National e-Infrastructure. In addition, this work has made use of the Infinity Cluster hosted by Institut d'Astrophysique de Paris, and was granted access to the HPC resources of TGCC (Très Grand Centre de Calcul), Irene-Joliot-Curie supercomputer, under the allocations A0170415682, SS010415380 and A0190415682. This research utilized the Sunrise HPC facility supported by the Technical Division at the Department of Physics, Stockholm University. We acknowledge the National Academic Infrastructure for Supercomputing in Sweden (NAISS), partially funded by the Swedish Research Council through grant agreement no. 2022-06725, for awarding this project access to the LUMI supercomputer, owned by the EuroHPC Joint Undertaking and hosted by CSC (Finland) and the LUMI consortium. The resources were provided through the LUMI Sweden Fall 2024 call under project NAISS~2024/8-6.

JJ, GL, and LD acknowledge support from the Simons Foundation through the Simons Collaboration on "Learning the Universe". This work was made possible by the research project grant "Understanding the Dynamic Universe," funded by the Knut and Alice Wallenberg Foundation (Dnr KAW 2018.0067). Additionally, JJ acknowledges financial support from the Swedish Research Council (VR) through the project "Deciphering the Dynamics of Cosmic Structure" (2020-05143) and GL acknowledges support from the CNRS IEA programme \dquotes{Manticore}. AL acknowledges support from the Swedish National Space Agency (Rymdstyrelsen) under Career Grant Project Dnr 2024-00171.

This work was done as part of the  \href{https://www.aquila-consortium.org/}{Aquila Consortium}.

\section*{Data Availability}

The initial conditions, posterior resimulations, and reduced data products from \manticoredeep\ are available on reasonable request to the corresponding author. Full public release, including gridded density and velocity fields, halo catalogues, and HEALPix convergence maps, will follow publication via the \href{https://www.cosmictwin.org/}{Manticore Project website}.



\bibliographystyle{mnras}
\bibliography{example} 


\appendix

\section{Galaxy Catalogue Inputs}
\label{sect:input_galaxy_catalogs}

\begin{figure*}
\centering
\begin{subfigure}[t]{0.49\textwidth}
    \caption{\tmpp}
    \centering
    \includegraphics[width=\textwidth]{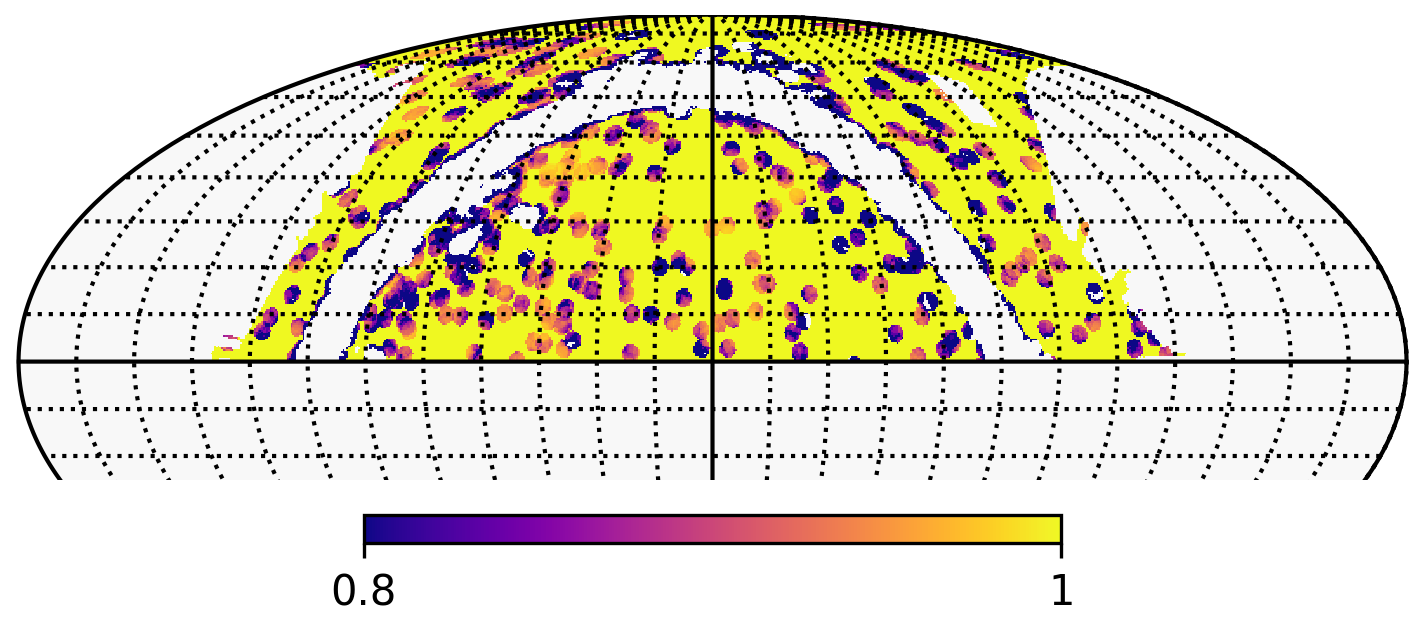}
    \label{fig:mask_2mplus}
\end{subfigure}
\hfill
\begin{subfigure}[t]{0.49\textwidth}
    \caption{2dFGRS}
    \centering
    \includegraphics[width=\textwidth]{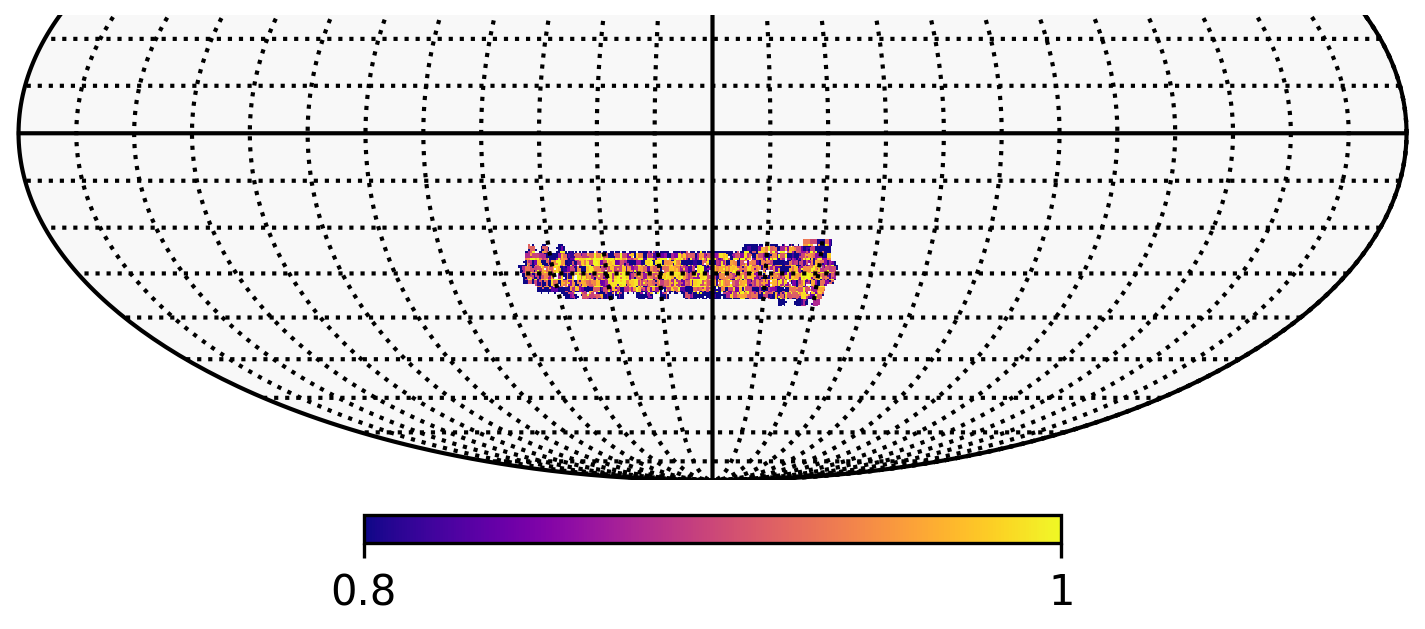}
    \label{fig:mask_2dfgrs}
\end{subfigure}

\vspace{-0.2em} 

\begin{subfigure}[t]{0.49\textwidth}
    \caption{6dFGRS}
    \centering
    \includegraphics[width=\textwidth]{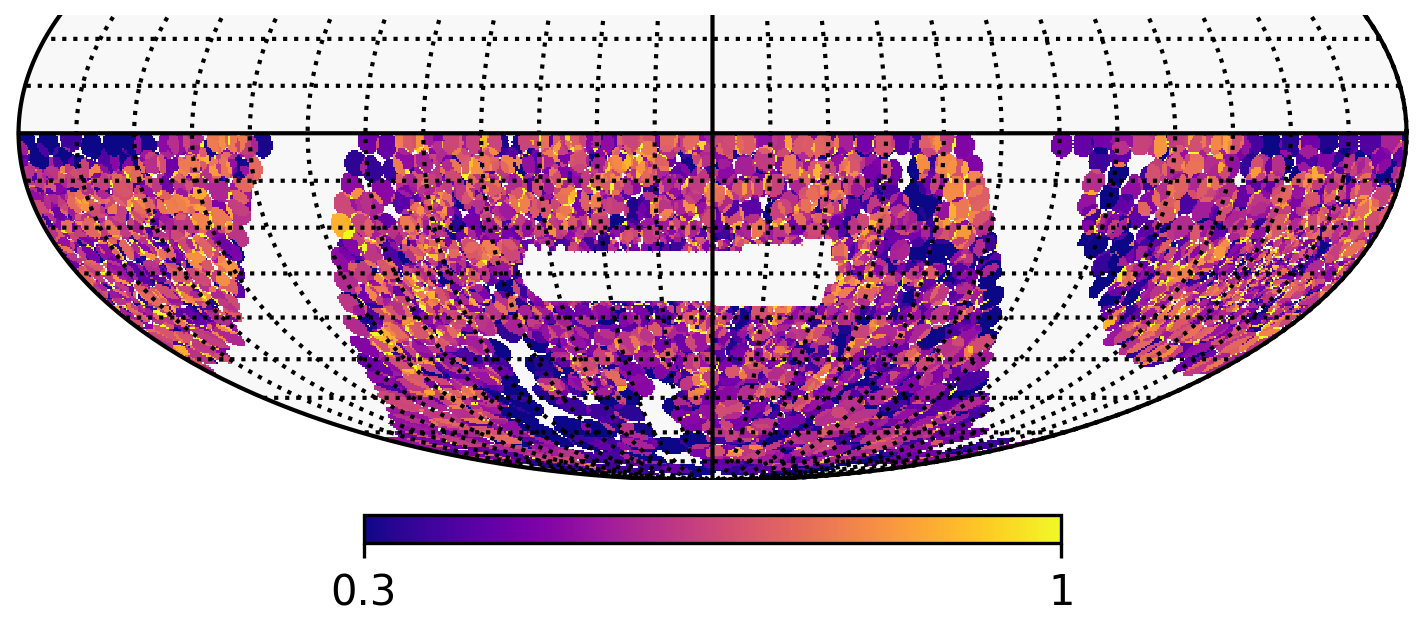}
    \label{fig:mask_6dfgrs}
\end{subfigure}
\hfill
\begin{subfigure}[t]{0.49\textwidth}
    \caption{CMASS-North}
    \centering
    \includegraphics[width=\textwidth]{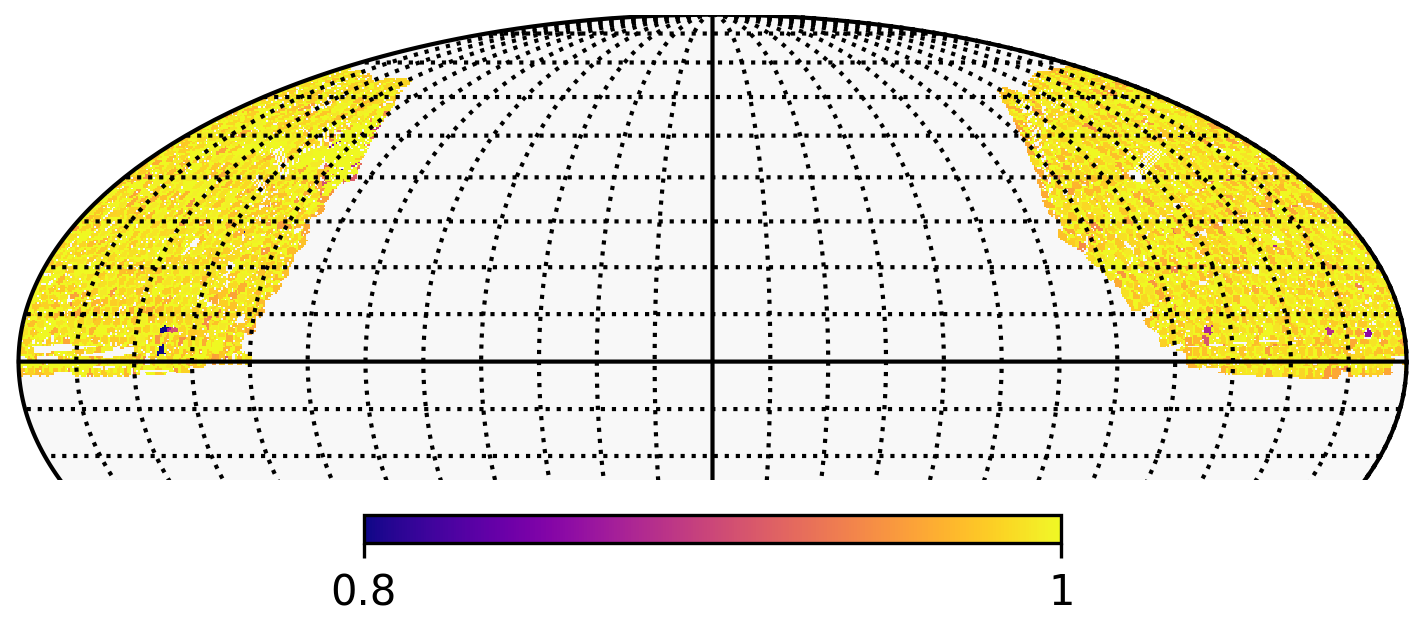}
    \label{fig:mask_cmass_north}
\end{subfigure}

\vspace{-0.2em} 

\begin{subfigure}[t]{0.49\textwidth}
    \caption{CMASS-South}
    \centering
    \includegraphics[width=\textwidth]{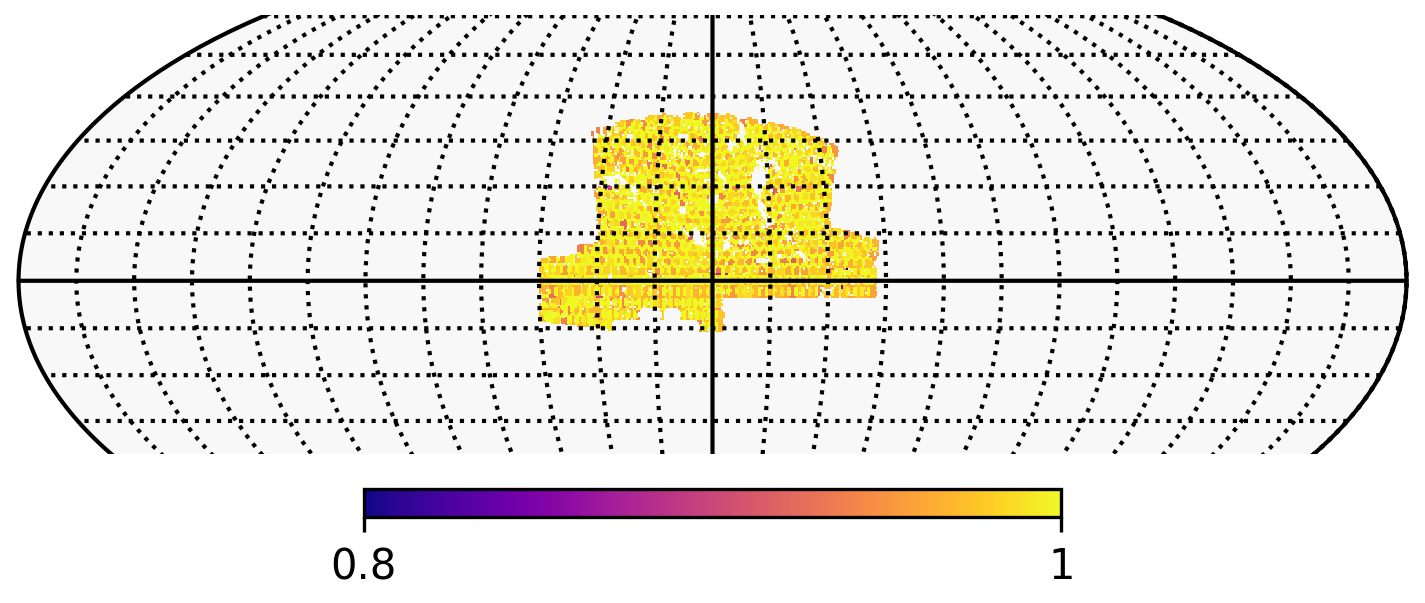}
    \label{fig:mask_cmass_south}
\end{subfigure}
\hfill
\begin{subfigure}[t]{0.49\textwidth}
    \caption{LOWZ-North}
    \centering
    \includegraphics[width=\textwidth]{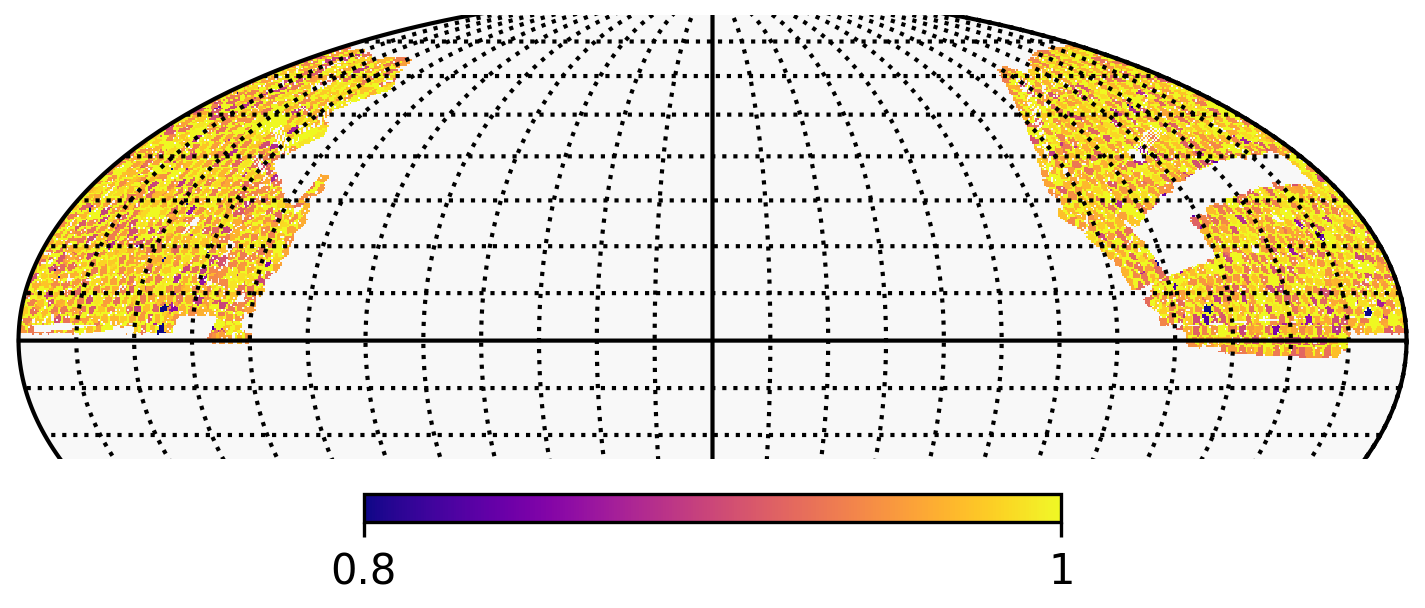}
    \label{fig:mask_lowz_north}
\end{subfigure}

\vspace{-0.2em} 

\begin{subfigure}[t]{0.49\textwidth}
    \caption{LOWZ-South}
    \centering
    \includegraphics[width=\textwidth]{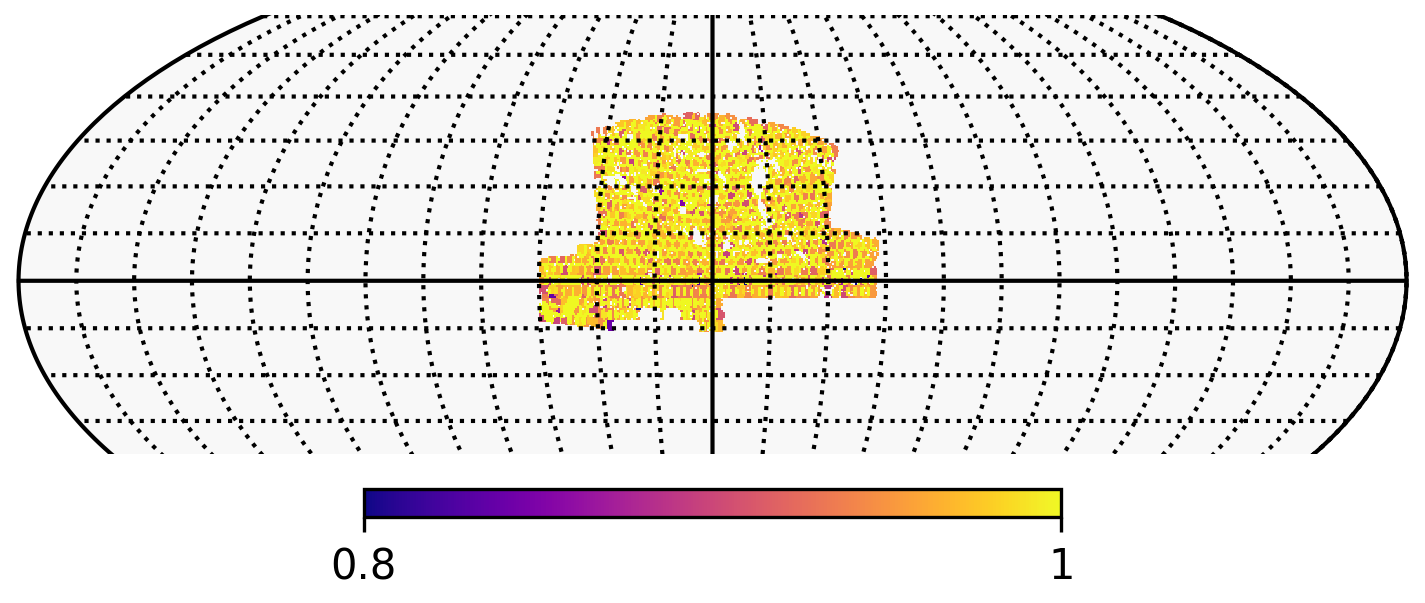}
    \label{fig:mask_lowz_south}
\end{subfigure}
\hfill
\begin{subfigure}[t]{0.49\textwidth}
    \caption{SDSS-Main}
    \centering
    \includegraphics[width=\textwidth]{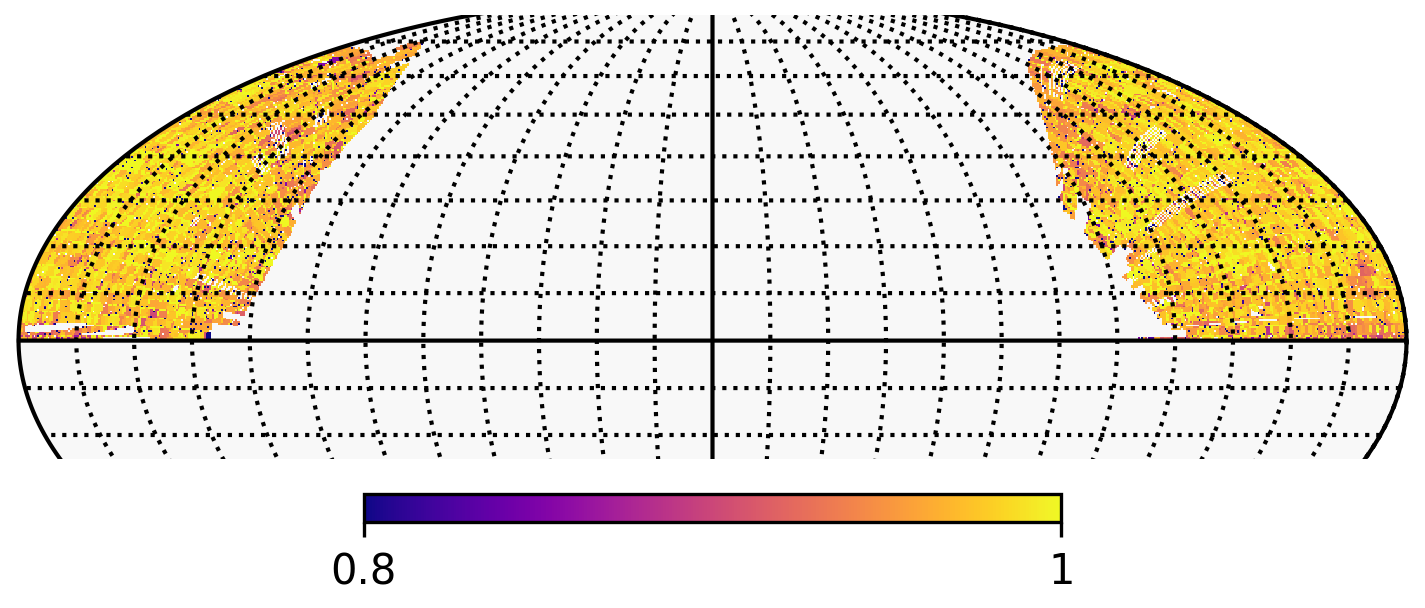}
    \label{fig:mask_sdss_main}
\end{subfigure}

\caption{Angular completeness masks used in the \manticoredeep inference, shown in the Equitorial coordinate system. Colour indicates the fractional spectroscopic completeness at each sky position after all quality cuts. Note the adjusted colour scale for 6dFGS due to its more variable fibre completeness.}

\label{fig:angular_masks}
\end{figure*}

\begin{figure}
    \includegraphics[width=\columnwidth]{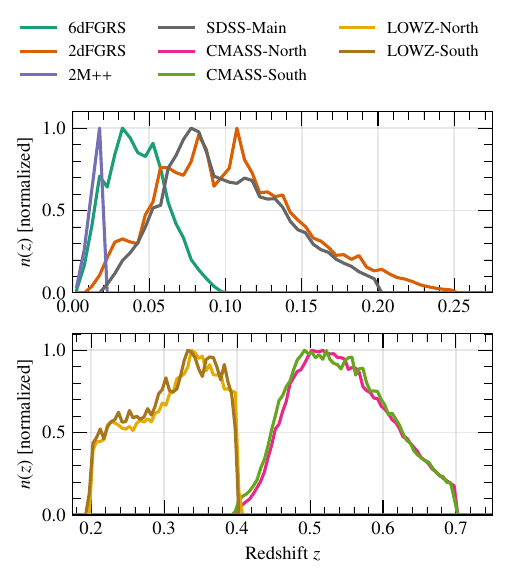}
    \caption{Normalised redshift distributions, $n(z)$, for galaxies used in the \manticoredeep\ inference. The top panel shows low-redshift surveys (\tmpp, 6dFGRS, 2dFGRS, and SDSS Main), while the bottom panel shows the BOSS samples (LOWZ North/South and CMASS North/South). Each distribution is normalised independently to highlight the redshift coverage and internal structure of each survey.}
    \label{fig:redshift_distributions}
\end{figure}

\begin{figure}
    \includegraphics[width=\columnwidth]{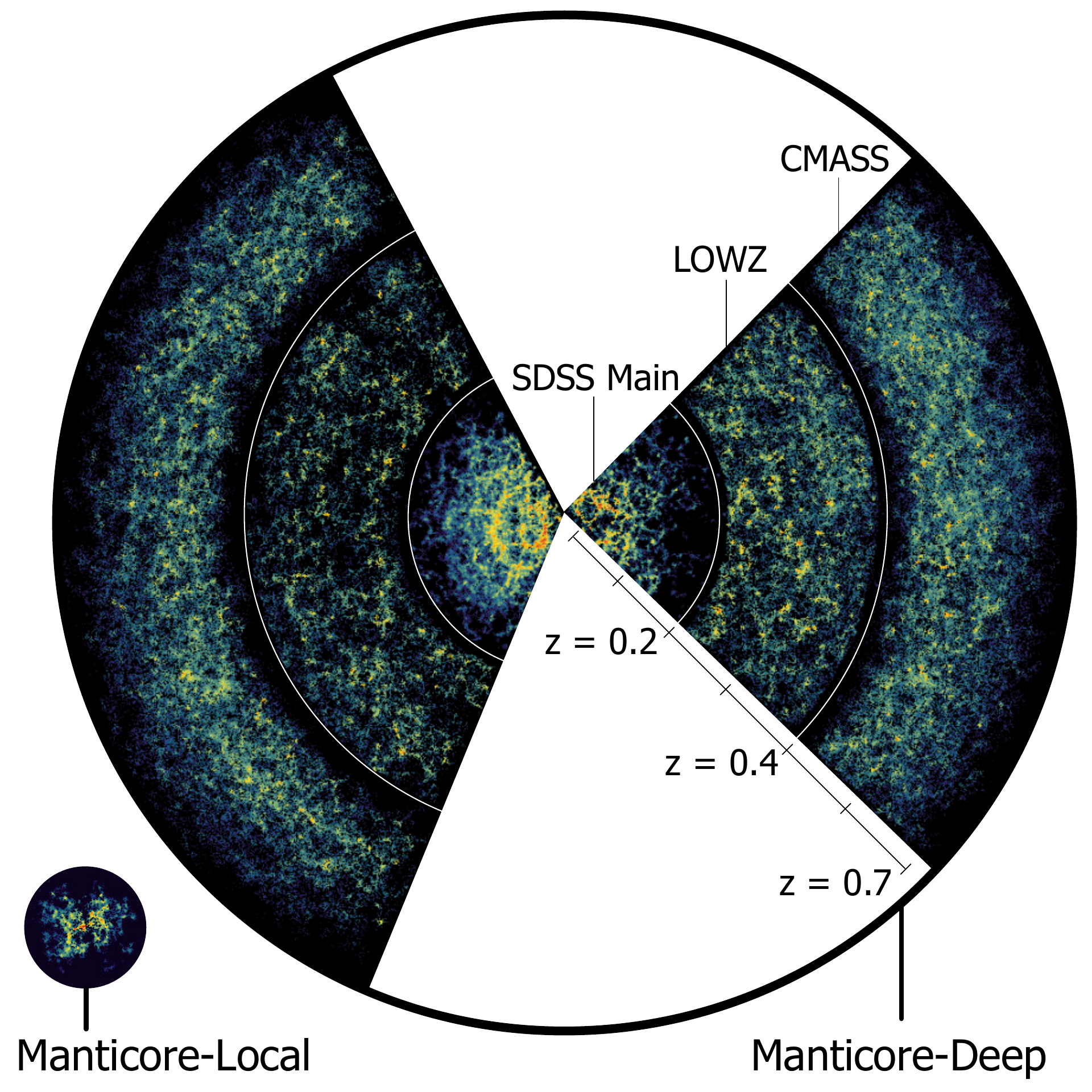}
    \caption{Galaxy distribution (coloured by galaxy density) used in the \manticoredeep\ inference within a narrow declination slice ($|\delta| < 10^\circ$). Radial rings mark redshifts $z=0.2$, $0.4$, and $0.7$. The inset shows the \manticorelocal\ volume for comparison.}

    \label{fig:galaxy_surveys}
\end{figure}

We construct the input dataset for \manticoredeep\ by combining multiple spectroscopic redshift surveys spanning $z \approx 0$ to $z \approx 0.7$, collectively providing sky coverage across both hemispheres. The \tmpp catalogue provides near-infrared coverage at low redshift, SDSS Main and 2dFGRS offer optical coverage at low-to-intermediate redshifts, 6dFGS provides complementary southern-sky coverage, and the BOSS surveys (LOWZ and CMASS) extend the dataset to higher redshifts. Our forward-modelling Bayesian framework treats each survey independently, allowing survey-specific selection functions, completeness masks, and galaxy populations. All galaxies are assigned uniquely to a single survey, ensuring strictly volume-disjoint samples without duplication in overlapping regions. Each survey is subdivided into independent subcatalogues, referred to as \squotes{\borg galaxy subcatalogues}, defined by narrow intervals in either absolute magnitude (for low-redshift surveys) or comoving distance (for BOSS), each assigned independent bias parameters. This captures population-dependent clustering variations while maintaining statistical modularity; galaxy bias modelling is described in detail in \citet{Jasche2019,McAlpine2025}. A summary of applied selection criteria and catalogue properties is provided in \cref{tab:survey_summary}, with survey-specific details in subsequent sections. The angular completeness masks, redshift distributions, and galaxy layouts are shown in \cref{fig:angular_masks,fig:redshift_distributions,fig:galaxy_surveys}.

\begin{table*}
\centering
\caption{
Summary of galaxy catalogues used in \manticoredeep. Each row lists the apparent and absolute magnitude cuts ($m$ and $M$), redshift cut ($z$), and, where applicable, comoving radial distance cut ($r$), as applied in the inference. For the low-redshift surveys (2M++, 6dFGS, 2dFGRS, SDSS Main), the radial selection function is modelled using a Schechter luminosity function with shape parameter $\alpha$ and characteristic magnitude $M^*$. For the BOSS samples (LOWZ and CMASS), the radial selection is not modelled via a luminosity function; instead, it is derived empirically from the data by binning galaxies in equally spaced comoving-distance intervals, with one bin per \borg\ galaxy subcatalogue. $N_{\rm catalogues}$ denotes the number of these subcatalogue partitions per survey (equally spaced in absolute magnitude for the low-$z$ samples, or in comoving distance for LOWZ/CMASS). $N_{\rm gal}$ gives the total number of galaxies used from each sample after applying all cuts.
}
\label{tab:survey_summary}
\begin{tabular}{lcccccccc}
\toprule
Survey & $m$ Range & $M$ Range & $z$ Range & $r$ Range [\mpch] & $N_{\rm catalogues}$ & $N_{\rm gal}$ & $\alpha$ & $M^*$ \\
\midrule
\tmpp             & $8.0 \leq m_{K} \leq 11.5$         & $-25 \leq M_{K} \leq -21$       & $0.00 \leq z \leq 0.02$   &---                  & 8  & 9,534     & $-0.94$ & $-23.28$ \\
2dFGRS           & $15.5 \leq m_{b_J} \leq 19.0$        & $-21 \leq M_{b_J} \leq -15$       & $0.01 \leq z \leq 0.25$   &---                  & 12 & 147,439   & $-1.21$ & $-19.66$ \\
6dFGRS            & $12.0 \leq m_{r_F} \leq 15.5$        & $-22 \leq M_{r_F} \leq -17$       & $0.00 < z < 0.175$   &---                  & 10 & 45,179    & $-1.21$ & $-20.98$ \\
SDSS-Main        & $14.5 \leq m_r \leq 17.5$             & $-22 \leq M_{^{0.1}r} \leq -19$   & $0.02 \leq z \leq 0.2$    &---                  & 12 & 390,759   & $-1.23$ & $-20.73$ \\
LOWZ-North       &---                                  &---                             & $0.2 \lesssim z \lesssim 0.4$   & $550 \leq r \leq 1000$   & 6  & 192,328   &---    &---\\
LOWZ-South       &---                                  &---                             & $0.2 \lesssim z \lesssim 0.4$   & $550 \leq r \leq 1000$   & 6  & 87,739    &---    &---\\
CMASS-North      &---                                  &---                             & $0.4 \lesssim z \lesssim 0.7$   & $1100 \leq r \leq 1600$  & 6  & 578,447   &---    &---\\
CMASS-South      &---                                  &---                             & $0.4 \lesssim z \lesssim 0.7$   & $1100 \leq r \leq 1600$  & 6  & 213,205   &---    &---\\
\midrule
Total   &---&---&---&---& 66 & 1,664,630 &---&---\\
\bottomrule
\end{tabular}
\end{table*}

\subsection{\tmpp Galaxy Catalogue}

The \tmpp galaxy catalogue~\citep{Lavaux2011} is a composite dataset incorporating the 2MASS Redshift Survey \citep[2MRS,][]{Huchra2012}, Sloan Digital Sky Survey Data Release Seven \citep[SDSS-DR7,][]{Abazajian2009}, and Six-Degree-Field Galaxy Redshift Survey Data Release Three \citep[6dFGRS,][]{Jones2009}. Photometry is derived from the Two-Micron-All-Sky-Survey (2MASS) Extended Source Catalogue \citep[2MASS-XSC,][]{Skrutskie2006} in the near-infrared $J$, $H$, and $K_S$ bands, minimally affected by dust extinction and stellar population differences, providing robust stellar mass tracers. For our inference, we construct a bright, volume-limited subsample with apparent magnitude limits of $8.0 \leq m_K \leq 11.5$ and absolute magnitudes in the range $-25 \leq M_K \leq -21$, restricted to redshifts $z \leq 0.02$, yielding a total of 9,534 galaxies. Galaxy magnitudes are recalculated using the $K_S$-band apparent magnitude within a circular isophote at 20 mags arcsec$^{-2}$ and corrected for Galactic extinction, cosmological surface brightness dimming, and stellar evolution as outlined by \citet{Lavaux2011}. We explicitly exclude overlapping volumes with the SDSS Main Galaxy Sample and 6dFGS to prevent double-counting, assigning each galaxy uniquely to a single input catalogue; this accounts for the reduction in input galaxies from the \tmpp catalogue relative to the \manticorelocal inference.

The radial selection function is modelled using a Schechter luminosity function in the $K$-band with parameters $\alpha = -0.94$ and $M^* = -23.28$, consistent with \citet{Lavaux2011}. The \tmpp galaxies are partitioned into eight independent \borg galaxy subcatalogues with equal spacing in absolute magnitude ($\Delta M_{K} = 0.5$).

\subsection{6dF Galaxy Survey (6dFGS)}

The 6dF Galaxy Survey (6dFGS) is a wide-area redshift survey covering most of the southern sky, specifically at declinations $\delta < 0^\circ$ and Galactic latitudes $|b| > 10^\circ$ \citep{2004MNRAS.355..747J}. It targeted galaxies primarily selected in the near-infrared $K$-band from 2MASS, but included additional redshift targets from optical SuperCOSMOS photometry. For our analysis, we construct a sample from the optically selected subset, using apparent and absolute magnitude cuts in the SuperCOSMOS photographic red band, $r_F$, which provides uniform spectroscopic completeness over our redshift and magnitude range of interest. The $r_F$ band is defined using UK Schmidt photographic plates digitised by SuperCOSMOS, corresponding to the IIIaF emulsion and RG630 filter, with an effective wavelength near 650–700\,nm. We restrict to galaxies in the range $12 \leq m_{r_F} \leq 15.5$, corresponding to the bright end of the survey’s photometric dynamic range (with a completeness limit near $r_F \sim 16.75$), impose absolute magnitude cuts of $-22 \leq M_{r_F} \leq -17$, and restrict to the redshift interval $0 < z < 0.175$, yielding a sample of 45,179 galaxies. The radial selection function is modelled using a Schechter luminosity function in the $r_F$ band with parameters $\alpha = -1.21$ and $M^* = -20.98$ \citep{jones2006}, and the sample is partitioned into 10 independent \borg galaxy subcatalogues with equal spacing in absolute magnitude ($\Delta M_{r_F} = 0.5$).

\subsection{2dF Galaxy Redshift Survey (2dFGRS)}

The 2dF Galaxy Redshift Survey \citep[2dFGRS;][]{2001MNRAS.328.1039C} is a large spectroscopic redshift survey targeting galaxies selected in the photographic $b_J$ band (centred near 450\,nm, slightly bluer than Johnson $B$), calibrated from UK Schmidt plates. The full 2dFGRS covers approximately 1,500 square degrees, divided across two declination strips and 100 random fields; for this work, we restrict to the Southern Galactic Cap strip centred at $\delta = -30^\circ$, which comprises the larger and more contiguous of the two main survey regions. The targeting completeness varies across the footprint due to fibre collisions and field coverage, but is well characterised within the Southern strip used here. We select galaxies with apparent magnitudes in the range $15.5 \leq m_{b_J} \leq 19.0$, within the spectroscopic completeness limit of $b_J \sim 19.45$, impose absolute magnitude cuts of $-21 \leq M_{b_J} \leq -15$, and restrict to redshifts $0.01 < z < 0.25$, yielding a final sample of 147,439 galaxies. The radial selection function is modelled using a Schechter luminosity function in the $b_J$ band with parameters $\alpha = -1.21$ and $M^* = -19.66$ \citep{2002MNRAS.336..907N}, and the sample is partitioned into 12 independent \borg galaxy subcatalogues with equal spacing in absolute magnitude ($\Delta M_{b_J} = 0.5$).

\subsection{SDSS Main Galaxy Sample}

The Sloan Digital Sky Survey (SDSS) Main Galaxy Sample~\citep{strauss2002} is a spectroscopic survey targeting galaxies brighter than $r = 17.77$, selected from photometry in the SDSS $ugriz$ system, covering thousands of square degrees primarily in the Northern Galactic Cap. We use data from the ``Sample dr72'' variant (Safe0 variant) of the NYU Value-Added Galaxy Catalogue\footnote{\url{http://sdss.physics.nyu.edu/vagc/}} \citep[NYU-VAGC;][]{2005AJ....129.2562B}, based on the final data release (DR7) of SDSS \citep{Abazajian2009}. From this catalogue we construct a flux-limited sample with spectroscopic redshifts $0.02 < z < 0.2$, $r$-band Petrosian apparent magnitudes corrected for Galactic extinction in the range $14.5 \leq m_r \leq 17.5$, and absolute magnitudes $-22 \leq M_{^{0.1}r} \leq -19$. The absolute magnitudes are $K$-corrected to $z=0.1$ using the code of \citet{BLANTON2003A,BLANTON2007} and the luminosity evolution model of \citet{BLANTON2003}, yielding a sample of 390,759 galaxies.

The NYU-VAGC provides a detailed spectroscopic completeness mask covering an effective survey area of approximately 6,437~deg$^2$, divided into subareas (``polygons'') each assigned a completeness value representing the fraction of photometric targets with usable spectra; the average completeness across our sample is approximately 0.92 (see \cref{fig:angular_masks}). We model the radial selection function using a Schechter luminosity function in the $r$-band with parameters $\alpha = -1.23$ and $M^* = -20.73$ \citep{BLANTON2003}, and apply completeness weights to correct for spectroscopic incompleteness from fibre collisions \citep{strauss2002}. The sample is partitioned into 12 independent \borg galaxy subcatalogues, equally spaced in absolute magnitude with bin widths of $\Delta M_{^{0.1}r} = 0.25$.

\subsection{BOSS LOWZ and CMASS Samples}

The Baryon Oscillation Spectroscopic Survey (BOSS) formed part of SDSS-III \citep{2013AJ....145...10D}, targeting luminous galaxies in two primary spectroscopic samples: LOWZ, which extends the SDSS Main Galaxy Sample to higher redshift and lower luminosities, and CMASS, selected to maintain an approximately constant stellar mass threshold with redshift. We use data from the final data release DR12, covering approximately 10,000~deg$^2$ in the Northern and Southern Galactic caps. The LOWZ sample covers $z \approx 0.2$–$0.4$, containing 192,328 galaxies in the Northern Galactic Cap (NGC) and 87,739 in the Southern Galactic Cap (SGC); CMASS spans $z \approx 0.4$–$0.7$, with 578,447 galaxies in NGC and 213,205 in SGC. We adopt an empirical $N(z)$ description without apparent or absolute magnitude cuts, binning galaxies into comoving radial shells with no volume overlap between the LOWZ and CMASS samples.

LOWZ galaxies are binned into six comoving distance intervals between $550$ and $1000$~\mpch, while CMASS galaxies are similarly binned into six intervals between $1100$ and $1600$~\mpch, defining a total of 24 \borg subcatalogues (12 for LOWZ and 12 for CMASS, each divided between NGC and SGC). Completeness, fibre collision corrections, and systematic effects were addressed using large-scale structure catalogues from the BOSS galaxy clustering working group \citep{Anderson2014, Reid2015} and \textsc{mangle} software to create corresponding \textsc{healpix} maps at $N_{\mathrm{side}}=2048$ (see \cref{fig:angular_masks}). This empirical approach incorporates the BOSS samples into our inference framework without relying on a parametric model for the selection function.

\section{MCMC convergence and sampling diagnostics}
\label{sect:appendix_convergence}

\begin{figure}
    \centering
    \includegraphics[width=\columnwidth]{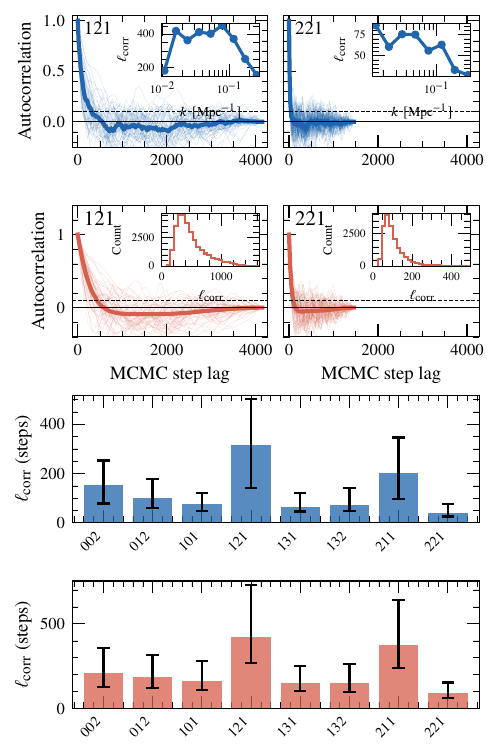}
    \caption{MCMC sampling diagnostics for the \manticoredeep\ tile inferences. \textit{Top row:} autocorrelation functions of individual modes of the final-density power spectrum for two representative subvolumes, tile 121 and tile 221, chosen to span the range of data content (and hence mixing efficiency) across the inferred tiles. Tile 121 contains 208\,313 galaxies (predominantly SDSS Main) and exhibits the longest correlation lengths of the eight representative subvolumes; tile 221 contains 39\,823 galaxies (predominantly 2dFGRS) and exhibits the shortest. Faded lines show individual modes, bold curves the median, and the dashed horizontal line marks the autocorrelation threshold of 0.1 used to define the correlation length $\ell_{\mathrm{corr}}$. Insets show the correlation length as a function of wavenumber $k$. \textit{Second row:} the same, for individual voxels of the initial white-noise field within the data-constrained regions; insets show the distribution of voxel correlation lengths. \textit{Bottom two rows:} median correlation length for power-spectrum modes (blue) and density voxels (red) across eight representative subvolumes spanning the range of survey footprints; error bars span the 16th--84th percentile range.}
    \label{fig:mcmc_diagnostics}
\end{figure}

The computational cost of the \manticoredeep\ inference ($0.5$--$1.5\times10^{6}$ CPU hours per tile) prohibits running multiple chains per subvolume, so cross-chain diagnostics such as the Gelman--Rubin statistic employed for \manticorelocal\ \citep{McAlpine2025} are not available here. Each of the 27 data-containing tiles is instead sampled by a single chain, following the structured burn-in procedure established in \citet{McAlpine2025}, and convergence is assessed through the internal diagnostics described below.

During burn-in the negative log-likelihood improves steadily as the sampler moves from its initial random state towards regions of high posterior probability; convergence to the typical set is signalled by the transition to a regime of stochastic but bounded fluctuations. This stabilisation is characteristic of high-dimensional HMC sampling, where the chain explores a thin high-probability shell rather than continuing towards the peak of the distribution, and we use it to define the end of the burn-in phase for each tile. Judged by this criterion, burn-in times across the 27 data-containing subvolumes range from 2500 to 4500 MCMC steps. This variation tracks the data density of the subvolume, as do all of the sampling timescales discussed in this appendix: burn-in time, correlation length, and total runtime. Data densities span two orders of magnitude across the inferred tiles, ranging from $2.5\times10^{-6}$ to $3.1\times10^{-4}$ galaxies $(h^{-1}\,\mathrm{Mpc})^{-3}$ (tiles 123 and 111, respectively), with a median of $3.2\times10^{-5}$. The most demanding subvolumes are the densest, those nearest the observer that fall under the SDSS Main footprint---the input catalogue with the highest galaxy counts and data density---which are also evolved to the lowest redshifts, where the density field is most non-linear. Likelihood stabilisation alone is, however, a necessary but not sufficient indicator of convergence: the likelihood can plateau while individual degrees of freedom continue to evolve slowly. We therefore complement it with autocorrelation measurements of the inferred field itself.

Specifically, we compute the per-step autocorrelation of two sets of quantities for the post--burn-in chain of each subvolume: individual voxels of the initial white-noise field within the data-constrained regions, and individual modes of the power spectrum of the final density field. The former are the sampled initial conditions themselves---the high-dimensional block that dominates the correlation time of the chain---while the latter, being a deterministic transform of the white-noise field, probe their mixing at the level of the evolved field from which all downstream science in this work derives. We define the correlation length $\ell_{\mathrm{corr}}$ of a voxel or mode as the first MCMC step lag at which its normalised autocorrelation falls below 0.1; this matches the diagnostic suite applied to \manticorelocal\ \citep{McAlpine2025}.

\Cref{fig:mcmc_diagnostics} summarises these diagnostics. The top two rows show the autocorrelation functions for two representative subvolumes, tile 121 (208\,313 galaxies, predominantly SDSS Main) and tile 221 (39\,823 galaxies, predominantly 2dFGRS), which bracket the range of mixing behaviour: tile 121 is the slowest-mixing of the eight representative subvolumes and tile 221 is the fastest, consistent with data-rich tiles imposing stronger constraints on the sampler and hence exhibiting longer correlation times. The bottom two rows show the median correlation lengths across eight representative subvolumes spanning the range of survey footprints. Median correlation lengths for power-spectrum modes range from approximately 50 to 400 steps across tiles, and for white-noise voxels from approximately 50 to 400 steps. As for \manticorelocal, the power-spectrum modes decorrelate at scale-dependent rates, with the slowest modes reaching correlation lengths of up to 400 steps in the most data-rich tiles.

Each tile chain is continued until \numrealisations\ independent posterior samples have been collected, running for approximately 750 to 6000 post--burn-in steps depending on the per-tile correlation length, corresponding to 15 correlation lengths per chain for the median white-noise voxel. The \numrealisations\ stored posterior realisations per tile are drawn at per-tile thinning intervals of 50--400 steps, matched to the measured correlation length of each subvolume, yielding effectively independent samples; equivalently, the effective sample size $N_{\mathrm{steps}} / (1 + 2\sum_{\ell}\rho_{\ell})$, with $\rho_{\ell}$ the autocorrelation at lag $\ell$, gives a median of approximately 15 per tile. We note additionally that each full-volume realisation concatenates independently drawn samples from 27 separate chains (\cref{eq:assembly}), so any residual intra-chain correlation in the slowest-mixing tiles affects only a small fraction of the parent volume and does not correlate across tile boundaries.

\section{Composite \borg--\textit{Planck} mask construction and validation}
\label{sect:appendix_mask}

\begin{figure}
    \centering
    \begin{subfigure}{\columnwidth}
        \includegraphics[width=\linewidth]{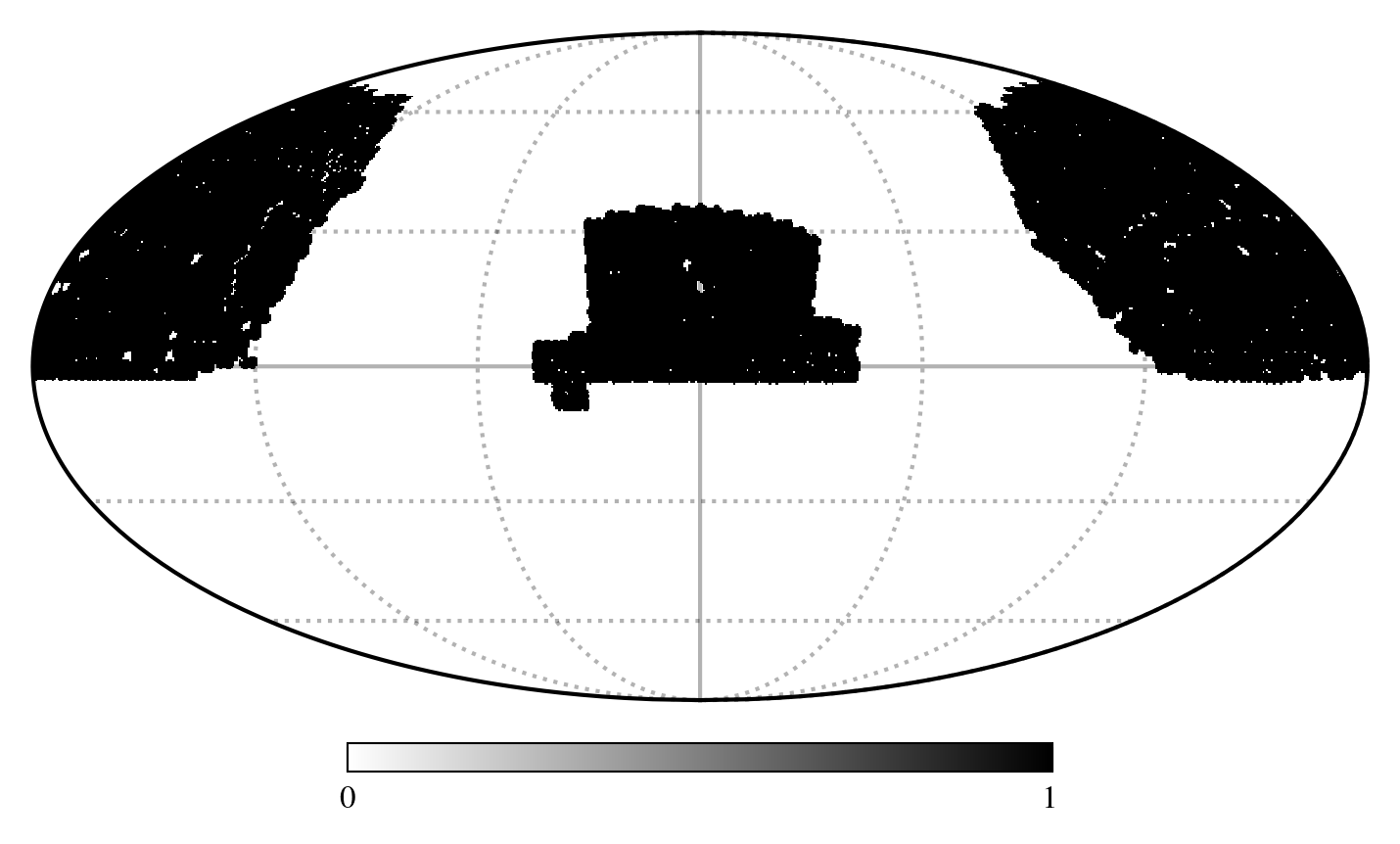}
        \caption{Raw joint \borg\ survey footprint.}
        \label{fig:borg_mask_raw}
    \end{subfigure}
    \\
    \begin{subfigure}{\columnwidth}
        \includegraphics[width=\linewidth]{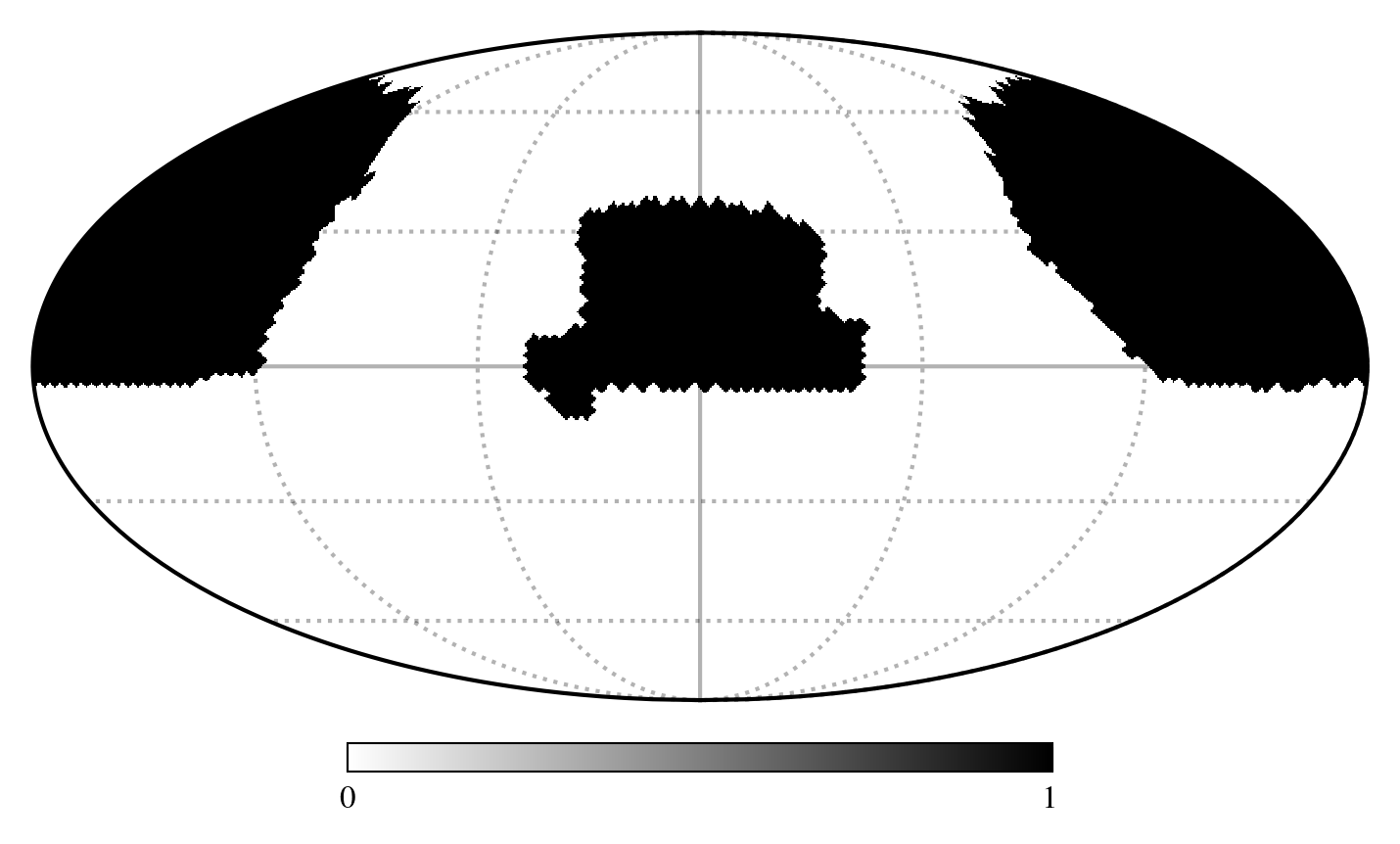}
        \caption{Hole-filled joint \borg\ survey footprint.}
        \label{fig:borg_mask_filled}
    \end{subfigure}
    \caption{Joint \borg\ survey footprint formed from the BOSS CMASS northern and southern masks, shown before (\textit{top}) and after (\textit{bottom}) filling of small-scale holes via downgrade to $\texttt{NSIDE}=32$ and upgrade back to $\texttt{NSIDE}=2048$.
    The raw footprint, multiplied by the \textit{Planck} lensing mask, defines the conservative composite \borg--\textit{Planck} mask used for the fiducial CMB lensing cross-correlation of \cref{sec:cross_corr}; the filled footprint defines the variance-reduced comparison mask.
    The raw footprint is also used for the kSZ stacking analysis of \cref{sec:ksz}, where pixel-space averaging is unaffected by small-scale mask structure.}
    \label{fig:borg_masks}
\end{figure}

The conservative CMB lensing cross-correlation of \cref{sec:cross_corr} is computed through a raw composite \borg--\textit{Planck} sky mask.
We also evaluate a variance-reduced composite mask built from a hole-filled \borg\ survey footprint, constructed as follows.
The joint \borg\ survey footprint is formed as the logical union of the BOSS CMASS northern and southern \borg\ input masks at $\texttt{NSIDE}=2048$ (top panel of \cref{fig:borg_masks}), which contains a large number of small-scale holes corresponding to pixels and narrow strips excluded by the survey pipeline.
We regularise this footprint by downgrading it to $\texttt{NSIDE}=32$, where isolated holes are absorbed by pixel averaging, upgrading the result back to $\texttt{NSIDE}=2048$, and thresholding the fractional coverage values above $0.5$ to obtain a binary mask (bottom panel of \cref{fig:borg_masks}).
The filled binary footprint is multiplied pixel-wise by the \textit{Planck} 2018 lensing mask and apodised with a $0.2^\circ$ Gaussian kernel prior to the \textsc{NaMaster} pseudo-$C_\ell$ estimation described in \cref{sect:cmb_methodology}.

The hole-filling is motivated by the impact of many sub-degree holes on the pseudo-$C_\ell$ estimator, which both spreads hole power to larger angular scales during apodisation and erodes effective sky area.
Holding every other pipeline choice fixed, the raw composite mask yields a conservative detection of \lensingsnr\ against the \textit{Planck} lensing map, compared to \lensingsnrfilled\ for the composite mask built from the filled \borg\ footprint; this roughly $50\%$ swing in recovered S/N from a single footprint-processing choice is our motivation for treating the raw-footprint result as the fiducial detection and documenting the filled-footprint comparison separately.

\begin{figure}
    \centering
    \includegraphics[width=\columnwidth]{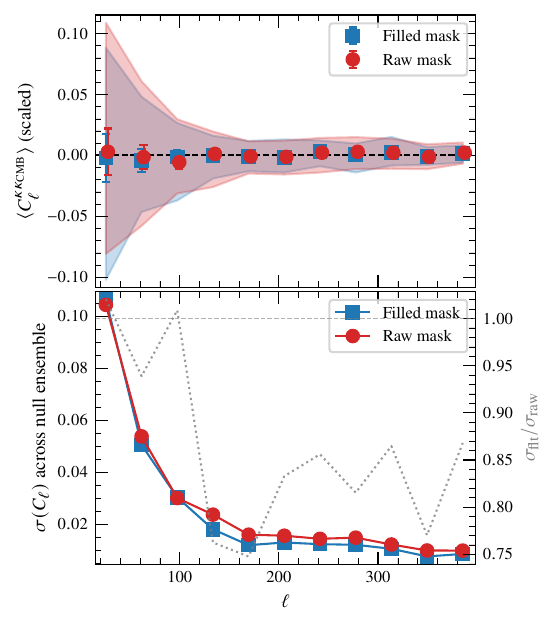}
    \caption{Null consistency test comparing composite masks built from the filled (blue) and raw (red) \borg\ survey footprints, both applied to the same ensemble of 30 rotated control maps generated from the \ncontrolsims independent low-resolution \randomlcdm\ simulations and cross-correlated with the \textit{Planck} lensing map.
    \textit{Top:} null-ensemble mean bandpowers with $1\sigma$ errors on the mean and $16$--$84$th percentile bands; both pipelines are consistent with zero at every bandpower and mutually consistent at $0.19\sigma$.
    \textit{Bottom:} per-bandpower uncertainty $\sigma(C_\ell)$ for each mask (left axis) and their ratio $\sigma_\mathrm{flt}/\sigma_\mathrm{raw}$ (right axis); using the filled \borg\ footprint reduces $\sigma(C_\ell)$ by $\sim\!14\%$ on average, with the improvement concentrated at $\ell\gtrsim 130$.}
    \label{fig:null_mask}
\end{figure}

To determine whether filling the \borg\ footprint reduces estimator noise or artificially introduces signal in unconstrained maps, we apply the filled-footprint and raw-footprint composite masks to the \ncontrolsims independent low-resolution \randomlcdm\ control simulations.
For this mask-operation null test only, each control convergence map is also evaluated under three random sky rotations, giving 30 rotated control maps in total.
These rotations provide different alignments between the unconstrained density field, the survey mask, and the \textit{Planck} lensing map, but they should not be interpreted as 30 statistically independent cosmological simulations.
Because the two mask pipelines act on the same rotated control maps, any filled-versus-raw difference can arise only from the mask operation itself.
The null-ensemble mean cross-spectrum is consistent with zero at every bandpower for both masks (worst per-bin $|\langle C_\ell\rangle|/\sigma_{\langle C_\ell\rangle}=1.39$ and $1.20$ for the filled-footprint and raw-footprint pipelines, respectively, across 11 bins), and the filled-footprint-vs-raw-footprint difference of the null means, evaluated through \cref{eq:sn_kappa}, gives $0.19\sigma$ with a bootstrap $p$-value of $0.99$: neither pipeline introduces a detectable bias relative to the other.
What differs between the two pipelines is the per-bandpower uncertainty: the composite mask built from the filled \borg\ footprint yields a $14\%$ mean reduction in $\sigma(C_\ell)$, concentrated at $\ell\gtrsim 130$ and approaching unity for the three lowest bandpowers (\cref{fig:null_mask}), consistent with the expected behaviour of a mode-coupling regularisation rather than a systematic---the largest angular scales are insensitive to the small-scale hole structure that distinguishes the two footprint choices.

Two caveats are worth noting.
The \randomlcdm\ ensemble randomises the matter field but shares the mask geometry, so the null test above only shows that the filling operation does not create a spurious signal when applied to unconstrained density fields under these rotated sky alignments.
It does not exclude the possibility that the survey-footprint regions admitted by filling overlap survey systematics (stellar density, extinction, seeing, depth) that are correlated with the constrained reconstruction and could therefore increase the measured CMB lensing signal in the posterior maps.
For this reason, we quote the raw-footprint composite-mask result as the conservative fiducial detection.
The filled-footprint-vs-raw-footprint null comparison bounds any footprint-processing contribution to the detection statistic at $0.19\sigma$ for these unconstrained controls, but it should not be read as a bound on all possible survey-systematics effects in the constrained reconstruction.
More broadly, the choice between the filled and raw \borg\ footprints is a pragmatic one, not a claim of correctness: the raw-footprint composite mask gives our conservative fiducial result, while the filled-footprint composite mask demonstrates the gain achieved when small-scale survey holes are regularised without producing a detectable bias in the available null test.

\section{Sensitivity of the kSZ detection}
\label{sect:appendix_ksz}

\begin{figure}
    \centering
    \includegraphics[width=\columnwidth]{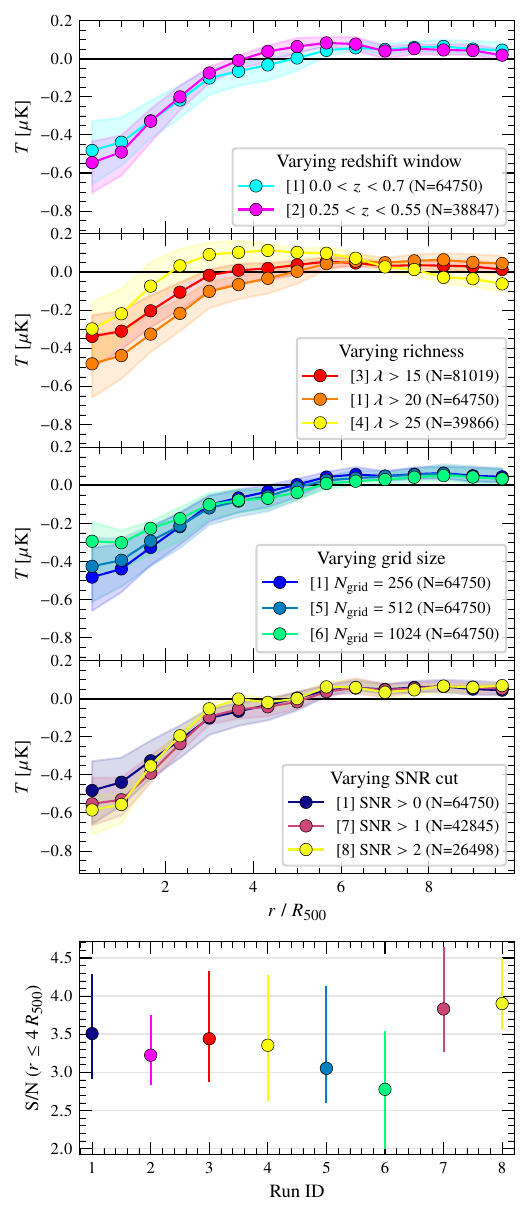}
    \caption{Sensitivity of the velocity-weighted kSZ stacked temperature profile to the analysis parameters.
    The top four panels show the median stacked profile across all \numrealisations\ posterior realisations (solid curves) with 10th--90th percentile bands (shaded regions) as a function of projected radius $r/R_{500}$.
    Each panel varies one parameter while holding the others at their fiducial values (Run~1, dark blue): redshift window (top), richness threshold $R_L$ (second), velocity grid resolution $N_\mathrm{grid}$ (third), and per-cluster velocity S/N cut (fourth).
    The number of WHL clusters surviving each selection is indicated in parentheses.
    The bottom panel summarises the detection S/N (integrated within $r \leq 4\,R_{500}$) for all eight configurations, with error bars spanning the 10th--90th percentile range across realisations.
    The velocity-permuted null distribution is omitted for clarity; it is unchanged from \cref{fig:ksz_detection}.
    Numerical values are tabulated in \cref{tab:ksz_sensitivity}.}
    \label{fig:ksz_parameter_exploration}
\end{figure}

We explore the sensitivity of the velocity-weighted kSZ detection (\cref{sec:ksz}) to four analysis parameters: the redshift window, the cluster richness threshold $R_L$, the velocity grid resolution $N_\mathrm{grid}$, and a per-cluster velocity S/N cut.
In each case we vary one parameter while holding the others at their fiducial values (Run~1: $N_\mathrm{grid} = 256$, $R_L > 20$, $0 < z < 0.7$, no S/N cut), which corresponds to the main result presented in \cref{fig:ksz_detection}.
\Cref{fig:ksz_parameter_exploration} shows the median stacked profile across all \numrealisations\ posterior realisations for each configuration, together with the integrated detection S/N evaluated within $r \leq 4\,R_{500}$, consistent with the fiducial analysis.
The velocity-permuted null distribution is omitted for visual clarity; it is unchanged from \cref{fig:ksz_detection}.
The per-cluster velocity S/N, defined in \cref{sec:ksz} as $|\langle v_\mathrm{r} \rangle|/\sigma_v$, quantifies the consistency of the velocity assignment across posterior realisations; a high S/N indicates that the inference assigns a similar velocity to a given cluster across independent MCMC samples.
\Cref{tab:ksz_sensitivity} summarises the detection significance for all eight configurations.
These variations are intended as a robustness and interpretation study rather than a trials-corrected optimisation of the detection significance; the fiducial configuration remains the one used for the main analysis.

\textit{Redshift window} (Runs~1, 2).---Restricting from the fiducial range to the narrower window $0.25 < z < 0.55$ adopted by \citet{Tanimura2021} reduces the cluster count from $64{,}750$ to $38{,}847$ and lowers the median detection significance from $3.5\sigma$ to $3.2\sigma$.
Despite fewer clusters, the narrower window shows a slightly deeper central decrement in the stacked profile, suggesting that intermediate redshifts benefit from stronger constraints in the \manticoredeep\ inference on average.
The full redshift range includes regions near $z \sim 0.7$ where the survey volume dominates but data constraints weaken, diluting the per-cluster velocity accuracy and broadening the realisation-to-realisation scatter.

\textit{Richness threshold} (Runs~3, 1, 4).---The richness cut has the most pronounced effect on the detection.
Lowering the threshold to $R_L > 15$ increases the sample to $81{,}019$ clusters but reduces the median S/N to $3.4\sigma$; raising it to $R_L > 25$ shrinks the sample to $39{,}866$ clusters with a similar S/N of $3.4\sigma$.
The fiducial $R_L > 20$ (S/N $= 3.5\sigma$) lies near the balance point between per-cluster signal strength and sample size.
At low richness, the inclusion of low-mass systems with small electron optical depths $\tau$ dilutes the kSZ signal per cluster.
At high richness, the surviving clusters are individually more massive but too few in number to overcome the statistical noise.
Both the amplitude and shape of the stacked profile change noticeably with the richness threshold, making this the parameter to which the detection is most sensitive.

\textit{Grid resolution} (Runs~1, 5, 6).---Increasing the velocity grid resolution from $256^3$ ($16$~\mpch\ per cell) to $512^3$ ($8$~\mpch) and $1024^3$ ($4$~\mpch) progressively weakens the signal, with the median S/N declining from $3.5\sigma$ to $3.1\sigma$ and $2.8\sigma$, respectively.
The cluster count is unchanged at $64{,}750$ in all three cases, since the grid resolution affects only the velocity assignment.
Finer cells resolve smaller-scale velocity fluctuations that are noisier in the posterior, whereas the kSZ signal is driven by bulk motions on scales larger than individual grid cells.
The $256^3$ grid therefore provides a velocity estimate on scales that are both relevant for the stack and robustly constrained by the inference; moving to finer grids introduces less constrained small-scale modes and weakens the evidence for coherent alignment.

\textit{Velocity S/N cut} (Runs~1, 7, 8).---Applying a per-cluster velocity S/N threshold increases the detection significance relative to the fiducial sample.
Requiring $\mathrm{S/N} > 1$ retains $42{,}845$ clusters and increases the median detection to $3.8\sigma$; a stricter cut of $\mathrm{S/N} > 2$ retains only $26{,}498$ clusters and gives a median significance of $3.9\sigma$.
These cuts preferentially select clusters in the well-constrained interior of the survey volume, where the posterior assigns velocities most consistently across realisations.
The improvement in detection significance despite reduced statistics demonstrates that removing objects with uncertain velocity assignments sharpens the coherent kSZ signal.
This motivates constructing an optimal estimator that weights each cluster by its per-realisation velocity S/N rather than applying a hard cut; however, robustly estimating the per-cluster velocity covariance requires many more posterior samples than the \numrealisations\ currently available, and we defer this to future work.

In summary, the kSZ detection significance is sensitive to the choice of cluster population: both the amplitude of the central temperature decrement and the overall shape of the stacked profile change with the selection criteria.
This complicates direct comparison of detection levels between studies; a fair comparison requires stacking the same cluster sample with identical selection cuts.
We advocate for standardised cluster selections in future kSZ benchmark analyses.

\begin{table}
    \centering
    \caption{Detection significance of the velocity-weighted kSZ signal for each parameter configuration explored in \cref{fig:ksz_parameter_exploration}. The S/N is computed by integrating the stacked profile within $r \leq 4\,R_{500}$ and comparing to the velocity-permuted null distribution. The S/N cut column gives the minimum per-cluster velocity S/N required for inclusion in the stack. Columns give the median, 10th, and 90th percentile of the S/N across all \numrealisations\ posterior realisations. Run~1 is the fiducial configuration used in the main analysis.}
    \label{tab:ksz_sensitivity}
    \setlength{\tabcolsep}{3pt}
    \small
    \begin{tabular}{ccccccccc}
        \hline
        Run & $N_\mathrm{grid}$ & S/N & $R_L$ & $z$ range & $N_\mathrm{cl}$ & \multicolumn{3}{c}{S/N} \\
        & & cut & cut & & & Med. & P10 & P90 \\
        \hline
        1 & 256 & $>0$ & $>20$ & $0.0$--$0.7$  & $64{,}750$ & 3.5 & 2.9 & 4.3 \\
        2 & 256 & $>0$ & $>20$ & $0.25$--$0.55$ & $38{,}847$ & 3.2 & 2.8 & 3.8 \\
        3 & 256 & $>0$ & $>15$ & $0.0$--$0.7$  & $81{,}019$ & 3.4 & 2.9 & 4.3 \\
        4 & 256 & $>0$ & $>25$ & $0.0$--$0.7$  & $39{,}866$ & 3.4 & 2.6 & 4.3 \\
        5 & 512 & $>0$ & $>20$ & $0.0$--$0.7$  & $64{,}750$ & 3.1 & 2.6 & 4.1 \\
        6 & 1024 & $>0$ & $>20$ & $0.0$--$0.7$ & $64{,}750$ & 2.8 & 2.0 & 3.5 \\
        7 & 256 & $>1$ & $>20$ & $0.0$--$0.7$  & $42{,}845$ & 3.8 & 3.3 & 4.6 \\
        8 & 256 & $>2$ & $>20$ & $0.0$--$0.7$  & $26{,}498$ & 3.9 & 3.6 & 4.5 \\
        \hline
    \end{tabular}
\end{table}

\section{Local cluster comparison between \manticorelocal and \manticoredeep}
\label{sect:appendix_clusters}

One of the validation tests performed in \manticorelocal \citep{McAlpine2025} was to identify 14 prominent galaxy clusters in the local Universe and compare their properties with observations, using the Local Universe Model (LUM) matching framework of \citet{Pfeifer2023}. Here, we examine whether \manticoredeep---despite its coarser spatial resolution and different combination of input surveys---recovers the same structures. We restrict the comparison to the nine most massive clusters from the \manticorelocal set, i.e., those with median \masscrit $\geq 3.5 \times 10^{14}$~\msolh in \manticorelocal, as lower-mass systems are less likely to be reliably recovered at the coarser resolution of \manticoredeep. These nine clusters are: Hercules (A2199), Hercules (A2147), Perseus (A426), Coma (A1656), Abell~119, Norma (A3627), Shapley (A3571), Abell~548, and Hercules (A2063).

\begin{figure}
    \centering
    \includegraphics[width=\columnwidth]{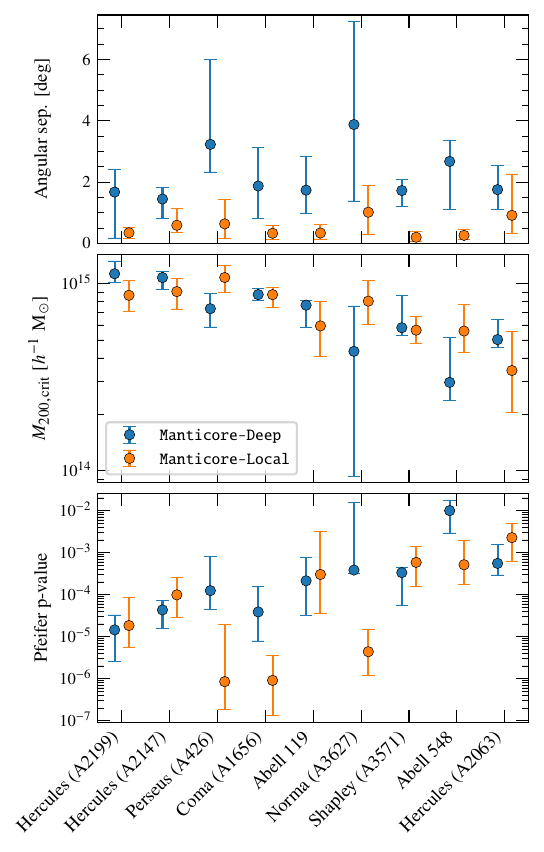}
    \caption{Comparison of prominent local clusters ($z \lesssim 0.05$) between \manticoredeep (this work) and \manticorelocal \citep{McAlpine2025}. \textit{Top}: angular separation between inferred and observed cluster positions. \textit{Middle}: halo mass \masscrit. \textit{Bottom}: \citet{Pfeifer2023} \pvalue. Points show the median across all \numrealisations\ posterior realisations; error bars span the \tentoninety percentile range. Only the nine clusters with median mass $\geq 3.5 \times 10^{14}$~\msolh in \manticorelocal are shown. Despite the lower resolution of \manticoredeep, there is general agreement between the two reconstructions across all three metrics.}
    \label{fig:cluster_comparison_three_panel}
\end{figure}

We apply the same LUM-based matching procedure as in \manticorelocal to the halo catalogues of the \manticoredeep resimulations. Because the \manticoredeep posterior comprises only \numrealisations resimulations, we report the median and \tentoninety percentile spread rather than full posterior distributions. The results are summarised in \cref{fig:cluster_comparison_three_panel}. The top panel shows the angular separation between inferred and observed cluster positions. \manticorelocal typically achieves separations of $0.5$--$1^{\circ}$, whereas \manticoredeep yields ${\sim}2$--$3^{\circ}$. These offsets are comparable to the voxel angular scale at the relevant distances (${\sim}1.5$--$4^{\circ}$ for $z \lesssim 0.05$, given the 4~\mpch cell size), indicating that the positional recovery is limited by the grid resolution rather than a systematic failure of the reconstruction. The middle panel compares inferred \masscrit values: the two reconstructions show general overlap, though \manticoredeep exhibits a larger spread. Perseus stands out as an exception, with posteriors that do not overlap as cleanly between the two runs; this may reflect its low Galactic latitude and proximity to the Zone of Avoidance, where the survey coverage entering \manticoredeep is less uniform.

The bottom panel displays the \citet{Pfeifer2023} \pvalue for each cluster. The \pvalue is systematically higher (less significant) in \manticoredeep, a direct consequence of the larger positional offsets, since the LUM \pvalue is a function of both the distance to the observed position and the halo mass. Despite this, detection significance remains comparable between the two reconstructions. Two factors contribute to the increased scatter in \manticoredeep relative to \manticorelocal. First, the input data differ substantially in the local volume. \manticorelocal relies exclusively on the all-sky \tmpp catalogue, which provides uniform and dense coverage at low redshift. \manticoredeep instead draws on a patchwork of surveys---\tmpp, SDSS, 6dFGS, and 2dFGRS---whose sky coverage, depth, and selection functions vary considerably. In regions where survey boundaries or overlap zones are suboptimal, the reconstruction is likely noisier. Second, \manticoredeep operates on a coarser inference grid (4~\mpch versus 2.66~\mpch for \manticorelocal), which limits the resolving power for individual structures. The convergence tests presented in the appendix of \citet{McAlpine2025} demonstrated that grid resolution affects mass recovery, although Perseus in particular was robust to resolution changes in that analysis---pointing to the input data composition as the more likely driver of the discrepancy here.

This comparison confirms that \manticoredeep successfully captures local, massive, individual cluster-scale structures despite its coarser grid and wider redshift coverage. Nevertheless, for detailed studies of the local Universe, \manticorelocal remains the recommended reconstruction.

\bsp	
\label{lastpage}
\end{document}